\documentclass{aa}  

\usepackage{graphicx}
\usepackage{txfonts}
\usepackage{caption}
\usepackage{lscape}
\usepackage[breaklinks=true]{hyperref} 

\hypersetup{
  colorlinks=true,   
  urlcolor=blue,     
  linkcolor=blue,     
  citecolor=blue
}

\newcommand{\msun}{\mbox{M$_{\odot}$}}

\newcommand{\lsun}{\mbox{L$_{\odot}$}}
\newcommand{\zsun}{\mbox{Z$_{\odot}$}}

\def\hii{\ion{H}{ii}}

\makeatletter
\renewcommand*\aa@pageof{, page \thepage{} of \pageref*{LastPage}}
\makeatother

\begin{document} 

  \title{Investigating episodic mass loss in evolved massive stars:}

 \subtitle{I. Spectroscopy of dusty massive stars in ten southern galaxies\thanks{Based on observations collected at the European Southern Observatory under ESO programmes 105.20HJ and 109.22W2.}}
 
   \author{A.Z. Bonanos\inst{\ref{noa}} 
        \and 
        F. Tramper\inst{\ref{leuven}, \ref{noa}}
        \and
        S. de Wit\inst{\ref{noa}, \ref{nkua}}
        \and
         E. Christodoulou\inst{\ref{noa}, \ref{nkua}}
         \and
        G. Mu\~noz Sanchez\inst{\ref{noa}, \ref{nkua}}
        \and \\
        K. Antoniadis\inst{\ref{noa}, \ref{nkua}}
        \and
        S. Athanasiou\inst{\ref{noa}, \ref{nkua}}
        \and
        G. Maravelias\inst{\ref{noa}, \ref{crete}}
        \and
        M. Yang\inst{\ref{china}, \ref{noa}}
        \and
        E. Zapartas\inst{\ref{crete}, \ref{noa}}}

   \institute{
        IAASARS, National Observatory of Athens, Metaxa \& Vas. Pavlou St., 15236, Penteli, Athens, Greece\label{noa}
        \and
 	Institute of Astronomy, KU Leuven, Celestijnenlaan 200D, 3001, Leuven, Belgium\label{leuven}
        \and
        Department of Physics, National and Kapodistrian University of Athens, Panepistimiopolis, Zografos, 15784, Greece\label{nkua}
        \and
        Institute of Astrophysics, FORTH, 71110, Heraklion, Greece\label{crete}
        \and
        National Astronomical Observatories, Chinese Academy of Sciences, Beijing 100101, People's Republic of China\label{china}
        }

   \date{}

 
  \abstract
   {Episodic mass-loss events such as giant eruptions in luminous blue variables or pre-supernova eruptions in red supergiants drastically alter the evolutionary path of a massive star, resulting in a rich and complex circumstellar environment and infrared excess. However, the incidence of these events, and hence their importance in massive star evolution, remains unknown.}
   {The ASSESS project (Episodic Mass Loss in Evolved Massive Stars: Key to Understanding the Explosive early Universe) aims to determine the role of episodic mass-loss in the evolution of massive stars. As a first step, we construct a catalog of spectroscopically identified dusty, evolved massive stars in ten southern galaxies for which {\it Spitzer} point-source catalogs are available. The resulting catalog will be used to identify stars which may have undergone an episodic mass-loss event. The target galaxies span a range of metallicities Z = 0.06--1.6 \zsun, allowing for the investigation of a potential metallicity dependence.}
   {We conducted multi-object spectroscopy of dusty massive star candidates in ten target galaxies using the Very Large Telescope (VLT). We obtained 763 spectra in WLM, NGC~55, NGC~247, NGC~253, NGC~300, NGC~1313, NGC~3109, Sextans~A, M83 and NGC~7793. The targets were selected using their {\it Spitzer} photometry, by prioritizing targets with a strong infrared excess, which indicates the presence of hot dust. We determined a spectral classification for each target. Additionally, we used archival images from the \textit{Hubble Space Telescope (HST)}, available for 150 of our targets, to provide a visual classification for 80 targets, as a star, cluster, or galaxy.}
   {We provide a catalog of 541 spectroscopically classified sources including 185 massive stars, of which 154 are newly classified massive stars. The catalog contains 129 red supergiants, 27 blue supergiants, 10 yellow supergiants, four luminous blue variable candidates, seven supergiant B[e] stars and eight emission line objects. Evidence for circumstellar dust is found in 24\% of these massive stars, based on their infrared colors. We report a success rate of $28\%$ for identifying massive stars among our observed spectra, while the average success rate of our priority system in selecting evolved massive stars was 36\%. Additionally, the catalog contains 21 background galaxies (including active galactic nuclei and quasars), 10 carbon stars and 99 \hii\ regions. We measured the line ratios [\ion {N} {ii}]/H$\alpha$ and [\ion {S} {ii}]/H$\alpha$ for 76 \hii\ regions and 36 other spectra with nebular emission-lines, thereby identifying eight sources with shocked emission.}
   {We present the largest catalog of evolved massive stars and in particular of red supergiants in nearby galaxies at low Z beyond the Local Group. The brightest and reddest of these are candidates for episodic mass loss. The fraction of dusty massive stars observed with respect to the initial selection is $\sim30\%$. We expect this catalog to trigger follow-up studies and to pave the way for a comprehensive study of the eruptive late stages of massive star evolution in the era of the \textit{James Webb Space Telescope} and the new survey telescopes (e.g. \textit{Euclid} mission, \textit{Nancy Grace Roman} Space Telescope, and \textit{Vera C. Rubin} Observatory).}


   \titlerunning{Spectroscopy of dusty, evolved massive stars in ten southern galaxies}
   \authorrunning{Bonanos et al.}

   \maketitle
%

\section{Introduction}\label{sec:intro}

The role of episodic mass loss in evolved massive stars is one of the outstanding open questions facing stellar evolution theory \citep{Smith2014}. While the upper limit to the masses of stars is thought to be 150~\msun\, \citep{Figer2005,Oey2005}, with numerous claims increasing it to 200--300~\msun\, \citep[based on R136a1;][]{Crowther2010, Banerjee2012, Brands2022,Kalari2022},
the masses of hydrogen-deficient Wolf-Rayet (WR) stars do not exceed 20~\msun\, \citep{Crowther2007}. Classical line-driven wind theory \citep{Kudritzki2000}, once thought to be responsible for removing the envelopes of massive stars, has been shown inadequate, both on theoretical grounds \citep[due to clumping,][]{Owocki1999} and estimations based on spectral lines \citep{Bouret2005,Fullerton2006,Cohen2014}, which require reductions in the mass-loss rates by a factor of $\sim$2-3. One alternative is the interaction of binary systems via Roche-Lobe overflow, which is predicted to occur in 70\% of massive stars and strip the envelopes in $\sim30\%$ of O stars in the Milky Way, given the high binary fraction ($\sim70\%$) of massive stars \citep{Sana2012, Duchene2013, Moe2017}. Episodic mass loss is possibly the dominant process that operates in all massive stars \citep[including the ones in binaries, or binary products, e.g.][]{Aghakhanloo2017,Beasor2020,Decin2023}, however, the physical mechanism responsible for it remains a mystery \citep[see, e.g., review by][]{Smith2014}.


The importance of episodic mass loss has come to the forefront in both the massive star and supernova (SN) communities. \textit{Spitzer} images have revealed numerous circumstellar shells surrounding massive, evolved stars in our Galaxy \citep{Gvaramadze2010,Wachter2010}. Episodes of enhanced mass loss have been recorded not only in luminous blue variables (LBVs), but also in extreme red supergiants \citep[RSGs, e.g. VY CMa, Betelgeuse;][]{Decin2006, Dupree2022} and yellow hypergiants \citep[e.g. IRAS 17163-3907;][]{Koumpia2020}. Moreover, many studies show evidence of circumstellar material (CSM) present around most normal Type II-P SN progenitors, at the moment of explosion \citep{Forster+2018,Morozova+2018}, pointing towards an abrupt increase of the mass loss rate synchronized with the explosion \citep{Davies+2022}. Such high mass-loss rates of SN progenitors are considered as a possible explanation of the ``RSG problem'' of not observing luminous RSGs as Type II-P SN progenitors \citep{Smartt2009,Smartt2015}. 
In cases where the pre-SN, hydrogen-rich CSM is dense enough, the interaction of the SN ejecta with it can be the main energy source of the event that would be identified as a IIn \citep[e.g.][]{Smith2017,Fox+2020}, for which LBVs have been identified as direct progenitors \citep[e.g.][]{Smith+2014}. This was the case of the remarkable SN2009ip progenitor that underwent a series of episodic mass loss events, which were identified as ``SN imposter'' events, before its final collapse \citep[e.g.][]{Mauerhan+2013,Smith+2022}. There is also evidence that at least a portion of superluminous SNe arise from interactions with CSM \citep[e.g.][]{Gal-Yam2012}, with mounting evidence suggesting that these events occur in low-metallicity host galaxies \citep[][and references therein]{Neill2011, Gal-Yam2019}.

Models of single-star evolution have adopted empirical, constant mass-loss prescriptions \citep{Meynet2015, Beasor2021} or, recently, time-averaged mass-loss rates \citep{Massey2023}, which highly influence the outcome. Recently, there have been improvements in the methodology of measuring RSG mass-loss rates depending on the stellar parameters and evolutionary phase, however, the resulting mass-loss rates differ significantly \citep{Beasor2020, Yang2023, Decin2023}.

The ASSESS project\footnote{http://assess.astro.noa.gr/} \citep{Bonanos2023} aims to tackle the role of episodic mass loss in massive stars by using the fact that mass-losing stars form dust and are bright in the mid-infrared (mid-IR). Physically, there are a number of ways a massive star can become a source of significant mid-IR emission. First, dust can form in a dense, but relatively steady stellar wind. In the most extreme cases, such as in the progenitors of the SN 2008S and the NGC300-OT 2008 transient \citep{Bond2009}, the wind is optically thick even in the near-IR and the source star is only seen in the mid-IR \citep{Prieto2008}. Second, a very massive star can have an impulsive mass ejection or eruption with dust forming in the ejected shell of material. Initially, the optical depth and dust temperatures are high, but then drop as the shell expands. The most famous example is the “great eruption” of $\eta$ Carinae in the 19th century \citep{Humphreys1994, Davidson1997, Smith2011b}, which ejected several solar masses of material. Third, the dust can be located in a circumstellar disk and emit over a broad range of temperatures, as observed in supergiant B[e] stars (sgB[e]) stars \citep{Kraus2019}.

While stars with significant mid-IR emission are intrinsically rare, many of the most interesting massive stars, such as $\eta$ Car or “Object X” in M33 \citep{Khan2011, Mikolajewska2015}, belong to this class. Therefore, we search for analogs of these interesting stars using mid-IR photometry of nearby galaxies along with the existing mid-IR “roadmaps” for interpreting luminous massive stars \citep{Bonanos2009,Bonanos2010}, which are based on known massive stars in the LMC and the SMC. These studies have identified LBVs, sgB[e], and RSGs among the brightest mid-IR sources, due to their intrinsic brightness and due to being surrounded by their own dust. Previous searches for dusty, evolved massive stars using mid-IR selection criteria have provided validation of the method \citep[see e.g.][]{Britavskiy2014,Britavskiy2015,Britavskiy2019,Kourniotis2017}.

The ASSESS team has collected published mid-IR photometric catalogs from \textit{Spitzer} of nearby galaxies within 5 Mpc with high star-formation rates: (a) seven dwarf galaxies within 1.5 Mpc from the DUSTiNGS project \citep{Boyer2015}: \object{IC 10}, \object{IC 1613}, \object{Phoenix}, \object{Peg DIG}, \object{Sextans A}, \object{Sextans B}, and \object{Wolf-Lundmark-Merlotte} (WLM), (b) 13 galaxies within 5 Mpc \citep{Khan2015,Khan2017}: \object{M31}, \object{M33}, \object{NGC 247}, \object{NGC 300}, \object{NGC 1313}, \object{NGC 2403}, \object{M81}, \object{M83}, \object{NGC 3077}, \object{NGC 4736}, \object{NGC 4826}, \object{NGC 6822}, and \object{NGC 7793}, and (c) five galaxies within 4 Mpc \citep{Williams2016}: \object{NGC 55}, \object{NGC 253}, \object{NGC 2366}, \object{NGC 4214}, and \object{NGC 5253}. The mid-IR photometry made available by the SAGE surveys of the LMC \citep{Meixner2006} and SMC \citep{Gordon2011} has been also searched for undetected, dust-obscured targets in our nearest neighbor galaxies. These catalogs contain mid-IR photometry of over 5 million point sources in 27 nearby galaxies, 19 of which have Pan-STARRS1 coverage \citep{Chambers2016}, providing an ideal dataset for a systematic study of luminous, dusty, evolved massive stars. We have compiled mid-IR photometric catalogs for these galaxies, including their counterparts in Pan-STARRS1 ($g,r,i,z,y-$bands) and other archival photometric catalogs (see Section~\ref{sec:photometry}) to construct the spectral energy distributions (SEDs) of evolved stars in these galaxies out to 24 $\mu$m. The single epoch $5\sigma$ depth of Pan-STARRS1 ranges from 22$^{nd}$ magnitude in $g-$band to 20$^{th}$ magnitude in $y-$band, corresponding to absolute magnitudes brighter than $-$6 in $g$ and $-$8 in $y$ at 3.5 Mpc, respectively, which include the most luminous, evolved targets. \citet{deWit2023} presented the results so far of applying our methodology to the Magellanic Clouds, yielding eight new RSG, a new yellow supergiant (YSG), a known sgB[e] and a candidate LBV.

Based on these catalogs, we have selected over 1000 luminous and red sources (selected by their colors in $[3.6]-[4.5$]) in these 27 galaxies and are conducting follow-up low-resolution spectroscopy. In this paper, we present the spectroscopic data obtained in ten southern galaxies and the resulting catalog of evolved massive stars. In Section~\ref{sec:observations}, we describe the target selection criteria and observations. In Section~\ref{sec:classification}, we present the spectral classification procedure and results for each galaxy. In Section~\ref{sec:discussion} we discuss the resulting catalog and in Section~\ref{sec:conclusions} we summarize the results.

\section{Observations and data reduction}
\label{sec:observations}

The aim of our observational campaign is to spectroscopically identify dusty massive stars in nearby, star-forming galaxies. This section describes the construction of photometric catalogs, and how these were used to identify potential targets. We then describe the multi-object spectroscopic (MOS) observations with the Focal Reducer and Low Dispersion Spectrograph (FORS2) on the VLT, and the data reduction process.

\subsection{Photometry catalogs}\label{sec:photometry}

To identify dusty massive stars we constructed multi-wavelength photometric catalogs for all nearby galaxies listed above, for which {\it Spitzer} \citep{werner2004} IR point-source catalogs are available. Here we describe these catalogs for the ten galaxies targeted in this paper: 
\object{WLM}, \object{NGC~55}, \object{NGC~247}, \object{NGC~253}, \object{NGC~300}, \object{NGC~1313}, \object{NGC~3109}, \object{Sextans~A},
\object{M83} and \object{NGC~7793}. These galaxies were chosen as they are visible from Paranal and contain a sufficient number of targets (see~\ref{sec:targets}) to justify MOS observations. Table~\ref{tab:galaxies} provides the properties of the galaxies, including the coordinates of each galaxy, the distance \citep[from Cosmicflows-2,][]{Tully2013}, and metallicity, as well as the radius $R$ (i.e. the size of the galaxy based on visual inspection) used to match the catalogs.

\begin{table*}
\centering
\caption{Properties of target galaxies.}\label{tab:galaxies}
\begin{tabular}{l c c c c r}
\hline\hline
Galaxy		& RA		& Dec	 	& $d$	& $Z$\tablefootmark{a}	& $R$\\
			& \small{(J2000)}	& \small{(J2000)}	& \small{(Mpc)}		& 	\small{($Z_{\sun}$)}		& \small{(\arcmin)} \\
\hline \\[-9pt]
WLM			& 00:01:58.16	& $-$15:27:39.3	& 0.96$\pm$0.05 & 0.14$^1$ & 9 \\
NGC 55		& 00:14:53.60	& $-$39:11:47.9	& 1.99$\pm$0.06 & 0.27$^{2}$	& 21 \\
NGC 247		& 00:47:08.55	& $-$20:45:37.4	& 3.52$\pm$0.06 & 0.40$^3$ 	& 14 \\
NGC 253		& 00:47:33.12 	& $-$25:17:17.6	& 3.56$\pm$0.08 & 0.72$^4$	& 21 \\
NGC 300		& 00:54:53.48	& $-$37:41:03.8	& 1.98$\pm$0.06 & 0.41$^5$ & 15 \\
NGC 1313  & 03:18:16.05	& $-$66:29:53.7	& 4.25$\pm$0.08	& 0.35$^6$	&  8\\
NGC 3109 & 10:03:06.88	& $-$26:09:34.5	& 1.37$\pm$0.06 & 0.21$^{7}$ &  13\\
Sextans A 	& 10:11:00.80	& $-$04:41:34.0	& 1.37$\pm$0.06 & 0.06$^{8}$ & 4	\\
M83			& 13:37:00.95	& $-$29:51:55.5	& 4.66$\pm$0.07 & 1.58$^{9}$	& 10 \\
NGC 7793	& 23:57:49.83	& $-$32:35:27.7	& 3.58$\pm$0.07	& 0.42$^{10}$	& 8 \\

\hline
\end{tabular}
\tablebib{
(1) \citet{Urbaneja2008}; (2) \citet{Hartoog2012}; (3) \citet{Davidge2021}; (4) \citet{Spinoglio2022}; (5) \citet{Kudritzki2008}; (6) \citet{Hadfield2007}; (7) \citet{Hosek2014}; (8) \citet{Kniazev2005}; (9) \citet{Hernandez2019}; (10) \citet{DellaBruna2021}.
}
\tablefoot{
\tablefoottext{a}{Metallicities are based on young clusters or massive stars, when available. An average value was reported in cases where multiple measurements or radial gradients exist.}
}
\end{table*}
\begin{table*}
\centering
\caption{Distribution of targets per galaxy and per photometric catalog. Mean coordinate offsets between {\it Spitzer} sources and the other surveys.}\label{tab:catalogs}
\begin{tabular}{l | r | c c c c | c c c }
\hline\hline
Galaxy	& {\it Spitzer}	& \textit{Gaia}	& PS1	& VHS	& Foreground & $d_{Gaia}$	& $d_{\mathrm{PS1}}$	& $d_{\mathrm{VHS}}$ \\ 
& & &  & &  & \small{(\arcsec)}	& \small{(\arcsec)}	& \small{(\arcsec)}   \\
\hline \\[-9pt]
WLM&		19969$^1$ &	384 &	3377 &	3612 &	98 & $0.37 \pm 0.17$	& $0.41 \pm 0.21$	& $0.30 \pm 0.23$ \\
NGC 55 &	8746$^2$ &	646	& 0		 &	0 &	139  & $0.31 \pm 0.23$	& --				& -- \\
NGC 247&	16658$^3$ &	550 &	2651 &	0 &	254 & $0.25 \pm 0.17$	& $0.30 \pm 0.21$	& -- \\
NGC 253&	9001$^2$ &	524 &	1715 &	0 &	309 & $0.33 \pm 0.17$	& $0.34 \pm 0.20$	& -- \\
NGC 300&	21739$^3$ &	1253 &	0 &		12002 &	275 & $0.29 \pm 0.19$	& --				& $0.27 \pm 0.19$  \\
NGC 1313&	 6156$^4$ &	 381& 0 &	0	 & 124	 & $0.37 \pm 0.18$ & -- & -- \\
NGC 3109 &	 9474$^4$ &	997 & 3046	 &	0	 &	371 & $0.23 \pm 0.17$ & $0.27 \pm 0.17$ & -- \\
Sextans A &		15876$^1$ &	204 &	910 &	457 &	19 & $0.31 \pm 0.18$	& $0.34 \pm 0.20$	& $0.33 \pm 0.20$ \\
M83 &	23331$^4$ &	1242 &	2580 &	4747 &	660  & $0.26 \pm 0.20$	& $0.31 \pm 0.23$	& $0.37 \pm 0.25$ \\
NGC 7793&		5617$^3$ &	279 &	926 &	0 &	69  & $0.40\pm0.21$	& $0.39\pm0.21$	& -- \\
\hline
\end{tabular}
\tablebib{
(1) \citet{Boyer2015}; (2) \citet{Williams2016}; (3) \citet{Khan2015}; (4) \citet{Khan2017}.
}

\end{table*}

{\it Spitzer} point-source catalogs containing sources detected in the IRAC \citep{fazio2004} 3.6 and 4.5 $\mu$m bands form the basis of our catalogs. Where available, these are supplemented by IRAC 5.8 and 8.0 $\mu$m and MIPS \citep{rieke2004} 24 $\mu$m magnitudes or their 3$\sigma$ upper limits (see Table~\ref{tab:catalogs} for references). These catalogs were cross-matched with optical and near-IR photometry, as well as astrometric data from the following surveys:

\begin{itemize}
\item Pan-STARRS1 (PS1): $g$ (481 nm), $r$ (617 nm), $i$ (752 nm), $z$ (866 nm), $y$ (962 nm)
\item VISTA Hemisphere Survey \citep[VHS;][]{mcmahon2012}: $J$ (1.25 $\mu$m), $K_s$ (2.15 $\mu$m)
\item {\it Gaia} DR2 \citep{gaia2016, gaia2018}: Proper motion in right ascension (R.A.), declination, and parallax.
\end{itemize}

A radius of 1\arcsec\, was used for cross-matching. Cases with multiple counterparts within 1\arcsec\ were flagged and no additional photometry was added. These targets constitute 0.1\% of the sample and are excluded from the priority target lists. For the sources that have a {\it Gaia} counterpart and a good astrometric solution ($astro\_excess\_noise < 1$~mas), we use the astrometric data to perform a conservative cleaning of foreground sources. Sources are flagged as foreground if they meet one of the following criteria:

\begin{itemize}
\item $pmra/pmra\_error > 3.6$
\item $pmdec/pmdec\_error > 3.9$
\item $parallax/parallax\_error > 3.6$
\end{itemize}

These cuts are based on values determined for stars in M31 and M33 \citep[three times the observed standard deviation, see][]{Maravelias2022}. Table~\ref{tab:catalogs} gives the total number of sources in each catalog, the number of matches in each of the additional photometric surveys, and the number of sources flagged as foreground contamination. It also provides the mean offset between coordinates from {\it Spitzer} and each of the other surveys.

\subsection{Target selection}\label{sec:targets}

We use the multi-wavelength photometric catalogs to construct target lists of potential dusty massive stars. The first criterion is an IR color of $m_{3.6} - m_{4.5} \geq 0.1$~mag. This selects targets with an IR excess and excludes the majority of remaining foreground stars (which mostly have $m_{3.6} - m_{4.5} \sim 0$~mag). Additionally, an absolute magnitude criterion of $M_{3.6} \leq -9.0$~mag (corresponding to $\log \rm L/ \lsun \sim 4.5$) and an apparent magnitude criterion of $m_{4.5} \leq 15.5$~mag are applied. The former excludes the majority of AGB stars \citep{yang2020}, while the latter prevents strong contamination by background IR sources \citep[i.e. background galaxies and quasars;][]{williams2015}. However, a small fraction of contaminants remains in our target lists (see Section~\ref{sec:discussion} for details).

\begin{table*}
\centering
\caption{Definition of the priority system and the criteria for each priority class.}\label{tab:priosystem}
\begin{tabular}{l | c c c c c c}
\hline\hline
Criterion		& P1		& P2	 	& P3	& P4	& P5 & P6\\
\hline \\[-9pt]
$m_{3.6} - m_{4.5}$ (mag)	& $\geq 0.5$	& $\geq 0.25$	& $\geq 0.5$ & $\geq 0.25$ & $\geq 0.1$  & $\geq 0.1$ \\
$M_{3.6}$ (mag)			& $\leq -9.75$	& $\leq -9.75$	& $\leq -9.75$ & $\leq -9.75$ & $\leq -9.0$ & $\leq -9.0$ \\
Optical/near-IR 			& Y	& Y	& N & N & Y & N \\
\hline
\end{tabular}

\end{table*}

\begin{table*}
\centering
\caption{Distribution of target priorities per galaxy.}\label{tab:prio}
\begin{tabular}{l r | r r r r r r | r c}
\hline\hline
Galaxy	&  \# Selected &	\multicolumn{6}{c}{\# Targets per Priority}  & \# Observed & Percentage \\
	&  Targets	& 1	& 2	& 3	& 4	& 5	& 6 & Targets &  \\
\hline \\[-9pt]
WLM			& 27	& 6		& 2		& 0		& 0		& 19	& 0	& 15 & 55.6	\\
NGC 55		& 156	& 0		& 0		& 75	& 54	& 0		& 27	& 71 & 45.5 \\
NGC 247		& 78	& 23	& 22	& 6		& 2		& 24	& 1		& 19 & 24.4 \\
NGC 253		& 283	& 32	& 50	& 97	& 48	& 27	& 29	& 99 & 35.0 \\
NGC 300		& 148	& 50	& 29	& 35	& 5		& 24	& 5	& 71 & 48.0	\\
NGC 1313	& 59	& 0 	& 0 	& 27	& 20	& 0		& 12&  8 & 13.6 	\\
NGC 3109	& 52	& 8 	& 4	    & 11	& 8		& 9 	& 12	& 7 & 13.6 \\
Sextans A	& 12	& 1		& 3		& 2		& 0		& 2		& 4	& 5 & 41.7	 \\
M83 		& 440	& 68	& 146	& 38	& 70	& 80	& 38	& 69 & 15.7 \\
NGC 7793	& 43	& 11	& 16	& 7		& 6		& 3		& 0	& 20 & 46.5	\\
\hline
\bf{Total}	&	\bf{1298}	& {\bf 199}	& {\bf 272}	& {\bf 298}	& {\bf 213}	& {\bf 188}	& {\bf 128}	& \bf{384}	& \bf{29.6} \\
\hline
\end{tabular}
\end{table*}

\begin{figure}
   	\resizebox{1.0\hsize}{!}{\includegraphics{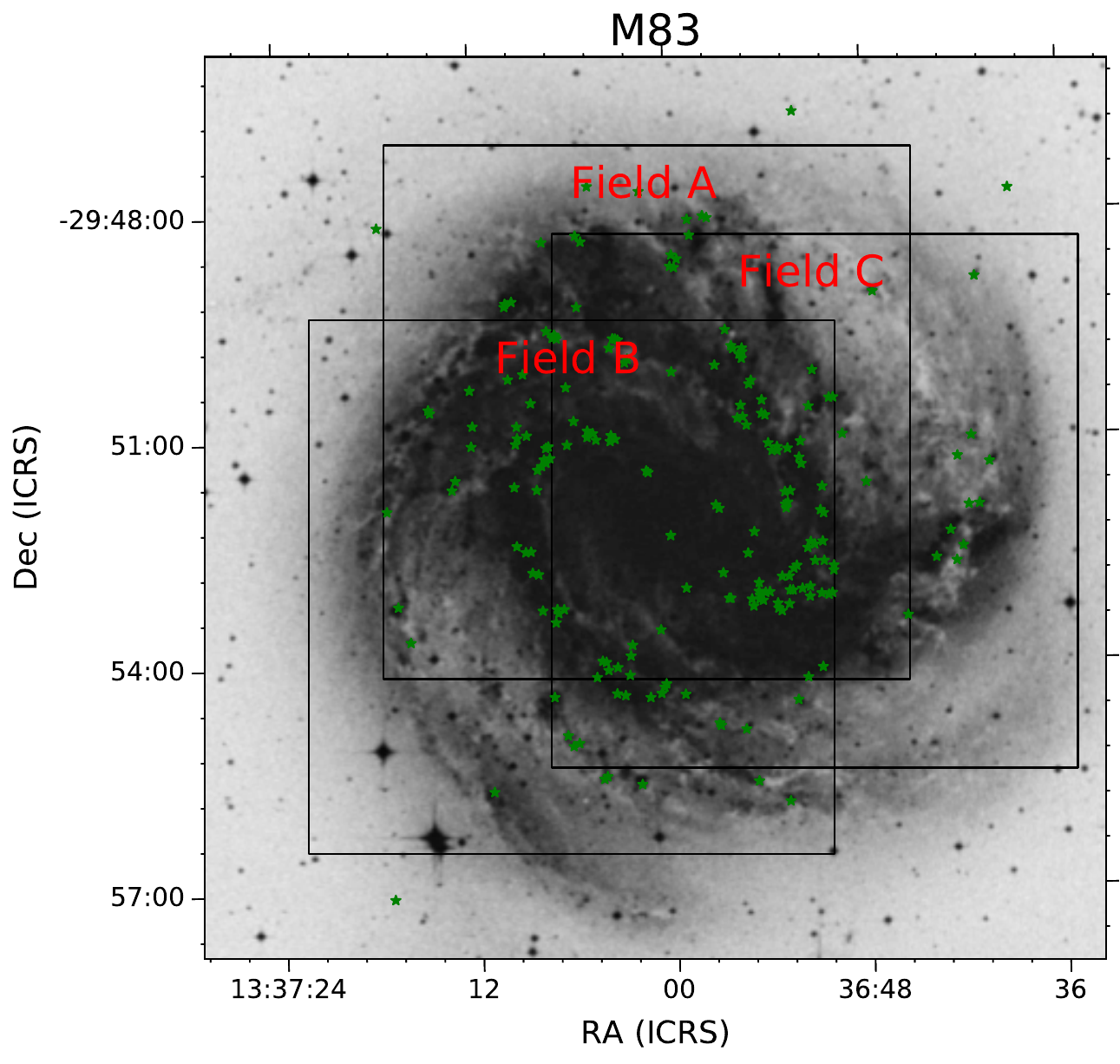}}
  	\caption{The three overlapping fields in M83. P1 and P2 targets are indicated. North is up and east to the left.}\label{fig:M83_Fields}
\end{figure}

The resulting target lists contain 1298 priority sources. We assign priorities 1 to 6 (P1 to P6) to each of the targets as described in Table~\ref{tab:priosystem}, using mid-IR colors, absolute magnitude criteria and finally, whether or not there are optical or near-IR counterparts. Priority 1 selects the brightest IR sources with a strong IR excess, which are most likely visible in the optical, while P2 targets are similar to P1 but with moderate IR excess. Table~\ref{tab:prio} presents the distribution of target priorities per galaxy for all the priority targets available (1298 sources), as well as the number and percentage of priority targets placed in slits and observed (384 targets or 30\% of selected targets). We note that there is a selection effect, as the priority targets sample more luminous IR sources with increasing distance. {\it Spitzer} color-magnitude diagrams (CMDs) of the target lists indicating the priority of each target and the magnitude criteria are shown in Appendix~{\ref{sec:ap_cmd}}. The fields for MOS observations were chosen to maximize the number of P1 and P2 targets that can be observed. This resulted in a total of 20 fields in the ten galaxies. In some cases (e.g. M83, see Figure~\ref{fig:M83_Fields}) the fields (strongly) overlap to allow as many priority targets as possible to be observed.

\subsection{Pre-imaging and mask design}
Pre-imaging of each field was obtained using the imaging mode of FORS2 to facilitate the design of the mask for the multi-object spectroscopy. Each field was observed for $3\times 10$s without a filter with the red detector. This prevents the saturation of our brightest targets, while still reaching deep enough for our faintest P1 and P2 targets (V $\sim$ 22 mag) to have sufficient signal-to-noise (S/N). The individual exposures were stacked using the IRAF \citep{iraf1, iraf2} task {\sc imcombine} before being loaded into the FORS Instrument Mask Simulator (FIMS)\footnote{http://www.eso.org/sci/observing/phase2/SMGuidelines/FIMS.html} software. 

We used rectangular slits of $1\arcsec \times 6\arcsec$. The slit length was chosen to provide enough coverage for sky subtraction, while maximizing the number of slits that can be placed on the mask. First, slits were placed on as many priority targets as could be fitted, as each slit blocks out the horizontal 6\arcsec\, strip at that location (as the spectrum gets dispersed on the detector in the horizontal direction). Any remaining space on the mask was filled in with other ``filler'' sources from the {\it Spitzer} catalogs ($F_C$), and, when space allowed, with random non-catalog filler objects ($F_{NC}$). An overview of the masks for each field, with the distribution of the total number of slits (956) per galaxy, field, and priority class is given in Table~\ref{tab:fields}. The pre-images with the location of the slits overlaid are shown in Appendix~\ref{sec_ap_masks}.

\begin{table*}
\centering
\small
\caption{Overview of masks.}\label{tab:fields}
\begin{tabular}{l c c c c c c c c c c c c c}
\hline\hline
			& \multicolumn{2}{c}{Field center} & & \multicolumn{10}{c}{Number of slits} \\
Galaxy			& RA	& Dec	& Rot	& P1	& P2	& P3	& P4	& P5	& P6	& {\bf Total$_{\mathbf{P}}$} & $F_C$\tablefootmark{a}	& $F_{NC}$\tablefootmark{b} & {\bf Total}\\
				& \small(J2000)	& \small(J2000)	& \small{(\degr)}\\
\hline \\[-9pt]
WLM &  &	&	& {\it 4}	& {\it 1}	& {\it 0} & {\it 0} & {\it 10} & {\it 0} & {\bf \emph{15}} & {\it 33} & {\it 2} & {\bf \emph{50}}\\
\quad {\it Field A} & 00:01:59.952	& $-$15:29:20.08	& 0	& 4		& 1		& 0		& 0		& 10	& 0	&{\bf 15}		& 33	& 2		& {\bf 50}	\\
\ \\

NGC 55 &  &	&	& {\it 0}	& {\it 0}	& {\it 40} & {\it 20} & {\it 0} & {\it 11} & {\bf \emph{71}} & {\it 73} & {\it 28} & {\bf \emph{172}}\\
\quad {\it Field A} & 00:14:37.183	& $-$39:11:16.76	& $-$70	& 0		& 0		& 12	& 6		& 0		& 4	& {\bf 22}		& 29	& 8		& {\bf 59}	\\
\quad {\it Field B} & 00:15:10.994	& $-$39:13:20.89	& $-$70	& 0		& 0		& 17	& 8		& 0		& 4	& {\bf 29}	& 13	& 12	& {\bf 54}	\\
\quad {\it Field C} & 00:15:44.088	& $-$39:14:39.01	& $-$70	& 0		& 0		& 11	& 6		& 0		& 3	& {\bf 20}		& 31	& 8		& {\bf 59}	\\
\ \\

NGC 247 &  &	&	& {\it 6}	& {\it 5}	& {\it 2} & {\it 0} & {\it 6} & {\it 0} & {\bf \emph{19}} & {\it 75} & {\it 35} & {\bf \emph{129}}\\
\quad {\it Field A} & 00:47:05.254	& $-$20:39:34.24	& 0	& 4		& 1		& 1		& 0		& 4		& 0	& {\bf 10}		& 32	& 23	& {\bf 65}	\\
\quad {\it Field B} & 00:47:09.110	& $-$20:46:27.98	& 0	& 2		& 4		& 1		& 0		& 2		& 0	& {\bf 9}		& 43	& 12	& {\bf 64}	\\
\ \\

NGC 253 &  &	&	& {\it 17}	& {\it 21}	& {\it 25} & {\it 20} & {\it 7} & {\it 9} & {\bf \emph{99}} & {\it 46} & {\it 25} & {\bf \emph{170}}\\
\quad {\it Field A} & 00:47:15.360	& $-$25:21:10.98	& 50	& 5		& 8		& 4		& 5		& 0		& 2	& {\bf 24}		& 16	& 12	& {\bf 52}	\\
\quad {\it Field B} & 00:47:37.284	& $-$25:16:55.02	& 50	& 4		& 6		& 13	& 9		& 4		& 4	& {\bf 40}		& 11	& 4		& {\bf 55}	\\
\quad {\it Field C} & 00:47:54.259	& $-$25:13:33.13	& 50	& 8		& 7		& 8		& 6		& 3		& 3	& {\bf 35}		& 19	& 9		& {\bf 63}	\\
\ \\

NGC 300 &  &	&	& {\it 28}	& {\it 13}	& {\it 18} & {\it 0} & {\it 12} & {\it 0} & {\bf \emph{71}} & {\it 100} & {\it 47} & {\bf \emph{218}}\\
\quad {\it Field A} & 00:54:53.117	& $-$37:38:42.22	& 0	& 7		& 5		& 3		& 0		& 5		& 0	&{\bf 20}	& 18	& 9		& {\bf 47}	\\
\quad {\it Field B} & 00:54:33.355	& $-$37:42:53.89	& 0	& 5		& 3		& 5		& 0		& 2		& 0	& {\bf 15}		& 36	& 7		& {\bf 58}	\\
\quad {\it Field C} & 00:54:57.074	& $-$37:43:46.49	& 0	& 7		& 3		& 9		& 0		& 3		& 0	& {\bf 22}	& 25	& 12	& {\bf 59}	\\
\quad {\it Field D}	& 00:55:28.788	& $-$37:42:49.28	& 0	&	9	& 2	& 1	& 0	& 2	& 0	& {\bf 14}	& 21	& 19	& {\bf 54} \\
\ \\
NGC 1313 &  &	&	& {\it 0}	& {\it 0}	& {\it 4} & {\it 2} & {\it 0} & {\it 2} & {\bf \emph{8}} & {\it 13} & {\it 0} & {\bf \emph{21}}\\
\quad {\it Field A} & 03:18:17.270 & $-$66:29:47.330	& 0	& 0		& 0		& 4		& 2		& 0		& 2	&{\bf 8}	& 13	& 0		& {\bf 21}	\\
\ \\

NGC 3109 &  &	&	& {\it 3}	& {\it 1}	& {\it 1} & {\it 0} & {\it 1} & {\it 1} & {\bf \emph{7}} & {\it 16} & {\it 0} & {\bf \emph{23}}\\
\quad {\it Field A} & 10:03:09.144 & $-$26:10:02.600	& 0	& 3		& 1		& 1		& 0		& 1		& 1	&{\bf 7}	& 16	& 0		& {\bf 23}	\\
\ \\

Sextans A 	&	&							&	& \it{1} & \it{2} & \it{0} & \it{0} & \it{1} & \it{1} & \bf{\emph{5}} & \it{19} & \it{0} & \bf{\emph{24}} \\
\quad {\it Field A} 	& 10:10:59.736	& $-$04:41:55.570	& 0	& 1	& 2	& 0	& 0	& 1	& 1	& \bf{5}	& 19	&	0	& \bf{24}\\
\ \\

M83 				&			&				&	& \it{21}	& \it{33}	& \it{2}	& \it{5}	& \it{7}	& \it{1}	& \bf{\emph{69}}	& \it{22}	& \it{0}	& \bf{\emph{91}}	\\
\quad {\it Field A} 	& 13:37:01.368	& $-$29:50:39.08	& 0	& 8	& 14 & 1	& 3	& 1	& 0	& \bf{27}	& 7	& 0	& \bf{34}\\
\quad {\it Field B} 	& 13:37:06.192	& $-$29:52:57.29	& 0	& 8	& 10	& 0	& 0	& 2	& 0	& \bf{20}	& 9	& 0	& \bf{29}\\
\quad {\it Field C} 	& 13:36:51.168	& $-$29:51:52.31	& 0	& 5	& 9	& 1	& 2	& 4	& 1	& \bf{22}	& 6	& 0	& \bf{28}\\
\ \\

NGC 7793 &  &	&	& {\it 7}	& {\it 7}	& {\it 1} & {\it 2} & {\it 3} & {\it 0} & {\bf \emph{20}} & {\it 26} & {\it 12} & {\bf \emph{58}}\\
\quad {\it Field A} & 23:57:47.880	& $-$32:35:24.61	& 0	& 7		& 7		& 1		& 2		& 3		& 0	& {\bf 20}	& 26	& 12	& {\bf 58}	\\
\ \\

\hline
{\bf Total}	&	&	&	& {\bf 87} & {\bf 83} & {\bf 93} & {\bf 49} & {\bf 47} & {\bf 25} & {\bf \emph{384}} & {\bf 423} & {\bf 149} & {\bf \emph{956}} \\
\hline
\end{tabular}

\tablefoot{
\tablefoottext{a}{Number of other non-priority sources from the \textit{Spitzer} catalog.}
\tablefoottext{b}{Number of non-priority sources that did not have a \textit{Spitzer} counterpart.}
}

\end{table*}

\subsection{Multi-object spectroscopy and data reduction}

The MOS observations for eight of the galaxies were obtained in 2020-2021 (ESO programme 105.20HJ) and for the remaining two galaxies (NGC~1313 and NGC~3109) in 2022 (ESO programme 109.22W2). The 1\arcsec\, slits with the GRIS\_600RI+19 grism yield a resolving power $R = \lambda/\Delta\lambda \sim 1000$. The nominal wavelength coverage of the chosen setup is 512$-$845 nm. The actual wavelength coverage varies from slit to slit, and depends on the horizontal location on the mask (e.g., for slits placed on the left of the mask, the bluest wavelengths are dispersed outside of the detector). Each field was observed for two observing blocks of $3 \times 900$s, yielding a total exposure time of 5400s per field. The observing log is provided in Appendix~\ref{sec:ap_obslog}.

The data were reduced using the FORS2 pipeline v5.5.7\footnote{http://www.eso.org/sci/facilities/paranal/instruments/fors/doc.html} under the EsoReflex \citep{esoreflex} environment. The pipeline applies the following procedures: bias correction, flat fielding, slit identification, flux calibration, sky subtraction, and extraction of the 1D spectra. In most cases, this produces good quality spectra. However, in cases with multiple objects on the slit, overlapping slits, strongly variable nebular emission, and/or vignetting on the top of the CCD, the sky subtraction performed by the pipeline leaves strong residuals. For this reason, we also performed the reduction without sky subtraction and manually selected the object and sky extraction regions from the 2D spectrum. For each slit, the automatically and manually extracted spectra were visually inspected, and the best reduction was chosen. In rare cases where slits strongly overlap, no spectra of sufficient quality could be extracted. In almost all cases these are filler targets. Additionally, the used grism causes a vertical shift, causing some of the spectra to fall off the top edge of the CCD or in the gap between the two CCDs. The total number of successfully extracted spectra is 763.

\section{Spectral classification and stellar content} \label{sec:classification}

The spectral classification of the observed sources was performed in two stages. Initially, sources were separated into broad classes (i.e.\ red, yellow, blue, emission-line object) based on the slope of the spectrum and the presence of emission lines. The wavelength range of our spectra is not significantly affected by extinction, therefore the color correlates with the effective temperature. Next, each spectrum was scrutinized and given a more robust classification, if allowed by the signal-to-noise ratio. The details of the classification procedure for each type of object are given in Sections~\ref{sec:class_red} to \ref{sec:class_em}. The catalog is presented in Section~\ref{sec:catalog}, and the stellar content of each target galaxy is described in Section~\ref{sec:class_galaxies}.

\begin{table*}
\centering
\caption{Distribution of classified targets\tablefootmark{a} per galaxy and spectral class.}\label{tab:specClass}
\begin{tabular}{l | l | l l l c c c r r r c c r}
\hline\hline
Galaxy	& Class. & RSG & BSG	& YSG & LBVc & sgB[e] & Em. & H \textsc{ii}\tablefootmark{b} & Clusters & Other & C-stars & Galaxies\tablefootmark{d} & Fgd  \\
Name	&  &  & 	&  &  &  & Obj. & &  &  Stars\tablefootmark{c} & &  &   \\
\hline \\[-9pt]
WLM			& 19      & - & -     & -     & - & 1 & 1 & -  & -  & 5 & 7     & 1 & 4\\
NGC 55		& 85      & 42& 4     & 1     & 2 & 1 & - & 15 & 1  & 12 & -    & 3 & 4\\
NGC 247		& 49 (3)  & 16& 3 (2) & 2 (1) & 1 & 1 & 1 & 10 & 2  & 5 & -     & 4 & 4\\
NGC 253		& 63 (1)  & 12& - (1) & 1     & - & 1 & 2 & 24 & 11 & 9 & -     & - & 3\\
NGC 300		& 105 (5) & 46& 7 (2) & 1 (2) & - & 2 & 2 & 10 & 3  & 8 & 1 (1) & 8 & 17\\
NGC 1313	& 15 (2)  & 1 & 1 (2) & -     & - & - & - & 5  & 3  & - & -     & - & 5\\
NGC 3109	& 15      & 1 & -     & 1     & 1 & - & - & 1  & -  & - & -     & 2 & 9\\
Sextans A	& 15 (2)  & 1 & 2 (2) & -     & - & - & 1 & -  & -  & 10 & 1    & - & -\\
M83		    & 46      & 1 & -     & -     & - & - & - & 28 & 5  & 1 & -     & - & 11\\
NGC 7793	& 28 (1)  & 9 & 1     & - (1) & - & 1 & 1 & 6  & 2  & 1 & -     & 3 & 4\\
\hline
\bf{Total}	& \bf{440 (14)} & \bf{129} & \bf{18 (9)} & \bf{6 (4)} & \bf{4} & \bf{7} & \bf{8} & \bf{99} & \bf{27} & \bf{51} & \bf{9 (1)} & \bf{21} & \bf{61}\\
\hline
\end{tabular}

\tablefoot{
\tablefoottext{a}{Candidates and uncertain classifications are denoted separately  from the secure classifications, in parenthesis.}\\
\tablefoottext{b}{The H~\textsc{ii} regions listed here are the cases in which the H~\textsc{ii} region is dominating the spectrum. In a few cases, where the spectrum revealed features of both an H~\textsc{ii} region and a RSG, it has been listed as a RSG.}\\
\tablefoottext{c}{O-stars, A-stars and hot, warm and cool stars are grouped under "Other Stars".}\\
\tablefoottext{d}{We group AGN, QSO and galaxies under "Galaxies".}
}

\end{table*}

\subsection{Classification of red and yellow objects}\label{sec:class_red}

We first inspected the red objects for the presence of molecular absorption bands. In cases where no features were seen (i.e. due to the object being faint or having poor S/N), the classification remained as previously assigned (i.e. red object). The majority of red objects exhibited molecular TiO absorption, indicating a star of M-type. To identify which of these are foreground stars, we calculated the shift of the spectral features due to the radial velocity of the star. When the radial velocity was not in agreement with the host galaxy, we classified them as foreground stars (fgd). As giant and dwarf stars would be far too faint to be detected at great distances, we assigned a supergiant classification whenever the radial velocity was in agreement with the radial velocity of the host galaxy. Given that the primary luminosity class diagnostic, the Ca~\textsc{ii} triplet, was out of range, the radial velocity confirmation served as one of the primary criteria for the RSG classification.

\begin{figure*}
   	\resizebox{1.0\hsize}{!}{\includegraphics{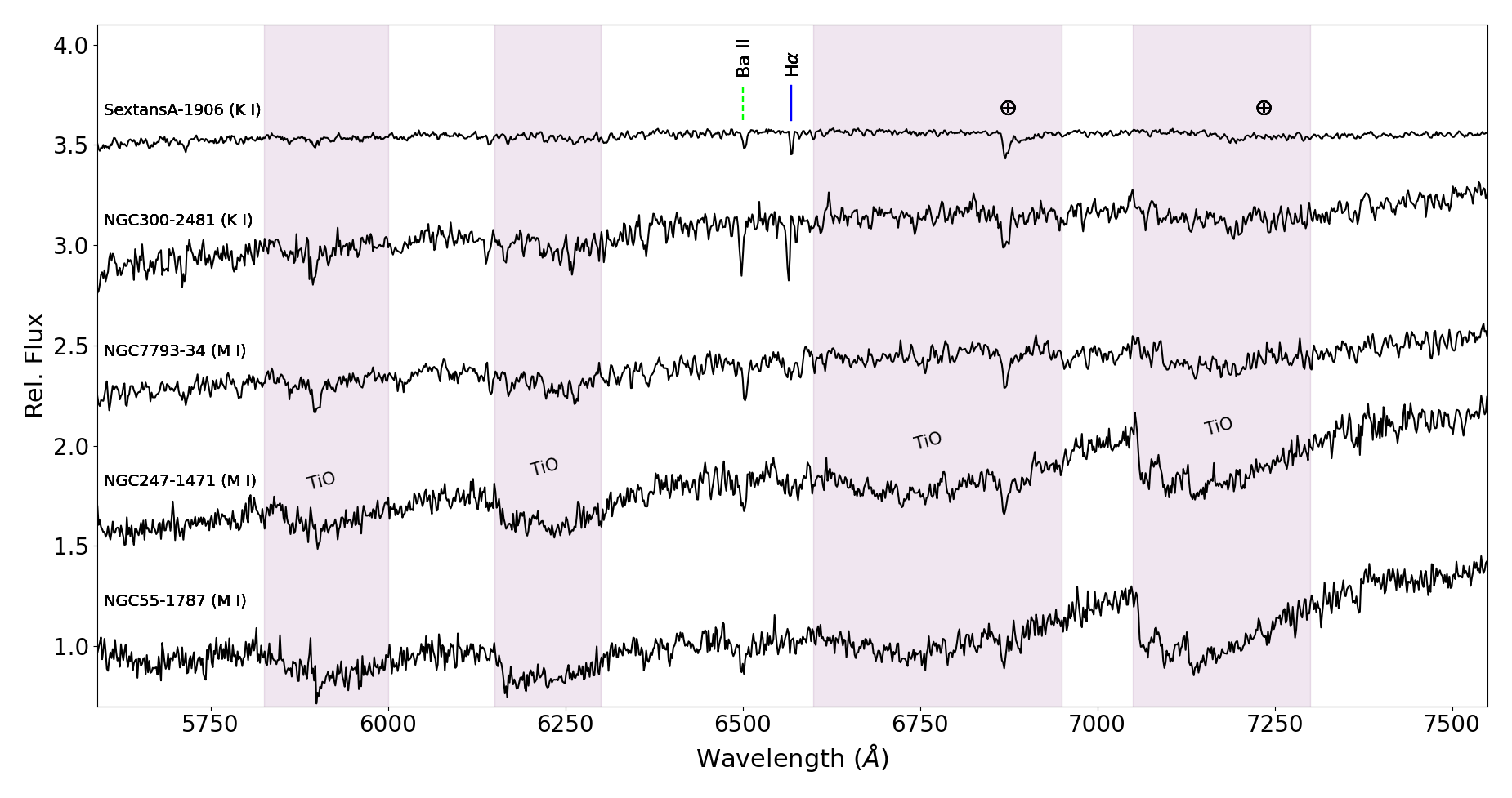}}
  	\caption{Representative spectra of targets classified as RSGs, following a temperature sequence from K-type (top) to M-type (bottom). Shaded areas correspond to regions with TiO absorption; telluric absorption is indicated with a $\oplus$ symbol.}\label{fig:RSGs_spectra}
\end{figure*}

\begin{figure*}
   	\resizebox{1.0\hsize}{!}{\includegraphics{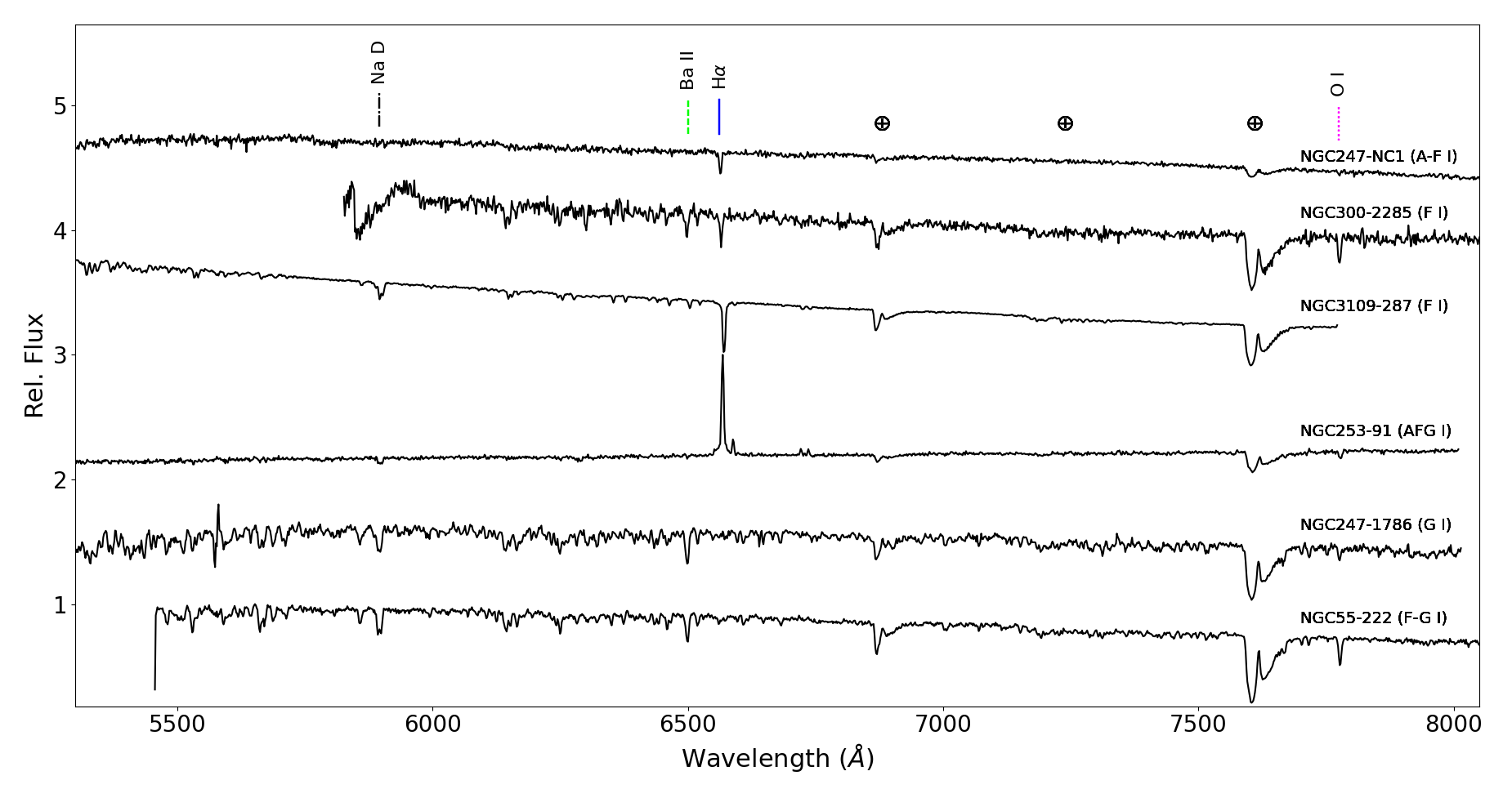}}
  	\caption{Spectra of all six targets classified as YSG. The strength of the oxygen triplet at $\lambda$7772 indicates a supergiant luminosity class. Other key spectral lines are indicated; telluric absorption is indicated with a $\oplus$ symbol.}\label{fig:YSGs_spectra}
\end{figure*}

Spectra showing traces of TiO absorption (i.e. at $\lambda\lambda$6150 and 7050) were classified as M-type red supergiants (M~I), whereas those without were classified as K-type (K~I). We checked all our M I sources for ZrO, but in the few cases it was detected it was much weaker than TiO, which is consistent with our luminosity class determination. Figure~\ref{fig:RSGs_spectra} presents representative examples of target spectra classified as RSG. For all classified RSGs, we have additionally verified that RSG atmospheric models (i.e. \textsc{marcs}) fit the observed spectra well (de Wit et al., in prep.). This forthcoming work will also include a more accurate spectral classification of the RSG into early or late-K and early or late-M. When molecular carbon features were present in a spectrum (i.e. the Swan C$_2$ bands or CN molecular bands), the object was classified as a carbon star (C-star). 

A few stars had a flat continuum and exhibited a variety of spectral lines. The vast majority of these objects were classified as yellow foreground dwarfs, due to their brightness and non-shifted spectral lines. However, six were classified as YSGs (with spectral types A-F, F or G), and four as YSG candidates (with luminosity class I-III). These showed a strong O~\textsc{i} $\lambda$7772 absorption triplet, which is sensitive to luminosity changes in the F0--K0 regime \citep{Kovtyukh2012}. Furthermore, {\it Gaia}~DR3 proper motions and parallaxes for these targets were small and the radial velocities were consistent with their host galaxy. Figure~\ref{fig:YSGs_spectra} presents all six spectra classified with certainty as YSG; sources of spectral types A, F or G, are collectively referred to as AFG.

\subsection{Classification of blue objects}\label{sec:class_blue}

\begin{figure*}
   	\resizebox{1.0\hsize}{!}{\includegraphics{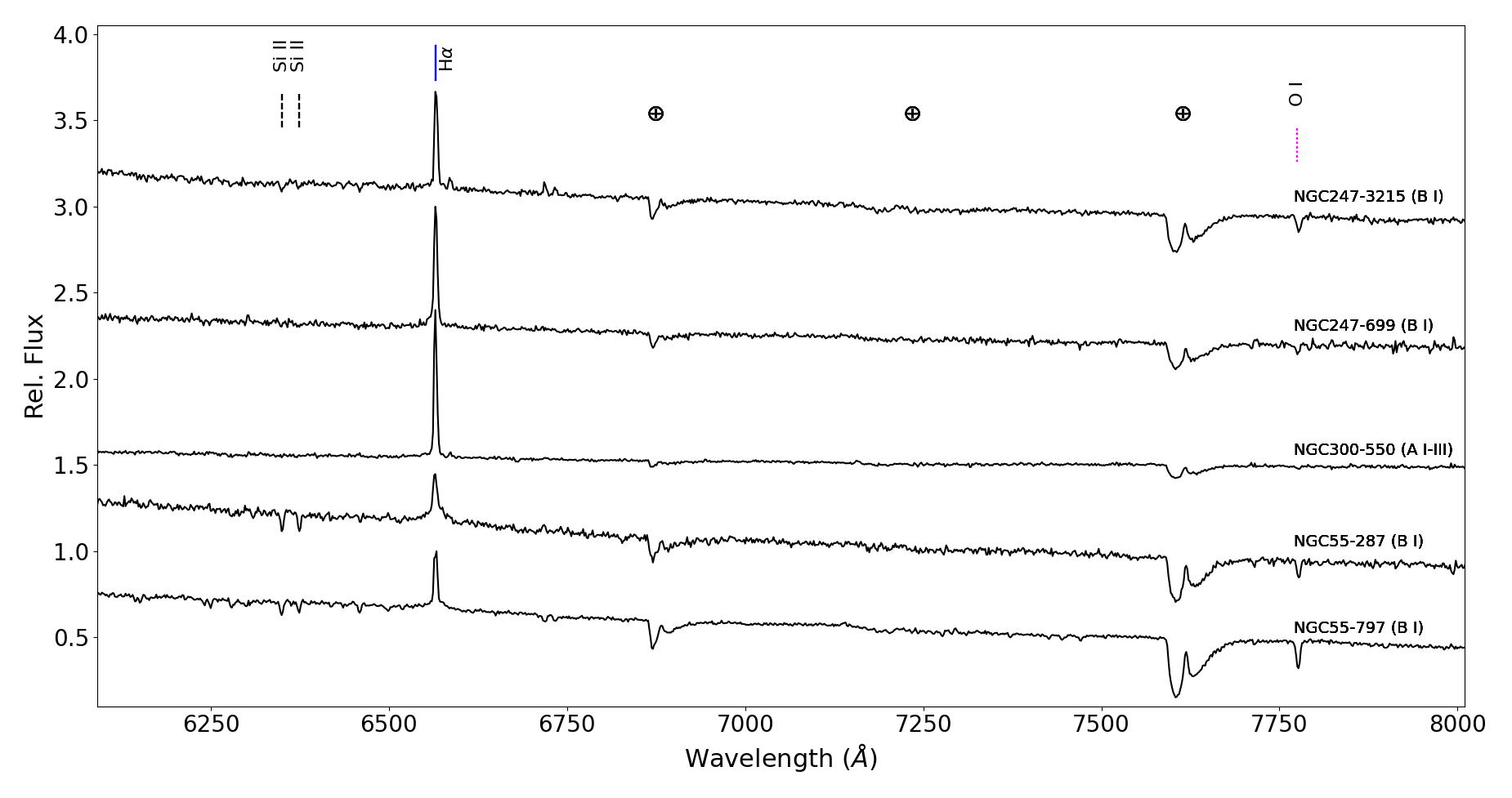}}
  	\caption{Representative spectra of BSG exhibiting H$\alpha$ emission. The H$\alpha$ profiles are a superposition of the nebular line profile broadened at the base by a circumstellar component, indicating significant mass-loss for these objects.} \label{fig:BSGs_spectra}
\end{figure*}

Initially, we inspected the blue objects for the presence of spectral features. If none were detected, the raw classification remained. The majority of blue objects were not classified further, as the spectra were of too poor quality to detect spectral lines. However, in cases where spectral lines were clearly detected, we were able to assign a spectral class. When only a broadened H$\alpha$ emission profile was identified in a spectrum, indicating an accelerating shell of mass surrounding a star, the object was classified as a BSG. Additionally, when Fe~\textsc{ii} emission was present in the region $\lambda\lambda$6200$-$6500, the object was classified as an LBV candidate. Ultimately, when [Fe~\textsc{ii}] ($\sim\lambda\lambda$5150$-$5550) and [O~\textsc{i}] $\lambda$6300 emission, originating from low-density regions in a circumstellar disk, were present, the object was classified as a sgB[e] star. \citet{Maravelias2023} presented the LBVs and sgB[e]. Blue objects with H$\alpha$ clearly in absorption along with other metallic lines, such as the O~\textsc{i} triplet ($\lambda$7772), Si~\textsc{ii} ($\lambda\lambda$6347, 6371) or He~\textsc{i} ($\lambda\lambda$5876, 6678, 7065), were classified as BSGc (of B or A type). When H$\alpha$ was in absorption, the supergiant class was only assigned to stars that showed a strong O~\textsc{i} $\lambda$7772 absorption triplet. In cases where a strong Si~\textsc{ii} doublet was observed \citep[indicating $T_{\rm eff} \geq 9000$;][]{Mashonkina2020}, sometimes paired with strong He~\textsc{i} absorption or a Paschen jump around $\lambda$8200, we assigned a B-type classification. In cases where there is mainly significant H$\alpha$ absorption, we assigned an A-type classification. Figure~\ref{fig:BSGs_spectra} presents examples of targets classified as BSG.

\subsection{Classification of emission-line objects}\label{sec:class_em}

\begin{figure*}
   	\resizebox{1.0\hsize}{!}{\includegraphics{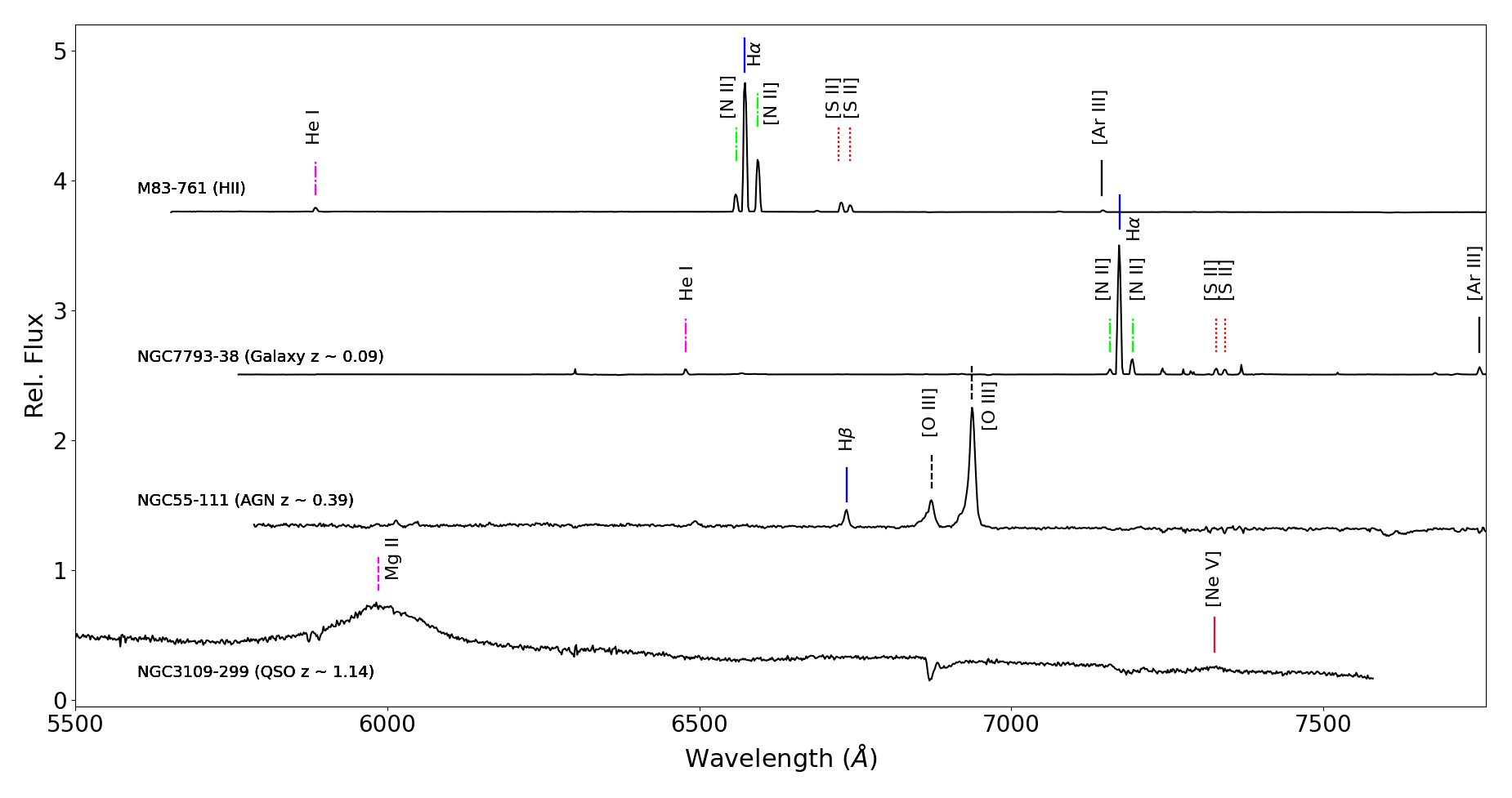}}
  	\caption{Example spectra of targets classified as \hii\ regions or galaxies (including AGN and quasars). Their redshifts were calculated from the emission lines indicated for each of these targets.}\label{fig:Conts_spectra}
\end{figure*}

We used the following procedure to robustly classify objects with a series of narrow emission lines. First, these objects were inspected for common transitions (e.g. [N~\textsc{ii}], [S~\textsc{ii}], He~\textsc{i}, and [Ar~\textsc{iii}]). In cases where a strong emission component was present, but we were not able to assign a robust classification, the target was classified as an "Emission Object" ("EmObj"). In the majority of cases, a doublet of [N~\textsc{ii}] ($\lambda\lambda$6548, 6583) and [S~\textsc{ii}] ($\lambda\lambda$6717, 6731) emission was present. If spectra contained only these five features, they were classified as Nebulae ("Neb."). This classification could either be singular or added to another classification (i.e. RSG + neb). Additionally, if He~\textsc{i} and [Ar~\textsc{iii}] (i.e. $\lambda$7136) emission was present in a nebular spectrum, we classified it as an H~\textsc{ii} region. For many cases of H~\textsc{ii} regions, a stellar continuum was clearly present. These received an additional classification (H~\textsc{ii} + star or H~\textsc{ii} + star + H$\alpha$, in case the H$\alpha$ emission profile appeared broadened, revealing a mass-losing star in the center of the H~\textsc{ii} region). If a spectrum had emission lines along with a blue component and a red component, it was classified as a cluster with blue stars and at least one RSG. Some spectra exhibited extremely redshifted emission lines. If the spectrum resembled an extremely redshifted H~\textsc{ii} region, the object was classified as a background galaxy. One spectrum of a background galaxy featured a Balmer jump and was therefore classified as a Balmer jump galaxy. Lastly, if a redshifted spectrum portrayed strong asymmetric and broad emission lines, we classified it as an active galactic nucleus (AGN) or a quasar. We used characteristic emission line patterns for such objects (e.g. \ion{Mg}{ii} $\lambda$2798) to measure their redshift. We present an \hii\ region, a galaxy, an AGN and a quasar in Figure~\ref{fig:Conts_spectra}.

We proceeded to analyze the spectra of all the \hii\ regions, along with all the targets exhibiting nebular contamination in their spectrum to measure the flux of the lines H$\alpha$, [\ion {N} {ii}] $\lambda$6583, [\ion {S} {ii}] $\lambda\lambda$6716 and 6731. We selected the manually reduced spectra as the target and sky apertures were selected more carefully than in the automatized pipeline version. However, we visually inspected the 2D spectrum to avoid overlapped slits, inhomogeneous nebular emission in the spatial direction, or other artifacts that could compromise the measurement of the flux. We used the IRAF task {\sc splot} to measure the fluxes and their associated error, establishing a minimum error of 3\% in the flux when the signal of the nebular emission was much more powerful than the noise. The results are presented in Section~\ref{sec:hiiregions}.

\subsection{Catalog}
\label{sec:catalog}

Following the procedure described above, we classified all of the 763 successfully extracted FORS2 spectra. We also used the Mikulski Archive for Space Telescopes (MAST) portal to search for available imaging of our targets from the {\it Hubble Space Telescope} using the Hubble Legacy Archive (HLA). When available, photometry from the Hubble Source Catalog \citep{Whitmore2016} assisted in the identification of our targets. We found that 150 of our 541 classified targets had available imaging and performed a robust visual classification for 80 of these, providing extra information on whether the targets are point-sources (i.e.\ stars) or extended sources (i.e.\ clusters, background galaxies). We compared the spectral classifications against the visual classification from the \textit{HST} images and found them to be consistent; in three cases we revisited and improved the spectral classification\footnote{\textit{HST} images for NGC247-143, NGC247-157, and NGC300-274 indicate galaxies, rather than the approximate "star", "red" and "cool star" classifications from the spectra, which were found to be galaxies at redshifts $z=0.133, 0.156$ and 0.163, respectively.}.

Of the 763 successfully extracted spectra, 454 were robustly classified ($\sim60\%$) either as an evolved massive star, an extended object (i.e. H~\textsc{ii} region or cluster) or a contaminant (i.e. a galaxy or a foreground star), while 87 (or 11\%) were given a more general classification (i.e. nebula, blue object or red object). The remainder of the spectra (222 stars or 29\%) were not of sufficient quality to be assigned a spectral class. Table~\ref{tab:specClass} presents the distribution of classified targets per galaxy and spectral class, for both robust classifications (440) and candidates (14).

The number of evolved massive stars (including candidates) that were classified is 185 targets ($\sim24\%$). This is the largest catalog of evolved massive stars beyond the Local Group \citep[e.g.][]{Bibby2012, Grammer2015, Britavskiy2019, Humphreys2019}. Out of the 185 targets, the RSGs, LBV candidates and sgB[e] stars are robust classifications (see Table~\ref{tab:specClass}), whereas $30-40\%$ of the BSGs and YSGs are candidates. This is because their luminosity class could not be confirmed from the spectral range and resolving power of our spectra, mainly due to the lack of resolved spectral lines that are sensitive to changes in the surface gravity. The global contamination fraction in our sample is (N$_{\rm fgd}$+N$_{\rm C}$+N$_{\rm AGN}$+N$_{\rm gal}$+N$_{\rm QSO}$) / N$_{\rm spec} \sim12\%$, where N$_{\rm spec}$ is the 763 spectra. 

The resulting catalog of all 541 spectroscopically classified targets is presented in Table~\ref{tab:spec_catalog}. The targets are ordered by galaxy RA and by their $m_{4.5}$ magnitude within each galaxy. The catalog contains the following information: target ID, coordinates (in degrees, J2000), field, target priority, 17 bands of photometry from {\it Spitzer} IRAC and MIPS, Pan-STARRS1, VHS, and \textit{Gaia} DR2, our spectral classification, comments on the spectral classification, the classification from the \textit{HST} images (when available), along with the \textit{HST} filters available and comments on the \textit{HST} images. 

\begin{table*}
\centering
\caption{Catalog of spectral classifications.}
\label{tab:spec_catalog}
\small
\begin{tabular}{l l l c c l c}
\hline
\hline      
ID    & RA    & Dec.   & ... & Spectral Class.   & Classification Notes\tablefootmark{a} & ...\\
\hline
NGC300-59    & 13.73010  & $-$37.60996 & ... & M I   & TiO abs. & ... \\
NGC300-61    & 13.72192  & $-$37.71981 & ... & Galaxy    & Shifted neb. lines ($z\sim$ 0.15) & ...\\
NGC300-65    & 13.70976  & $-$37.64707 & ... & AGN   & 	Broad and shifted H$\alpha$, H$\beta$, [O~\textsc{iii}]~$\lambda$5007 em. ($z\sim$ 0.22) & ...\\
NGC300-66    & 13.65649  & $-$37.76726 & ... & AFG V   & RV $\sim$ 0, strong Na~\textsc{i} doublets, Ba~\textsc{ii}~$\lambda$6496 abs. & ...\\
NGC300-67    & 13.86639  & $-$37.73878 & ... & sgB[e]    & 	Fe~\textsc{ii}, [Ca~\textsc{ii}], [O~\textsc{i}] and strong H$\alpha$ em. & ...\\
NGC300-82    & 13.73466  & $-$37.66102 & ... & Galaxy    & Shifted neb. lines ($z\sim$ 0.11) & ...\\
NGC300-84    & 13.67848  & $-$37.71988 & ... & \hii\ region    & Neb. lines, [O~\textsc{i}], [Ar~\textsc{iii}] and He~\textsc{i} em. & ...\\
NGC300-114   & 13.72206  & $-$37.64396 & ... & M I   & TiO abs. & ...\\
NGC300-125   & 13.71701  & $-$37.67567 & ... & M I   & TiO abs. & ...\\
NGC300-138   & 13.60430  & $-$37.74157 & ... & AFG V   & Strong H$\alpha$, D1/D2 and Ba~\textsc{ii}~$\lambda$6496 abs. & ...\\
NGC300-145   & 13.72232  & $-$37.72957 & ... & \hii\ region    & Neb. lines, faint [Ar~\textsc{iii}] em. & ...\\
NGC300-150   & 13.88370  & $-$37.72683 & ... & M I   & TiO abs. & ...\\
NGC300-153   & 13.69379  & $-$37.73638 & ... & M I   & TiO abs. & ...\\
\hline
\end{tabular}
\vspace{0.3cm}

\tablefoot{
\tablefoottext{a}{``Neb. lines'' indicates the presence of the H$\alpha$, [N~\textsc{ii}], and [S~\textsc{ii}]  emission quintet.}\\
\small This table is available in its entirety in machine-readable and Virtual Observatory (VO) forms in the online journal. A portion is shown here for guidance regarding its form and content.
}

\end{table*}

\subsection{Comments on individual galaxies}\label{sec:class_galaxies}
In the following subsections, we present the results of our target classification in each galaxy.

\subsubsection{WLM} \label{subsec:WLM}
The WLM sample mostly consisted of stars without a priority or a low priority. No RSGs were found in this galaxy, despite recent studies on the RSG content of WLM (i.e. \citealt{Britavskiy2015,GonzalezTora2021}, as these targets were not in our priority system). 

We observed one sgB[e] (WLM-95), which is the first such source identified in a low-Z galaxy ($\sim0.14~\zsun$) presented by \citet[][source WLM-1]{Maravelias2023}. In addition to strong, broadened H$\alpha$ and weak Fe~\textsc{ii} emission lines, significant He~\textsc{i} $\lambda$5875 and $\lambda$6678 absorption is present. This object was previously classified as an emission line star by \citet{Massey2007} and \citet{Britavskiy2015}. We report no significant changes in the spectrum (over a period of eight years), as the presence of \ion{Fe}{ii} and [\ion{O}{i}] emission and absence of [\ion{Ca}{ii}] emission were observed in both of the spectra.   

No BSGs, YSGs or LBV candidates were identified in WLM. Finally, the majority of observed red stars contained strong carbon absorption bands typical of carbon-rich AGB stars. No H~\textsc{ii} regions were found.

\subsubsection{NGC~55}
We classified 42 RSGs in NGC~55, mostly of M~\textsc{I} type. De Wit et al. (in prep.) will provide a more detailed classification as well as stellar parameters through MARCS modeling. In addition, six stars with broadened H$\alpha$ emission were observed, the largest number found in a single galaxy in our sample. Four were classified as BSG (NGC55-260, NGC55-287, NGC55-797 and NGC55-NC6), two as LBVc (NGC55-736, NGC55-2924) and one as a sgB[e] star \citep[NGC55-178; see][]{Maravelias2023}. One additional BSG (NGC55-260) has H$\alpha$ in absorption, yet shows a strong O~\textsc{i} $\lambda$7772 triplet. We observed one YSG (NGC55-222) of spectral type F-G. 

The contamination by foreground stars is low for NGC~55, although we detected 17 extended objects. We also identified a background galaxy, an AGN at $z=0.39$ and a quasar at $z=1.158$, which is coincident with the X-ray source 4XMM 001547.5-391240. The AGN is presented in Figure~\ref{fig:Conts_spectra}.

\subsubsection{NGC~247}
We identified 16 M-type RSGs in NGC~247. As the median RSG type depends on metallicity \citep{Elias1985,Levesque2006}, we note that the RSG spectral types are in agreement with the previously estimated metallicity \citep[$\sim0.4$~\zsun;][similar to M33 and the LMC]{Davidge2021} of the galaxy (see Section~\ref{sec:hiiregions}).

Various other types of evolved massive stars were observed, including one LBVc (NGC247-1192) and one sgB[e] (NGC247-246), both described in \citet{Maravelias2023}. NGC~247 hosts a bright YSG (NGC247-1786). A blackbody fit to the available photometry revealed an effective temperature around 5600~K. We observed one additional YSG (NGC247-NC1), one candidate YSG (NGC247-3021) and five BSG or candidate BSG. The contamination by H~\textsc{ii} regions in NGC~247 is relatively high, on the order of $\sim20\%$. Some spectra contained both a bright blue component and at least one RSG, meaning these are most likely clusters (see Section~\ref{sec:class_em}).

\subsubsection{NGC~253}
We identified 12 RSGs and one sgB[e], NGC~253-739 \citep{Maravelias2023}. All of the RSGs were classified as M-types, except for one potential K-type (NGC253-3509). For this star, the detection of the TiO bands was hampered by the low S/N of the spectrum (although a best-fit \textsc{marcs} model gives a $T_{\rm eff}$ compatible with a K-type supergiant). The metallicity of NGC~253 (0.72~\zsun) is consistent with finding mainly M-type RSGs. We obtained a wide range of radial velocities for these RSGs, varying from 70 to 450 km s$^{-1}$. These values are in agreement with the kinematical maps measured for NGC~253 \citep{Hlavacek2011}. One spectrum (NGC253-91) revealed a promising YSG candidate, sharing similarities with an early K-type RSG, yet it was too hot to be modelled with \textsc{marcs} models. Finally, we identified one BSG candidate. The confusion by H~\textsc{ii} regions was more extreme in NGC~253 (38\%), compared to NGC~247 (20\%), which is a similarly distant and inclined galaxy. Eleven targets revealed a composite spectrum including at least one blue and one red component. Three targets were classified as foreground stars. 

\subsubsection{NGC~300}
We identified 46 RSGs in NGC~300, the majority of which were classified as M-types. Additionally, one BSG (NGC300-901) and two sgB[e] (NGC300-67 and NGC300-389) have been identified due to the presence of H$\alpha$ emission (for both the BSG and sgB[e] stars), along with Fe~\textsc{ii} and [O~\textsc{i}] $\lambda$6300 emission (for the sgB[e]). The sgB[e] NGC300-67 further displayed strong He~\textsc{i} lines in emission (see NGC300-1 in \citealt{Maravelias2023}). We observed six additional BSGs and one YSG showing H$\alpha$ in absorption but with the O~\textsc{i} $\lambda$7772 triplet being sufficiently strong. Ten targets were classified as H~\textsc{ii} regions, with a higher number of contamination coming from foreground stars (17). Finally, background galaxies, AGN and a quasar contributed to eight more contaminants.

\subsubsection{NGC~1313}
One RSG of spectral type M was identified in this galaxy (NGC1313-310), in agreement with expectations for the metallicity of the galaxy ($\sim0.6$~\zsun). Additionally, one BSG and two BSGc were detected, but no LBVs or sgB[e] stars. NGC~1313 was one of the farthest galaxies in our sample, and as such, about half of the observed targets were classified as extended objects (i.e. clusters and H~\textsc{ii} regions). Lastly, we observed five foreground stars.

\subsubsection{NGC~3109}
We identified one target as a RSG of early K-type (NGC3109-167). This is consistent with the bulk of the RSG types for this metallicity ($\sim0.2$~\zsun). One object (NGC3109-188) has been classified as an LBVc due to the strong circumstellar H$\alpha$ and Fe~\textsc{ii} emission features detected, along with He~\textsc{i} lines in absorption (see \citealt{Maravelias2023}). Apart from one non-priority YSG (NGC3109-287), most of the non-priority targets were classified as foreground stars. We identified a Balmer jump galaxy and a quasar at $z\sim1.14$, which is shown in Figure~\ref{fig:Conts_spectra}.

\subsubsection{Sextans~A}
One RSG was identified in Sextans~A (SextansA-1906), with an early K-type spectrum, which is compatible with the low metallicity of the galaxy (0.06~\zsun). The sample of Sextans~A mainly contained blue stars, the majority of which were classified as O stars. We classified SextansA-s1 and SextansA-s2 as supergiants (O-B I) based on their H$\alpha$ emission, although \citet{Lorenzo2022} classified them as O9.5II and O3-O5Vz, respectively. A few spectra contained broadened H$\alpha$ emission profiles, but we were unable to determine whether these are indeed BSGs. No LBVs, sgB[e], foreground stars or H~\textsc{ii} regions were observed. 

\subsubsection{M83} \label{subsec:M83}
We have identified one RSG (M83-682) of spectral type K, which is unexpected at supersolar metallicity (1.6 \zsun), with no further LBV, sgB[e], YSG or BSG candidates in this galaxy (see Table \ref{tab:specClass}). The lack of identified massive stars is mostly the result of the increased contamination due to crowding, and decreased S/N due to the large distance to the galaxy (see Section~\ref{sec:effectiveness}). The majority of spectra taken for M83 revealed characteristics of extended objects (i.e. H~\textsc{ii} regions and stellar clusters). Although some of the observed H~\textsc{ii} regions revealed a stellar continuum and a broadened H$\alpha$ emission profile, indicating the existence of a mass-losing blue star in the center, we were unable to determine its nature. We narrowed down the spectra of many objects to blue stars with some degree of nebular contamination, but the spectra generally lacked the quality to determine whether these are supergiants or LBV candidates.

\subsubsection{NGC~7793}
The sample in NGC~7793 contains nine RSGs showing clear molecular TiO absorption. We estimate the spectral type to be early M for the majority of these objects. One particular source of interest is NGC7793-111, for which we noticed strong circumstellar H$\alpha$, clear Fe~\textsc{ii} emission features in the left wing of H$\alpha$, and weak [Ca~\textsc{ii}] $\lambda\lambda$7291--7323 emission (although the spectrum suffers from various residuals and artifacts). The area around the [O~\textsc{i}] $\lambda$6300 line was highly contaminated, and therefore the distinction between an LBV and a sgB[e] classification could not be made. \citet{Maravelias2023} provided additional photometric information, which resulted in a sgB[e] classification. We further classified one BSG (NGC7793-721) and one YSG (NGC7793-30) candidate. Six H~\textsc{ii} regions and several background galaxies and foreground stars were found in the sample.

\section{Discussion}\label{sec:discussion}

To determine which of our classified targets have previous spectral classifications in the literature, we searched for all publications providing spectral types of massive stars in our ten target galaxies and compiled a catalog of these sources. Then we cross-matched our 541 targets against the known massive stars in our catalog and found 31 matches, all of which were among our 185 massive stars, yielding 154 newly classified massive stars.

Table~\ref{tab:matched_catalog} lists the IDs and coordinates of the matched sources from our work and the literature (in degrees, J2000), the newly determined and previous classifications, and a comment field listing the cross-matched ID from the literature. The matched sources include two LBV candidates, a sgB[e] and an Fe star (which we reclassify as sgB[e]; see \citealt{Maravelias2023} for more details), 10 RSG for which we provide a spectral type classification for the first time, a cool star\footnote{Reduction artifacts prevented a more accurate classification.} that has been previously classified as a RSG, three \hii\ regions in M83 that have been identified as Wolf-Rayet (WR) clusters, and several O, B and A supergiants for which more accurate spectral types are available in the literature. We note an interesting case, NGC3109-167 classified as KI in this work (as weak TiO bands were detected), which has a previous classification as G2I, therefore may exhibit spectral type variability. NGC~300-52 (K-M V) was previously classified as a WR candidate by narrow-band imaging \citep{Schild2003}, however, its proper motion from \textit{Gaia} DR3 confirms that it is a foreground source.

To visualize our results, we proceeded to construct CMDs and figures showing the distribution of the classified sources in each galaxy. In the following subsections, we present a discussion of the CMDs, the spatial distribution plots, and the effectiveness of our target selection method.

\subsection{Color-magnitude diagrams and spatial distribution plots}
\label{sec:discusscmd}

We constructed both cumulative CMDs and CMDs for the individual galaxies in the mid-IR (m$_{3.6}$ vs. $m_{3.6}-m_{4.5}$ and $J-m_{3.6}$), near-IR (K$_s$ vs. $J-K_s$) and optical ($G$ vs. $G_{BP}-G_{RP}$ and $g$ vs. $g-r$), presenting the evolved massive stars (RSG, BSG, YSG, LBV, and sgB[e]) classified in this work, to verify the robustness of our classifications. Figure~\ref{fig:AbsMagCMDs} presents four cumulative CMDs for all our evolved massive stars and \ion{H}{ii} regions in all ten galaxies. For the construction of these CMDs, we used the distances in Table~\ref{tab:galaxies} to convert to absolute magnitude. The CMDs for the individual galaxies are presented in Appendix \ref{sec:ap_cmdclass}. These show apparent magnitude vs. color and also include the background galaxies. The CMDs for NGC~300 (Figure~\ref{fig:NGC300_CMD}) and NGC~55 (Figure~\ref{fig:NGC55_CMD}) are noteworthy, as they contain the largest number of classified targets (110 and 85, respectively).

In the CMDs, we have grouped together the candidate supergiants with the confirmed supergiants. We also merged the Galaxies, AGN and quasars, labelling them as "Galaxies" in the CMDs. For the optical CMDs, we use photometry from \textit{Gaia} DR2 and Pan-STARRS1, when available. For the galaxies with near-IR data (e.g. VHS) we created CMDs similar to \cite{Bonanos2009}, which are used to separate the LBVs and sgB[e] from the rest of the sources. Unfortunately, both the optical and the \textit{JHK$_s$} photometry is scarce for our sample; the Araucaria Project \citep[e.g.][]{Karczmarek2021} will soon release near-IR photometric maps for some of our galaxies. The \textit{Spitzer} MIPS [24] band is not reliable for the galaxies studied in our survey because of the low resolution of MIPS ($6\arcsec$), which corresponds to large regions at distances of a few Mpc, therefore it was not used.

By examining the positions of evolved massive stars on the CMDs, we confirm the location of sgB[e] stars at around $J - m_{3.6} \sim 3.0$~mag, in agreement with the location of these stars in the respective CMDs in the Magellanic Clouds \citep{Bonanos2009, Bonanos2010}. The 
emission objects are found to be colocated with the LBVs and sgB[e], with the exception of WLM-1325 (see Figure~\ref{fig:WLM_CMD}) at M$_{3.6} \sim -6$~mag. In particular, we note the priority~1 emission line objects NGC300-309 and NGC7793-47 (Figures~\ref{fig:NGC300_CMD} and \ref{fig:NGC7793_CMD}), and speculate that they could be faint LBVs. We find the majority of RSG and YSG to be outside our priority system (see Section~\ref{sec:effectiveness}). We finally highlight the BSGs with mid-IR excess (see e.g. Figure~\ref{fig:NGC300_CMD}), which implies the presence of circumstellar material around them. Some massive stars do not appear as priority targets because they did not pass our $m_{4.5}$ criterion. We detect \ion{H}{ii} regions mainly in more distant and inclined (edge-on) galaxies, e.g. NGC~253 and M83, as crowding and contamination of the mid-IR photometry used to select targets increases with distance (see Section~\ref{sec:effectiveness}).

Given the large number of RSGs, we can explore trends and comment on outliers. We find that RSGs cluster at $J - K_{S} \sim 1.0$~mag and lie on a vertical branch spanning $\sim3$~mag in $m_{3.6}$ and $\sim2.5$~mag in $K_s$ (see Figure~\ref{fig:NGC55_CMD} for NGC~55 and Figure~\ref{fig:NGC300_CMD} for NGC~300), similar to the LMC \citep{Bonanos2009, Woods2011, Jones2017} and SMC \citep{Bonanos2010, Ruffle2015}. In all these galaxies, the bulk of the population is concentrated within $\sim2$~mag, between $-$9.5 and $-$11.5~mag in both $M_{3.6}$ and $M_{Ks}$.
Interestingly, the most luminous RSGs were selected as priority targets, as they have the highest mass-loss rates, and therefore dustier environments and are brighter in the mid-IR. We estimate that we probe RSG luminosities between log$(L/ \lsun) \sim 4.4-5.6$, from $M_{K_S}$ (using a bolometric correction BC$_K=$ 2.69~mag and adopting the distances from Table \ref{tab:galaxies}). As improved distances exist for some target galaxies, the location of RSGs on the CMDs and the Hertzsprung-Russell diagram will be revisited in de Wit et al. (in prep.).

We note that several RSGs have unusually blue mid-IR colors $m_{3.6} - m_{4.5} \lesssim -0.5$ mag, but normal optical colors and spectra, i.e. NGC55-790, NGC55-1040, NGC55-1047, NGC253-1534, NGC55-1734, NGC247-2912 and NGC247-3231 (see Figure~\ref{fig:NGC55_CMD} and~\ref{fig:NGC247_CMD}). We attribute their blue mid-IR colors to their faint mid-IR photometry (i.e. $m_{4.5}\sim17.5$~mag), which is likely contaminated in these edge-on galaxies, and to variability, as the 3.6~$\mu$m and 4.5~$\mu$m data were obtained in some cases at different epochs. Furthermore, the blue colors could be due to the CO absorption bands at 4.5~$\mu$m \citep{Verhoelst2009}. Similarly, NGC247-3683 and NGC247-2966 have blue optical colors, but a classification of "MI+Neb." The most likely explanation for the blue colors of these RSGs is that they are part of a system or cluster including at least one hot star. The blue component was not accessible and therefore undetected by our spectra, but would provide the necessary ionizing radiation to produce the observed emission lines.

Comments on specific targets are given below:

\begin{itemize}
   \item LBVs in NGC~55: NGC55-736 and NGC55-2924 are not priority targets as they do not satisfy the apparent magnitude selection criterion in $m_{4.5}$. NGC55-2924 does not satisfy the absolute magnitude criterion in $m_{3.6}$ either, which implies that our magnitude cuts exclude fainter LBVs. 

  \item NGC300-357 is the only candidate YSG found among our priority targets (P3; the spectrum of the object has H${\alpha}$ in emission).

  \item NGC1313-48 is a remarkable blue supergiant (BI), very luminous in the mid-IR, but lacking optical photometry (a P6 target), which is shown in the mid-IR CMD (Figure~\ref{fig:NGC1313_CMD}). It is located near the center of the galaxy (Figure~\ref{fig:NGC1313_class}).
\end{itemize}

We finally constructed spatial distribution diagrams (Figures~\ref{fig:WLM_class}-\ref{fig:NGC7793_class} in Appendix \ref{sec:ap_spclass}) for the evolved massive stars found in each galaxy of our sample using images from the Digitized Sky Survey 2 (DSS2). We find the evolved massive stars and the \hii\ regions to be mainly located in the plane of the host galaxy or its spiral arms, which is the expected location for such objects. These diagrams helped us correct the YSG classification for NGC~3109--1209 (due to its redshifted spectrum that matched the radial velocity of the target galaxy) to a foreground source, since it is not located on the disk of the host galaxy and has a high proper motion in \textit{Gaia} DR3.

\subsection{Effectiveness of the selection method}\label{sec:effectiveness}

As described in Section~\ref{sec:catalog}, out of the 956 slits, 763 were successfully extracted and of these, 454 were classified (i.e. 47\% of the total slits, or 59\% of the successfully extracted ones). We identified 185 evolved massive stars among these, which is 19\% of the total slits (956), 24\% of the successfully extracted targets (763), 28\% of the classified targets (541) and 41\% of the accurately classified targets (454). This moderate success rate is due to two factors: (i) only 384 out of 956 targets (195 out of the 541) were priority targets, and (ii) many targets, independent of priority class, suffered from reduction errors (e.g. vignetting, overlap, poor sky subtraction), which prevented a classification.

In Table~\ref{tab:PrioSpecClass}, we present the distribution of stars per galaxy and per massive star class for the total number of classified stars (541), the priority stars (195), the classified massive stars (185) and the massive stars that were in the priority system (44). The breakdown of priority massive stars by spectral type is also given, along with the success rates of recovering massive stars. The average success rate of selecting massive stars was $28\%$, with NGC~55 and NGC~300 achieving the highest rates ($\sim 50\%$). Excluding Sextans A, which only had 1 priority target, and M83 and NGC~253, which were highly contaminated by H~\textsc{ii} regions, the average success rate of our priority system in selecting evolved massive stars is 36\%, while for NGC~55, NGC~247, NGC~300 and NGC~3109 the success rate ranged from 38-42\%. We also find that $15-25\%$ of our classified RSG, BSG, LBVc and YSG were priority targets, whereas for sgB[e] stars the fraction is much higher at $86\%$. By comparing the CMDs with the selection of priority targets (Appendix~\ref{sec:ap_cmd}) to the CMDs with the classified sources, we find that many of the "filler" stars turned out to be RSG or YSG (see e.g. the CMD of NGC~300 in Figure~\ref{fig:CMD_NGC300} and in Figure~\ref{fig:NGC300_CMD}). In some galaxies, priority targets with color $m_{3.6} - m_{4.5} \geq 1.0$ mag were not placed in slits (see e.g. Figures~\ref{fig:CMD_NGC1313} and~\ref{fig:CMD_NGC7793}). This is due to the limitations imposed by MOS observations.

We revisit the cuts applied to limit contamination (see Section~\ref{sec:targets}), i.e. the magnitude cut at 4.5~$\mu$m to remove background galaxies, the color cut $m_{3.6} - m_{4.5} \geq 0.1$~mag to avoid yellow foreground dwarfs and \textit{Gaia} cleaning to remove foreground sources. The magnitude cut was quite successful, yielding a 4\% contamination by galaxies (as a fraction of the 454 classified targets), however, it limited us from selecting all but the most luminous massive stars in galaxies found at distances greater than 4~Mpc. The \textit{Gaia} cleaning resulted in a 13\% contamination by foreground stars. Future data releases of \textit{Gaia} will improve the removal of foreground sources. Our selection criteria in M83 yielded a large number of H~\textsc{ii} regions, as the \textit{Spitzer} IRAC photometry with its $\sim 2\arcsec$ resolution gets increasingly contaminated with distance. We note that most H~\textsc{ii} regions in M83 were P1 and P2 targets (i.e. with optical counterparts). Furthermore, as all targets have been observed with similar exposure times throughout the observing campaign, the robustness of the classifications generally drops with distance. Ideally, targets without optical counterparts (P3, P4 and P6) need to be followed up with near-IR spectroscopy. This is particularly important for sources with $m_{3.6} - m_{4.5} > 1.0$~mag.

The completeness of our study of dusty massive stars (i.e.\ the priority targets placed in slits and observed vs.\ the total number of available priority targets) in these galaxies ranges from 14\% (NGC 1313 and NGC 3109) to 56\% (WLM), with an average of 30\% (see last column of Table~\ref{tab:prio}). This does not take into account the number of sources we rejected when creating the original catalog ($\sim 0.1\%$) when deblending, as they are negligible. The completeness achieved, along with the small number of detected dusty, massive stars per spectral class, limits our ability to determine the frequency of episodic mass loss. However, these percentages can be used to estimate a conservative total number of dusty, massive stars in each galaxy, which will of course be limited by the galaxy inclination, as well as the angular resolution and sensitivity of \textit{Spitzer}. For example, using the completeness for each galaxy from Table~\ref{tab:prio} and the corresponding number of priority RSGs from Table~\ref{tab:PrioSpecClass}, we predict that six additional dusty RSGs exist in NGC~55, four in NGC~300 and one in NGC~247 and NGC~7793. These are of interest for characterizing dust-enshrouded RSGs \citep[e.g.][]{Beasor2022} in more distant galaxies, especially in NGC~55 and NGC~300, which are expected to contain 17 and 12 dusty RSGs, respectively.

\subsection{\hii\ regions}
\label{sec:hiiregions}

In addition to evolved massive stars, our catalog contains 99 \hii\ regions. The majority of them were found in M83 and NGC~253, where we identified 28 and 24 \hii\ regions, respectively (see Table~\ref{tab:specClass}). 
Since there have been previous studies of \hii\ regions in M83, we cross-matched the 28 \hii\ regions identified in that galaxy, using a $1\arcsec$ radius, and found three matches in \citet[][M83-127 with 132, M83-433 with 209, M83-1048 with 247]{Rumstay1983}, one match with \citet[][M83-993 with HII058]{Long2022} and one match with \citet[][M83-1048 with HII-13]{Winkler2023}. Cross-matches with the \hii\ regions reported by \citet{bresolin2002}, \citet{Bresolin2005}, and \citet{Bresolin2009} did not yield any matches within a $1\arcsec$ radius. We therefore present 24 new \hii\ regions in M83, which can be further studied to e.g.\ identify shocked material (e.g.\ supernova remnants) or to derive the abundances \citep[e.g.][]{Bresolin2016}. 

\begin{figure*}
   	\resizebox{1.0\hsize}{!}{\includegraphics{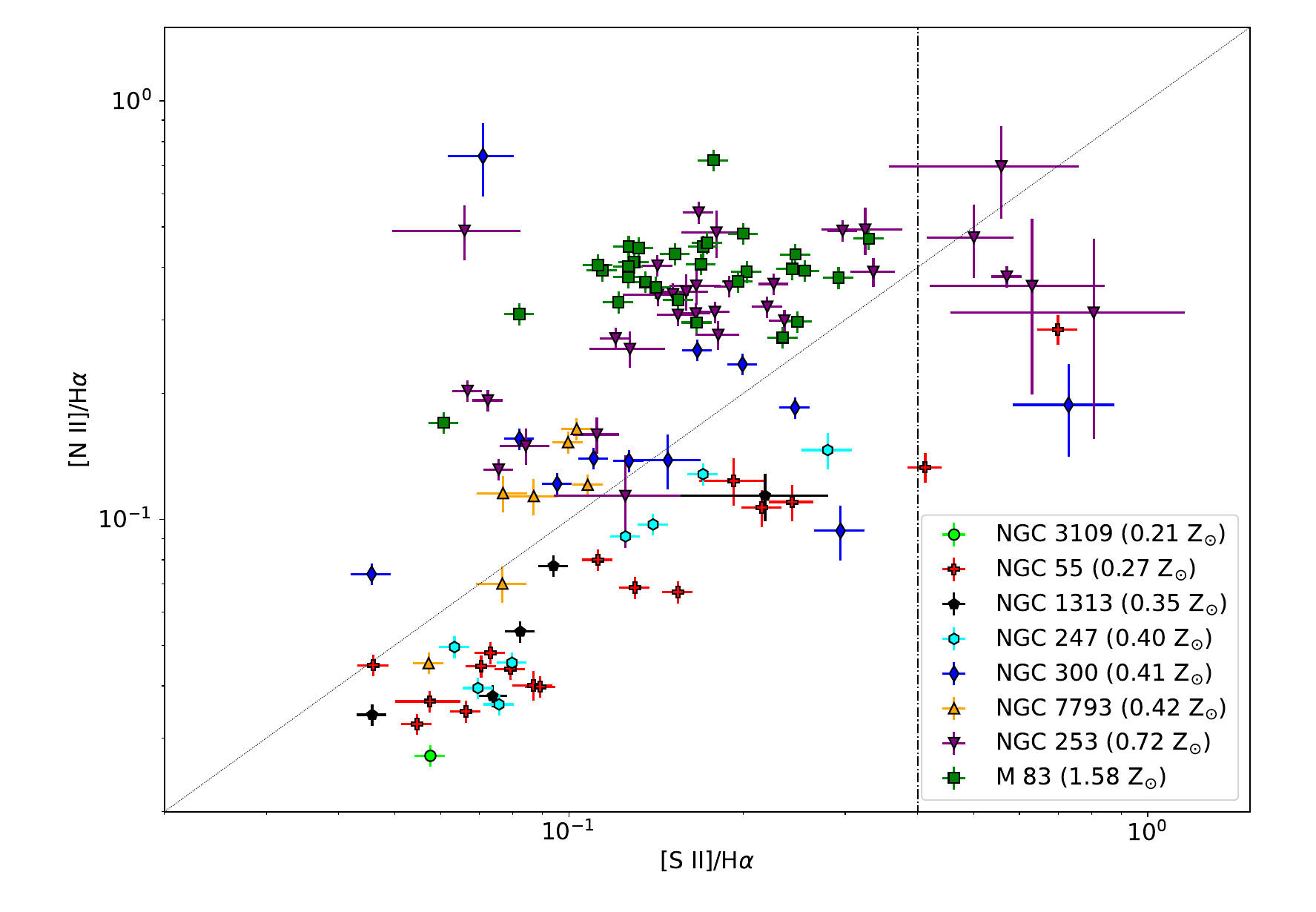}}
  	\caption{[\ion {N} {ii}]/H$\alpha$ vs. [\ion {S} {ii}]/H$\alpha$ for all \hii\ regions and spectra classified as Nebulae, found in eight of our target galaxies, which range from Z=0.2--1.6 \zsun. The vertical line indicates the value [\ion {S} {ii}]/H$\alpha=0.4$, above which the emission is shocked. The diagonal shows the 1:1 relation. }\label{fig:hnsratios}
\end{figure*}

Historically, the ratio [\ion {S} {ii}]/H$\alpha$ has been used as a diagnostic to distinguish between photo-ionized and shocked emission, using 0.4 as the boundary \citep{Mathewson1973}. However, recent works suggest using additional nebular line ratios to better assess the origin of the nebular emission \citep{Kopsacheili2020}. Following this approach, we plotted our emission-line flux ratio measurements (see Section~\ref{sec:class_em} and Table~\ref{tab:hii_catalog}) on the diagram [\ion {N} {ii}]/H$\alpha$ vs. [\ion {S} {ii}]/H$\alpha$ (see Figure~\ref{fig:hnsratios}), discovering eight targets with [\ion {S} {ii}]/H$\alpha$ higher than 0.4. Among these objects, we distinguish four "RSGs+nebula" (NGC55-889, NGC300-NC4, NGC253-452 \& NGC253-1162), two "blue+nebula" (NGC55-NC12 \& NGC253-NC4), a \hii\, region (NGC253-1657) and a cluster (NGC253-769). Further analysis is needed to reveal the origin of this shocked material (e.g. SN remnants in the field, shock interaction because of strong winds, etc.) and whether any of the four "RSGs+neb" contain bow shocks similar to those around Betelgeuse \citep{Mohamed2012, Mackey2012}, or IRC-10414 \citep{Gvaramadze2014}. Furthermore, we identified two targets with an enhanced [\ion {N} {ii}]/H$\alpha$ ratio with respect to their host galaxies. Their classification is ``composite'' (NGC253-1) and ``EmObj+neb'' (NGC300-NC21). In particular, the latter shows stellar \ion {N} {ii} emission in the series of $\lambda\lambda5668-5712$ and the lines $\lambda 6482$ and $\lambda 6611$. These features in emission could suggest a N-rich mass-losing star, which is enriching the gas around it and enhancing the relative abundance of nitrogen over hydrogen. 

The ratio [\ion {N} {ii}]/H$\alpha$ is a well-known indicator for metallicity in \hii\ regions, SNRs \citep{Leonidaki2013} as well as in star-forming galaxies \citep{Storchi-Bergmann1994}. In Figure~\ref{fig:hnsratios}, the measurements in each galaxy follow distinct lines depending on their metallicity, with the most metal poor ones located in the lower part of the diagram and the most metal-rich in the upper part. However, the trend does not hold for supersolar oxygen abundances \citep{Kewley2008}, which explains why the measurements for our highest metallicity (0.7 \zsun\, and 1.6 \zsun) galaxies overlap. The case of NGC~247 (0.4~\zsun) is noteworthy, as the line ratios overlap with those of NGC~55 (0.27~\zsun) and NGC~1313 (0.35~\zsun) especially for three \hii\ regions located in the outskirts of the galaxy, instead of those of NGC~300 and NGC~7793 ($\sim0.4\,\zsun$), implying a metallicity of $\sim0.3~\zsun$ for this galaxy. However, the identification of 16 M-type RSG \citep[a narrow effective temperature range of RSG was also noted by][]{Davidge2021} in this galaxy is consistent with a metallicity of 0.4~\zsun.

\subsection{Other objects of interest}

Carbon stars were identified only in nearby galaxies (seven in WLM\footnote{WLM-1120 corresponds to star 10 in \citet{Britavskiy2015}.}, one in Sextans A, two in NGC 300). The carbon star SextansA-12116 has been reported as a variable \citep{Boyer2015b}, extreme AGB star by \citet{Boyer2017}. In more distant galaxies it is likely that we have also observed such stars, but because of their lower luminosities compared to RSGs, we were not able to classify them since the S/N of their spectra was too low. Carbon stars are difficult to filter out because they occupy similar regions as the RSGs in most CMDs. Among the targets classified as galaxies, there are three quasars, two of which have an X-ray counterpart: NGC55-664 (4XMM J001547.5-391240), NGC300-349 (4XMM J005449.7-374000), NGC3109-299. All quasars have broad emission features and are at redshifts between 1--2.

Another noteworthy target (P1) is the ``emission object'' NGC~300-309, which turns out to be the supernova imposter SN~2010da or NGC~300 ULX-1 \citep{Binder2020}. It has been explained as an ultraluminous X-ray source with a RSG donor and a neutron star \citep{Heida2019}. We detected H$\alpha$ circumstellar emission, which remains strong at above 100 times the continuum, in agreement with the latest spectrum in 2015 reported by \citet{Villar2016}. Poor sky subtraction prevents us from detecting other lines.

\section{Summary}\label{sec:conclusions}

We have presented the results of a spectroscopic survey of luminous, evolved massive stars with dust in nearby (1$-$5~Mpc) southern galaxies, which were observed with FORS2 on VLT, as part of the ASSESS project. We developed a ``priority system'' by applying magnitude and color cuts on archival \textit{Spitzer} photometry to select dusty, evolved massive stars for multi-object spectroscopy in the galaxies \object{WLM}, \object{NGC~55}, \object{NGC~247}, \object{NGC~253}, \object{NGC~300}, \object{NGC~1313}, \object{NGC~3109}, \object{Sextans~A},
\object{M83} and \object{NGC~7793}. Out of the 763 objects observed, we obtained classifications for 541 of them, of which 454 were robust classifications; 87 were given a general classification and 222 were not of sufficient quality to allow for a classification. The robust classifications include 185 evolved massive stars, 99 \hii\ regions, 10 carbon stars, 21 galaxies (including quasars and AGN) and 61 foreground sources, yielding a contamination fraction of 12\%. Of the 185 evolved massive stars, 154 have spectral types reported for the first time. The evolved massive stars include 129 RSG, 27 BSG, 10 YSG, four LBVc, seven sgB[e] and eight emission line objects; $24\%$ of these are within the priority system, i.e. have clear signs of circumstellar dust. We further measured the [\ion {N} {ii}]/H$\alpha$ and [\ion {S} {ii}]/H$\alpha$ line ratios for all \hii\ regions and nebular emission-line spectra, which yielded eight shocked emission regions, four of which contain a RSG component.

The fraction of dusty massive stars observed with respect to those initially selected in these galaxies is on average $30\%$, due to the constraints imposed by the MOS observations. We report a success rate of $28\%$ for identifying massive stars among our observed spectra (which include ``filler'' sources outside the priority system), while the average success rate of our priority system in selecting evolved massive stars was 36\% (see Section~\ref{sec:effectiveness}), due to reduction errors (e.g. vignetting, overlapping spectra, poor sky subtraction). The efficiency of recovering priority massive stars vs. distance given the resolution of \textit{Spitzer} ($1.7\arcsec$ at 3.6~$\mu$m) starts dropping at 3.5 Mpc (Table~\ref{tab:PrioSpecClass}), which implies that the \textit{James Webb Space Telescope} with its $0.1\arcsec$ resolution will enable the photometric selection of dusty massive stars out to 60~Mpc. Furthermore, both the \textit{Euclid} mission and the \textit{Nancy Grace Roman} Space Telescope, with their wide-field capabilities in the optical and near-IR will provide deep photometry, which combined with the variability information from the upcoming \textit{Vera C. Rubin} Observatory’s Legacy Survey of Space and Time, will enable the identification and detailed study of all massive stars, including the dusty, evolved ones, in a large number of nearby galaxies. Moreover, machine-learning methods are expected to greatly improve the efficiency of identifying massive stars, such as the photometric classifier presented by \citet{Maravelias2022} with its overall accuracy of $\sim80\%$, and therefore to increase the sample of dusty, evolved massive stars.

The resulting catalog is the largest catalog of evolved massive stars and of RSG beyond the Local Group. This catalog can serve as a valuable resource for follow-up studies of specific targets or types of objects, such as detailed modeling of RSG (de Wit et al., in prep.). Follow-up spectroscopy is currently underway for LBVs, certain blue and yellow supergiants and selected RSGs. A forthcoming paper will present the results of our observations of dusty, evolved targets in galaxies located in the northern sky.


\begin{acknowledgements}
The authors acknowledge funding support from the European Research Council (ERC) under the European Union’s Horizon 2020 research and innovation program (Grant agreement No. 772086). EZ also acknowledges support from the Hellenic Foundation for Research and Innovation (H.F.R.I.) under the ``3rd Call for H.F.R.I. Research Projects to support Post-Doctoral Researchers'' (Project No: 7933). We thank Elias Koulouridis for helping with the classification of the quasars, Despina Hatzidimitriou for helpful discussions and the referee for a careful reading and a very prompt report with suggestions that improved the manuscript.

Based on observations collected at the European Southern Observatory under ESO programme 105.20HJ and 109.22W2. This work is based in part on observations made with the \textit{Spitzer Space Telescope}, which is operated by the Jet Propulsion Laboratory, California Institute of Technology under a contract with NASA. This work has made use of data from the European Space Agency (ESA) mission {\it Gaia} (\url{https://www.cosmos.esa.int/gaia}), processed by the {\it Gaia}
Data Processing and Analysis Consortium (DPAC,
\url{https://www.cosmos.esa.int/web/gaia/dpac/consortium}). Funding for the DPAC has been provided by national institutions, in particular the institutions participating in the {\it Gaia} Multilateral Agreement. Based on observations made with the NASA/ESA Hubble Space Telescope, and obtained from the Hubble Legacy Archive, which is a collaboration between the Space Telescope Science Institute (STScI/NASA), the Space Telescope European Coordinating Facility (ST-ECF/ESAC/ESA) and the Canadian Astronomy Data Centre (CADC/NRC/CSA).  The Digitized Sky Surveys were produced at the Space Telescope Science Institute under U.S. Government grant NAG W-2166. The images of these surveys are based on photographic data obtained using the Oschin Schmidt Telescope on Palomar Mountain and the UK Schmidt Telescope. The plates were processed into the present compressed digital form with the permission of these institutions. This research made use of Astropy\footnote{\url{http://www.astropy.org}}, a community-developed core Python package for Astronomy \citep{astropy2013, astropy2018} and Photutils, an Astropy package for detection and photometry of astronomical sources \citep{photutils}.

\end{acknowledgements}

\begin{landscape}
\begin{table}
\small
\centering
\caption{Catalog of matched evolved stars.}
\label{tab:matched_catalog}
\begin{tabular}{l r r r r l l l}
\hline
\hline      
ID    & RA   & Dec.  & RA Lit. & Dec. Lit. & Spectral Class. & Spec. Class Lit.  & Comments\\
\hline
WLM-95		&	0.50967	  &	$-$15.46217   &	0.50967	     &	$-$15.46217     &	sgB[e]		     &	Em. line star&	WLM-23 in \citet{Britavskiy2015} \\
WLM-1120	&	0.49258	  &	$-$15.49978   &	0.4926	     &	$-$15.4998      &	C		     &	Carbon star  &	\\
WLM-1234	&	0.48392	  &	$-$15.44022   &	0.4840	     &	$-$15.44016667  &	A-star		     &	A7 Ib	     & A6 in \citet{Bresolin2006} \\
WLM-1285	&	0.47412	  &	$-$15.48049   &	0.47403	     &	$-$15.4805      &	K-M V		     &	M1-3 II, Fgd &	\\
NGC55-75	&	3.87163	  &	$-$39.23562   &	3.871625     &	$-$39.23561111  &	K I		     &	RSG	     &  \\
NGC55-87	&	3.88813	  &	$-$39.22835   &	3.888166667  &	$-$39.22833333  &	M I		     &	RSG	     &  \\
NGC55-146	&	3.91958	  &	$-$39.24735   &	3.919583333  &	$-$39.24727778  &	M I		     &	RSG	     &  \\
NGC55-200	&	3.98479	  &	$-$39.26915   &	3.984625     &	$-$39.26905556  &	M I + Neb.	     &	RSG	     &  \\
NGC55-324	&	3.90853	  &	$-$39.24726   &	3.908458333  &	$-$39.24727778  &	Cool star	     &	RSG	     &	RSG30 in \citet{Patrick2017} \\
NGC55-332	&	3.98038	  &	$-$39.24992   &	3.980333333  &	$-$39.25002778  &	M I		     &	RSG	     &  \\
NGC55-NC15	&	3.99023	  &	$-$39.26145   &	3.990458333  &	$-$39.26152778  &	K I		     &	RSG	     &  \\
NGC247-246	&	11.75904  &	$-$20.79448   &	11.75908333  &	$-$20.794425    &	sgB[e]	      	     &	sgB[e]	     &  \\
NGC300-52	&	13.71927  &	$-$37.69690    &	13.719125    &	$-$37.69697222  &	K-M V		     &	WR candidate & \#20 in \citet{Schild2003} \\
NGC300-125	&	13.71701  &	$-$37.67567   &	13.71695833  &	$-$37.67558333  &	M I		     &	RSG	     &  \\
NGC300-266	&	13.60651  &	$-$37.66517   &	13.60641667  &	$-$37.66516667  &	K I		     &	RSG	     &  \\
NGC300-273	&	13.70238  &	$-$37.63394   &	13.70233333  &	$-$37.63388889  &	M I		     &	RSG	     &  \\
NGC300-346	&	13.72130	  &	$-$37.65789   &	13.72133333  &	$-$37.65783333  &	M I		     &	RSG	     &  \\
NGC300-499	&	13.69984  &	$-$37.62761   &	13.69983333  &	$-$37.62755556  &	K I		     &	RSG	     &  \\
NGC300-NC3	&	13.88864  &	$-$37.71326   &	13.88866667  &	$-$37.71299444  &	B-A I-III	     &	B8	     & \#13 in \citet{Bresolin2002b} \\
NGC300-NC20	&	13.88312  &	$-$37.69420    &	13.88316667  &	$-$37.69398056  &	B I		     &	A2	     &	\#10 in \citet{Bresolin2002b} \\
NGC3109-167	&	150.74825 &	$-$26.15352   &	150.7480833  &	$-$26.15345556  &	K I		     &	G2I	     & \#14 in \citet{Evans2007} \\
SextansA-J101057&	152.73880  &	$-$4.68204    &	152.7388333  &	$-$4.681944444  &	O-star		     &	B0I	     &  \\
SextansA-J101058&	152.74168 &	$-$4.67437    &	152.7417083  &	$-$4.674444444  &	O-star		     &	B0.5I	     &  \\
SextansA-OB321	&	152.75271 &	$-$4.67896    &	152.75275    &	$-$4.678972222  &	O-star		     &	O9.7 I((f))  &	\\
SextansA-OB326	&	152.72419 &	$-$4.68687    &	152.7242083  &	$-$4.686944444  &	Hot star	     &	O7.5 III((f))&	\\
SextansA-OB421	&	152.74668 &	$-$4.66334    &	152.7467083  &	$-$4.663361111  &	O-star		     &	B1 I	     &  \\
SextansA-OB521	&	152.77228 &	$-$4.71108    &	152.7724167  &	$-$4.711138889  &	O-star		     &	O9.5 III-V   &	\\ 
SextansA-OB622	&	152.75987 &	$-$4.67072    &	152.7599167  &	$-$4.670722222  &	O-star		     &	B0 I	     &  \\
M83-79		&	204.25611 &	$-$29.85718   &	204.2559167  &	$-$29.85716667  &	\hii\ + H$\alpha$ star&	WR cluster   &	\#74 in \citet{Hadfield2005} \\
M83-721		&	204.24452 &	$-$29.80163   &	204.2446667  &	$-$29.80169444  &	\hii\ + H$\alpha$ star&	WR cluster   &	\#64 in \citet{Hadfield2005} \\
M83-780		&	204.17111 &	$-$29.86555   &	204.1710	     &	$-$29.86552778  &	\hii\ + H$\alpha$ star&	WR	     &	\#2 in \citet{Hadfield2005} \\
\hline
\end{tabular}
\vspace{0.3cm}

\end{table}
\end{landscape}
\begin{table*}
\centering
\caption{Distribution of classified, priority (Prio) and massive stars (MS) per galaxy and spectral class.}\label{tab:PrioSpecClass}
\begin{tabular}{l | r | r | r | r | r l l l l l | r | r}
\hline\hline
Galaxy	& (1)   & (2)  & (3)  & (4) & RSG & BSG	& YSG & LBVc & sgB[e] & Em. & (4)/(2) & (3)/(1)\\
Name	&  Class. & Prio &  MS & MS Prio &  & 	&  &  &  & Obj. &  & \\
\hline \\[-9pt]
WLM			& 22	&4 & 2 & 1  & - & -  & -  & - & 1 & - & 25\% & 9\% \\
NGC 55		& 101 & 37 & 50 & 14 & 11& 2  & -  & - & 1 & - & 38\% & 50\% \\
NGC 247		& 63 & 10 & 27 & 4  & 2 & -  & -  & - & 1 & 1 & 40\%  & 43\% \\
NGC 253		& 79 & 45 & 17 & 4  & 2 & -  & -  & - & 1 & 1 & 9\% & 22\% \\
NGC 300		& 126 & 31 & 62 & 13 & 8 & 1  & 1  & - & 2 & 1 & 42\% & 49\% \\
NGC 1313	& 17 & 7 & 4 & 1  & - & 1  & -  & - & - & -  & 14\% & 24\% \\
NGC 3109	& 17 & 5 & 3 & 2  & 1 & -  & -  & 1 & - & - & 40\%  & 18\% \\
Sextans A	& 17 & 1 & 6 & 0  & - & -  & -  & - & - & -  & 0\% & 35\% \\
M83		    & 59 & 42 & 1 & 1  & 1 & -  & -  & - & - & - & 2\% & 2\% \\
NGC 7793	& 40 & 13 & 13 & 4  & 2 & -  & 1  & - & - & 1 & 31\% & 33\% \\
\hline
\bf{Total}	& \bf{541} & \bf{195} & \bf{185} & \bf{44} & \bf{27} & \bf{4} & \bf{2} & \bf{1} & \bf{6} & \bf{4} & \bf{36\%}\tablefootmark{a} & \bf{28\%} \\
\hline
\end{tabular}

\tablefoot{
\tablefoottext{a}{This average percentage excludes NGC~253, Sextans~A and M83.}
}

\end{table*}

\begin{figure*}
    \centering
    \includegraphics[width=\columnwidth]{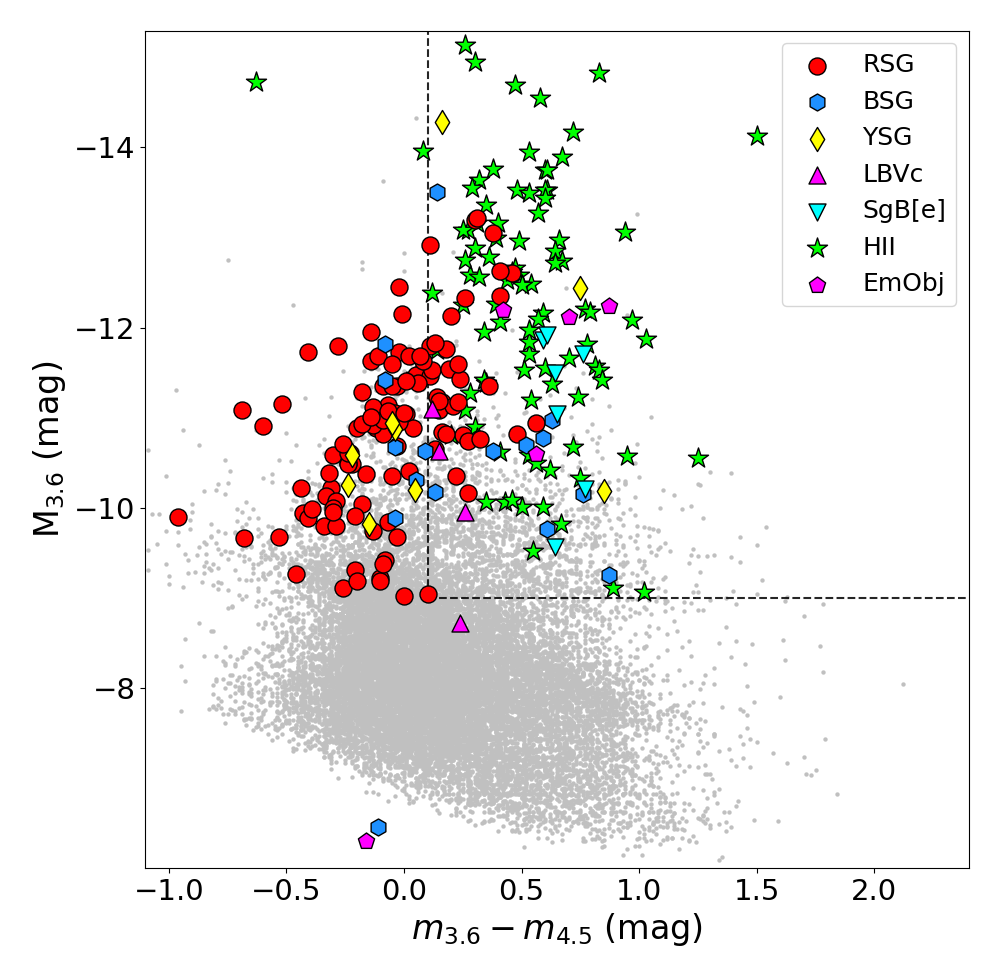}
    \includegraphics[width=\columnwidth]{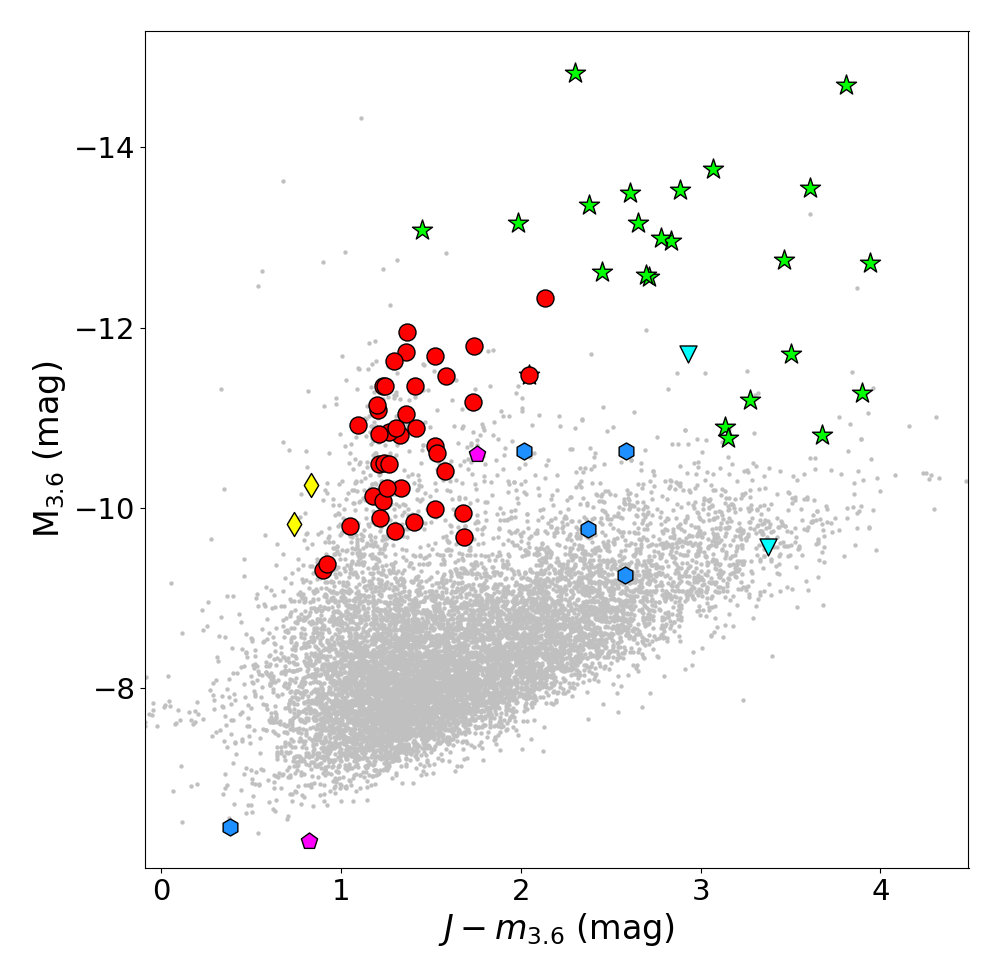}
    \includegraphics[width=\columnwidth]{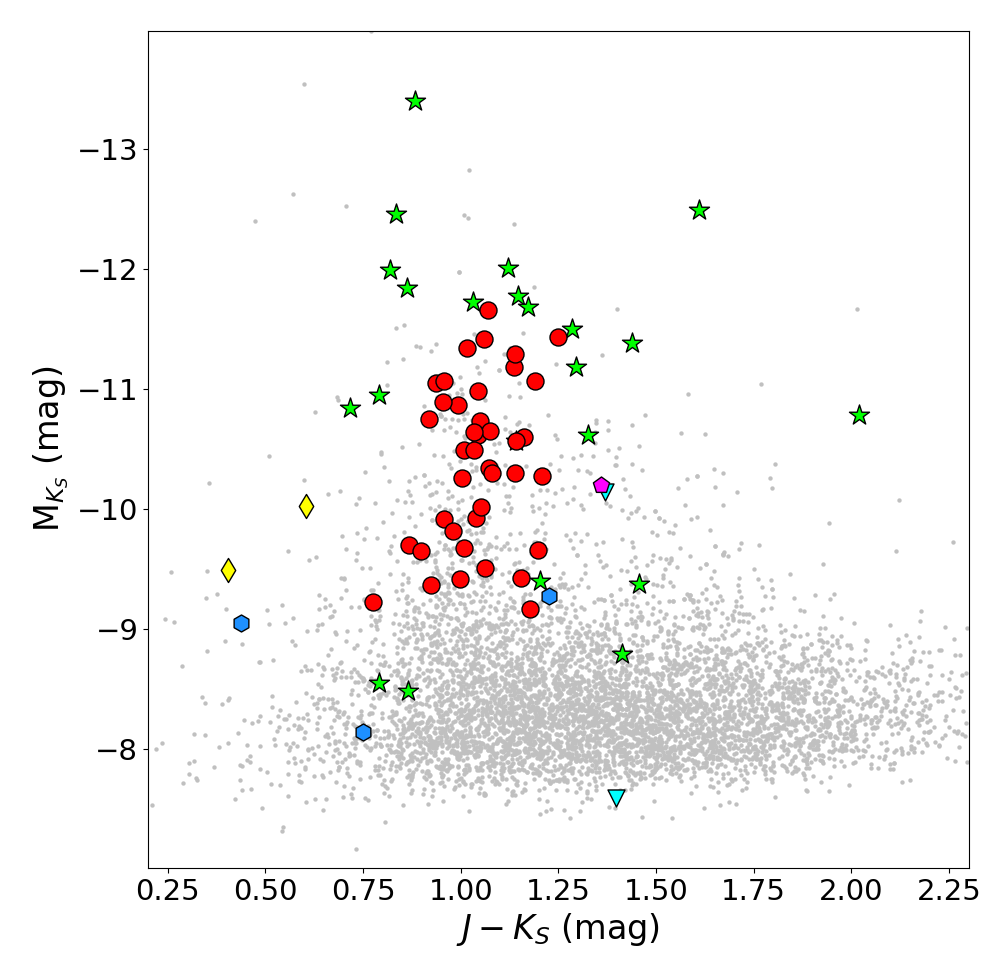}
     \includegraphics[width=\columnwidth]{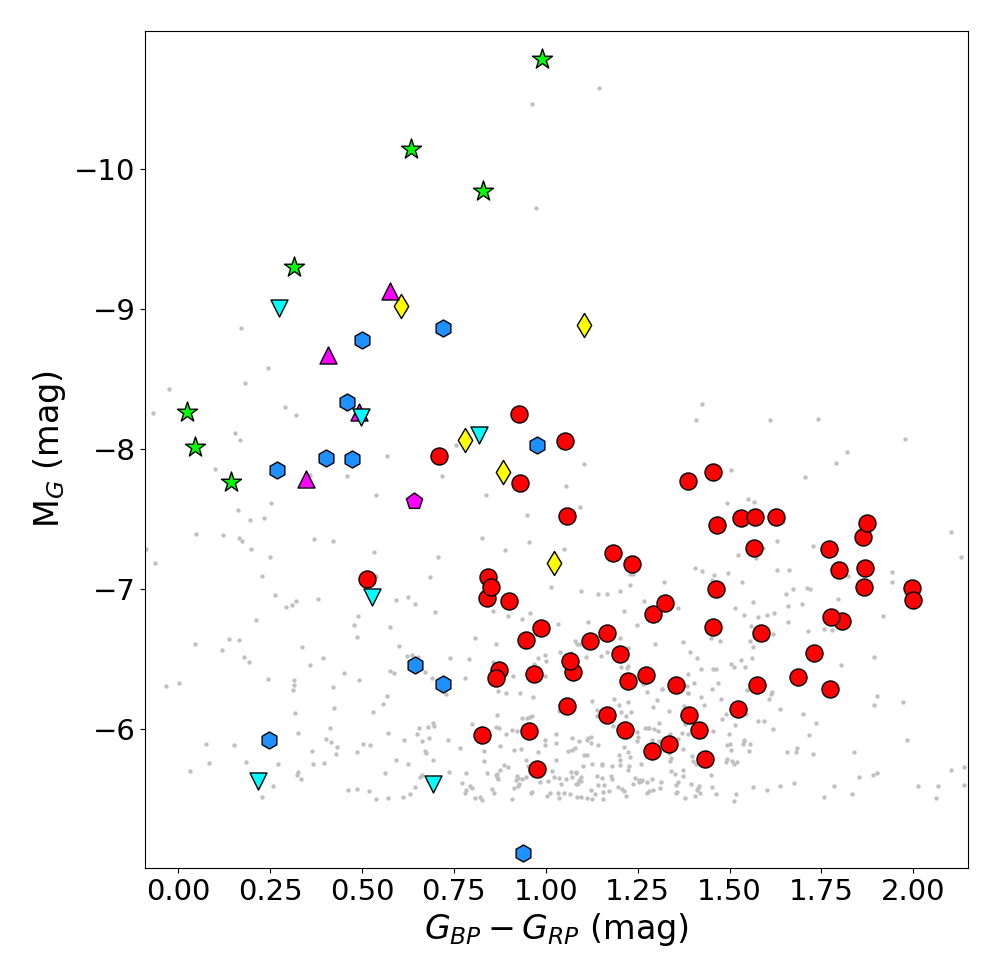}
    \caption{Infrared and optical cumulative absolute magnitude CMDs for all evolved massive stars and \ion{H}{ii} regions. Background sources are from NGC~300. The dashed lines in the first panel indicate the criteria used for target selection; all our priority targets are located within the box at the top right defined by these lines.}
    \label{fig:AbsMagCMDs}\end{figure*}

%
\bibliographystyle{aa} 
\bibliography{assessVLT.bib} 
%

\clearpage

\appendix
\onecolumn


\clearpage
\section{Target selection from {\it Spitzer} color-magnitude diagrams} \label{sec:ap_cmd}

\begin{figure}[!ht]
\centering
\begin{minipage}{.5\textwidth}
  \centering
  \includegraphics[width=.85\textwidth]{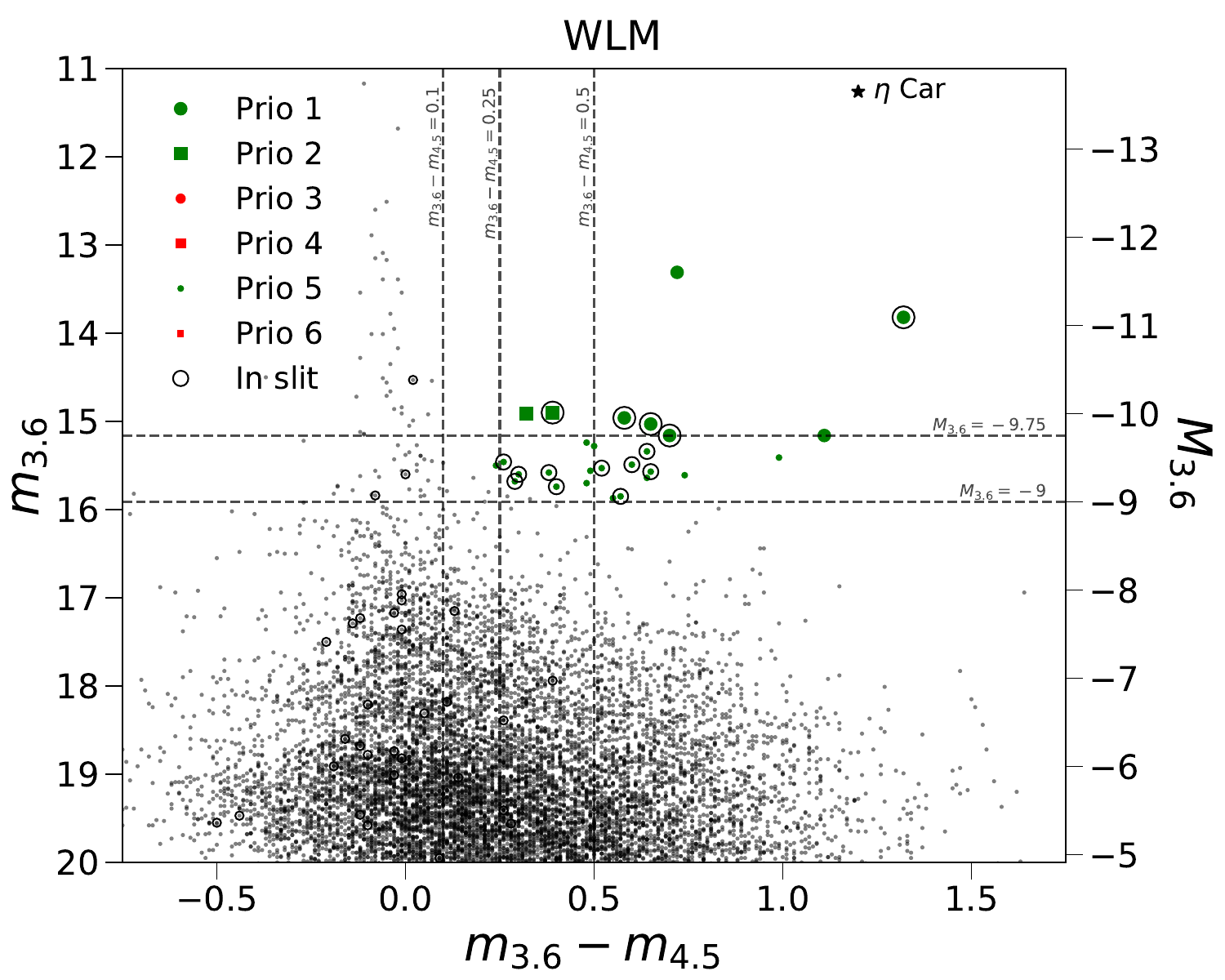}
  \captionof{figure}{{\it Spitzer} CMD of WLM.}
  \label{fig:CMD_WLM}
\end{minipage}%
\begin{minipage}{.5\textwidth}
  \centering
  \includegraphics[width=.85\textwidth]{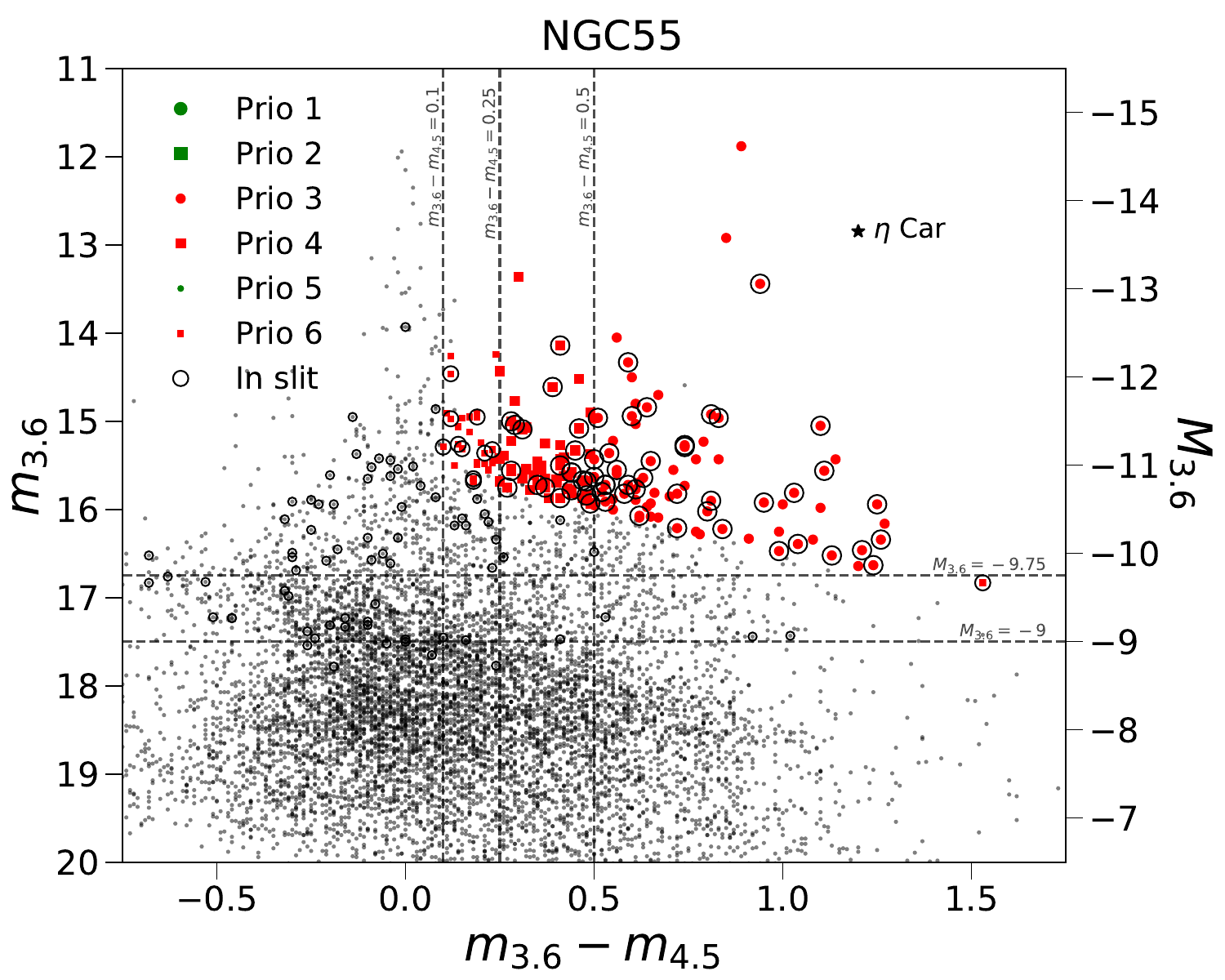}
  \captionof{figure}{{\it Spitzer} CMD of NGC~55.}
  \label{fig:CMD_NGC55}
\end{minipage}
\end{figure}

\begin{figure}[!h]
\centering
\begin{minipage}{.5\textwidth}
  \centering
  \includegraphics[width=.85\textwidth]{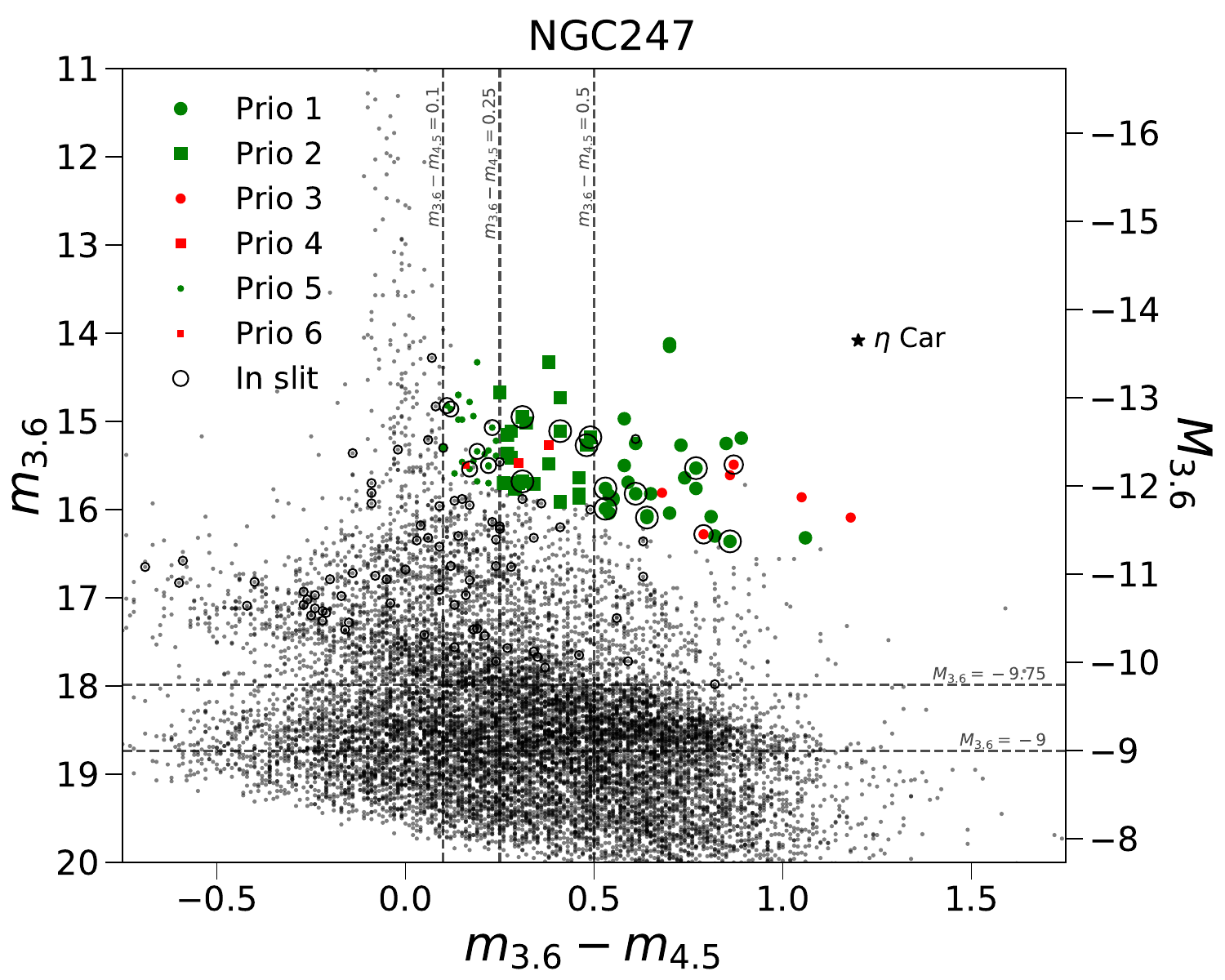}
  \captionof{figure}{{\it Spitzer} CMD of NGC 247.}
  \label{fig:CMD_NGC247}
\end{minipage}%
\begin{minipage}{.5\textwidth}
  \centering
  \includegraphics[width=.85\textwidth]{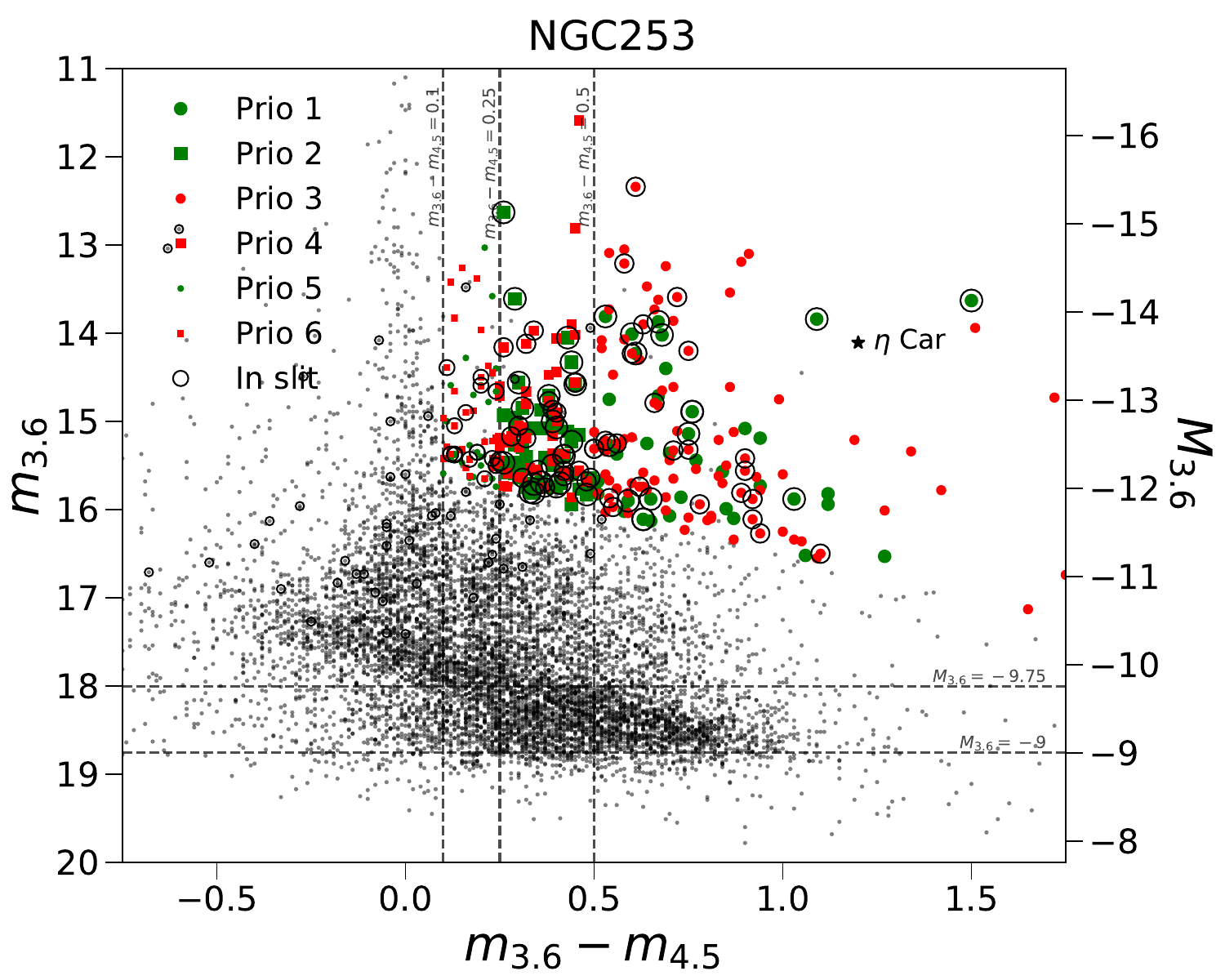}
  \captionof{figure}{{\it Spitzer} CMD of NGC~253.}
  \label{fig:CMD_NGC253}
\end{minipage}
\end{figure}

\begin{figure}[!h]
\centering
\begin{minipage}{.5\textwidth}
  \centering
  \includegraphics[width=.85\textwidth]{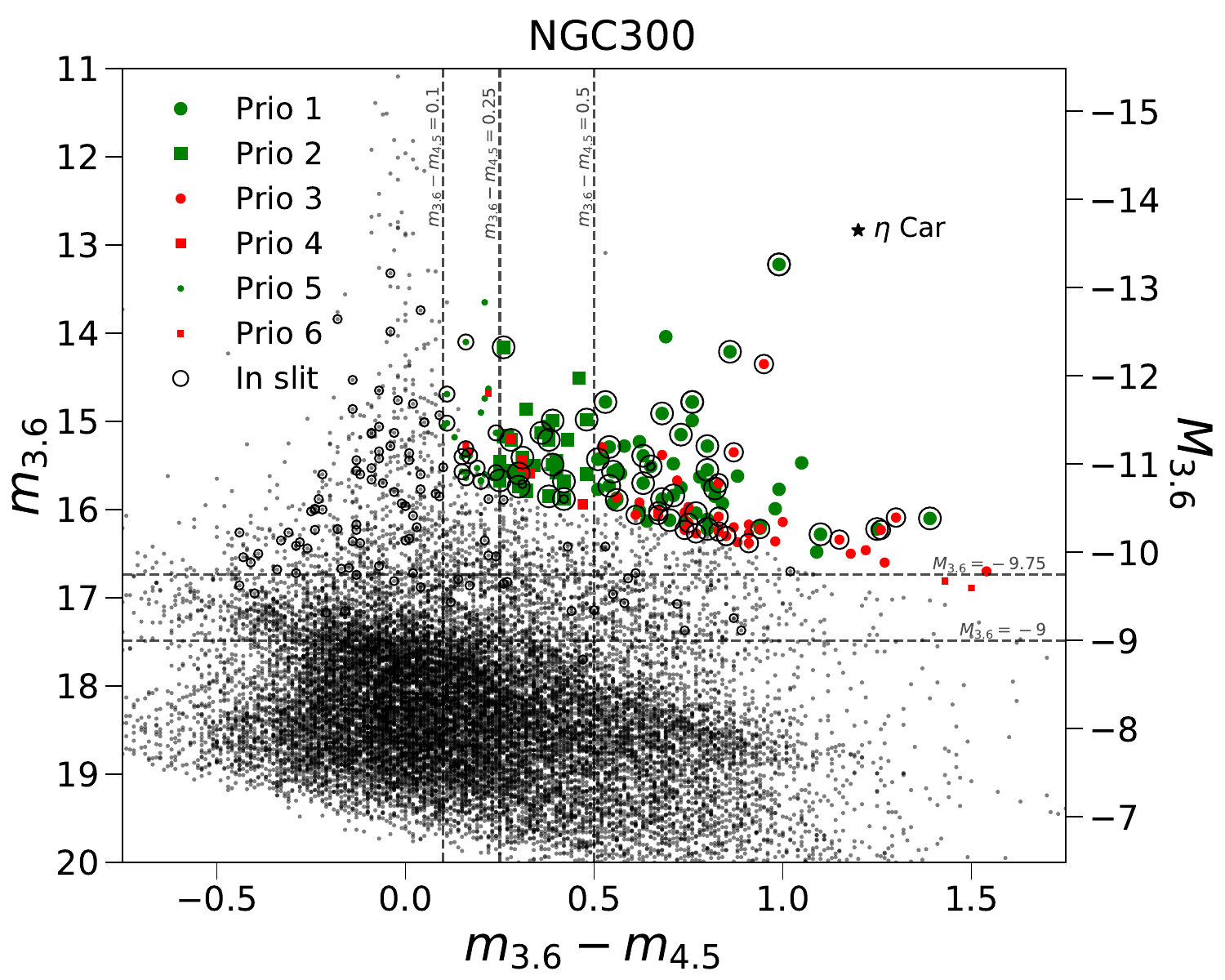}
  \captionof{figure}{{\it Spitzer} CMD of NGC 300.}
  \label{fig:CMD_NGC300}
\end{minipage}%
\begin{minipage}{.5\textwidth}
  \centering
  \includegraphics[width=.85\textwidth]{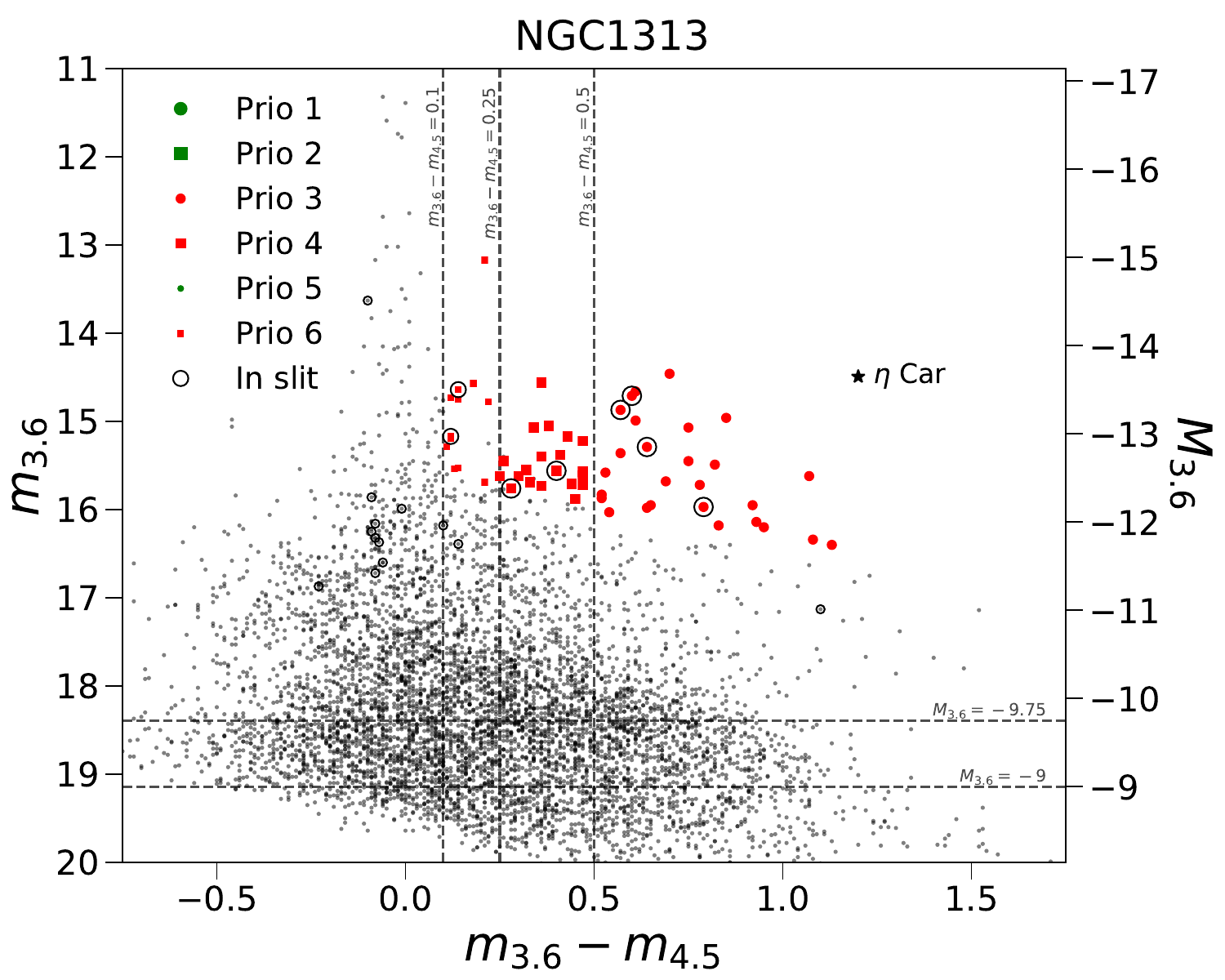}
  \captionof{figure}{{\it Spitzer} CMD of NGC 1313.}
  \label{fig:CMD_NGC1313}
\end{minipage}
\end{figure}

\begin{figure}[!h]
\centering
\begin{minipage}{.5\textwidth}
  \centering
  \includegraphics[width=.85\textwidth]{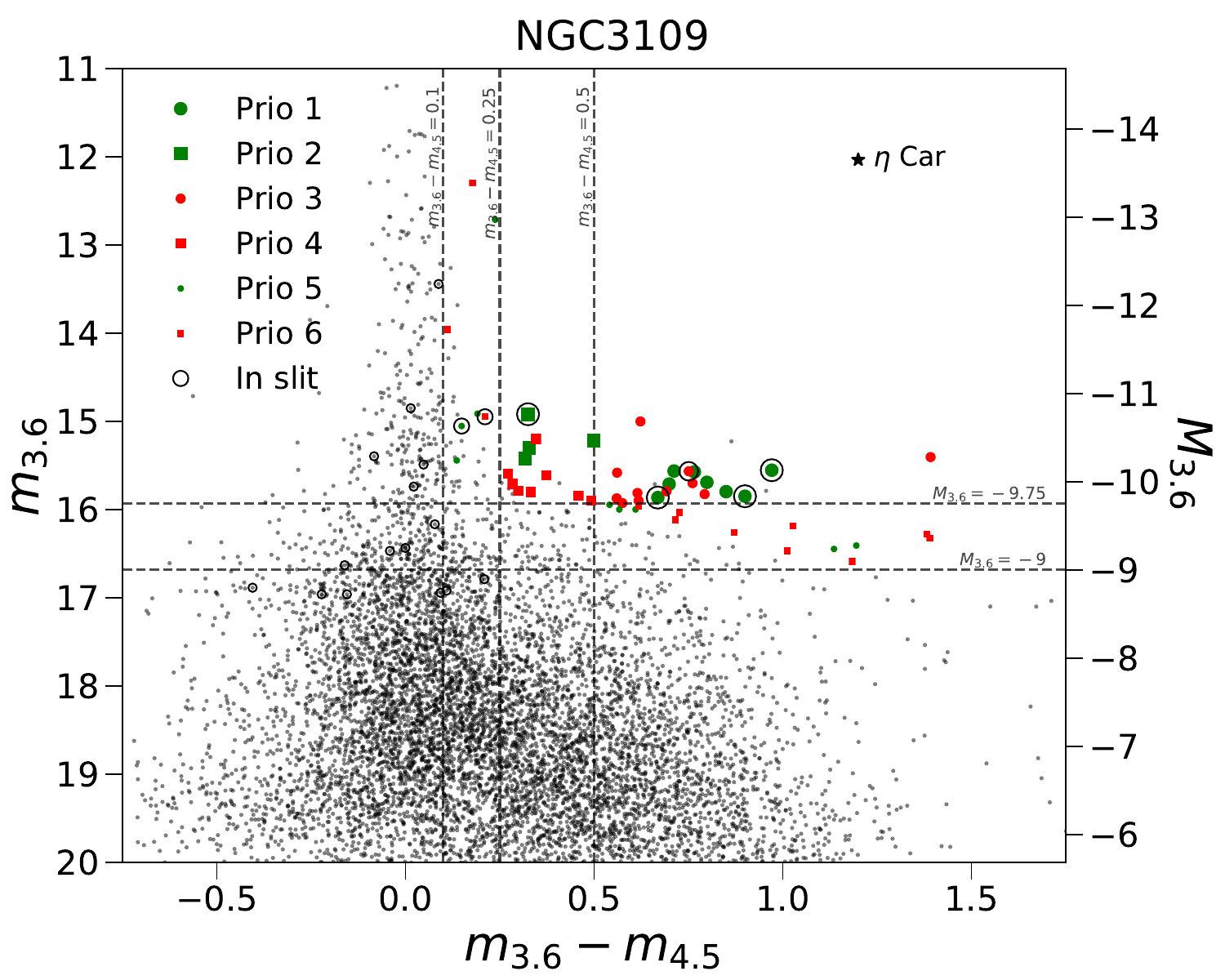}
  \captionof{figure}{{\it Spitzer} CMD of NGC 3109.}
  \label{fig:CMD_NGC3109}
\end{minipage}%
\begin{minipage}{.5\textwidth}
  \centering
  \includegraphics[width=.85\textwidth]{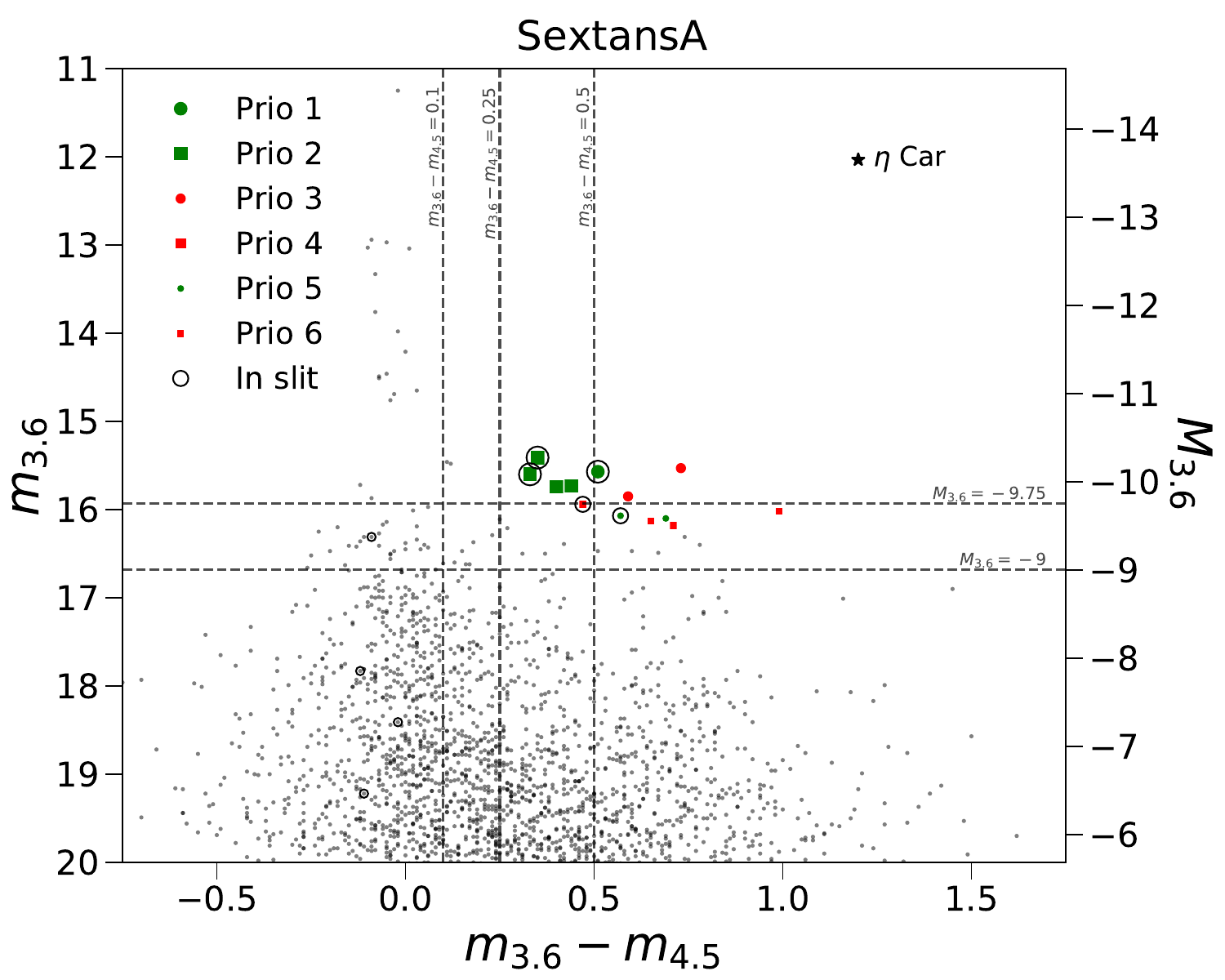}
  \captionof{figure}{{\it Spitzer} CMD of Sextans A.}
  \label{fig:CMD_SextansA}
\end{minipage}%
\end{figure}

\begin{figure}[!h]
\centering
\begin{minipage}{.5\textwidth}
  \centering
  \includegraphics[width=.85\textwidth]{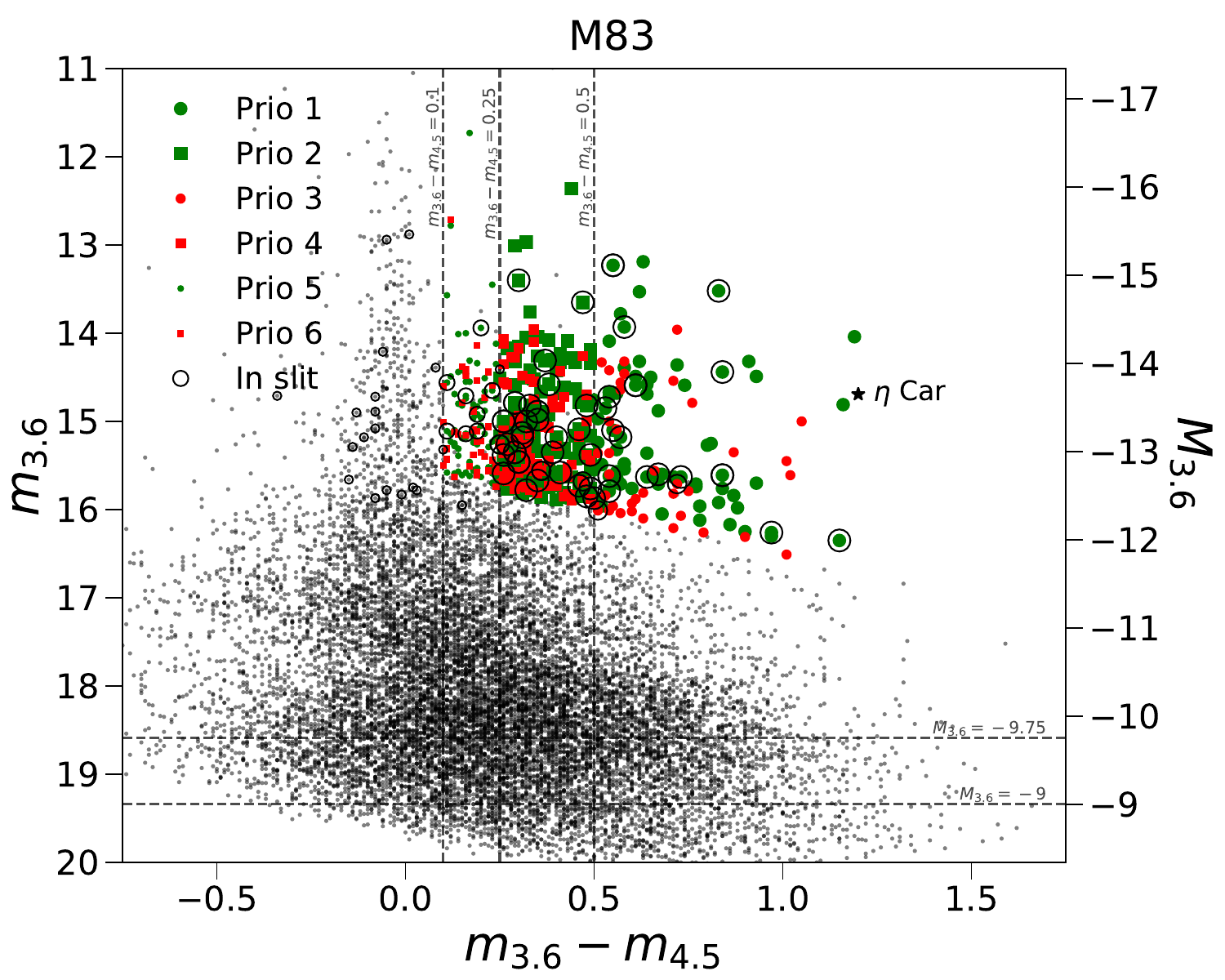}
  \captionof{figure}{{\it Spitzer} CMD of M83.}
  \label{fig:CMD_M83}
\end{minipage}%
\begin{minipage}{.5\textwidth}
  \centering
  \includegraphics[width=.85\textwidth]{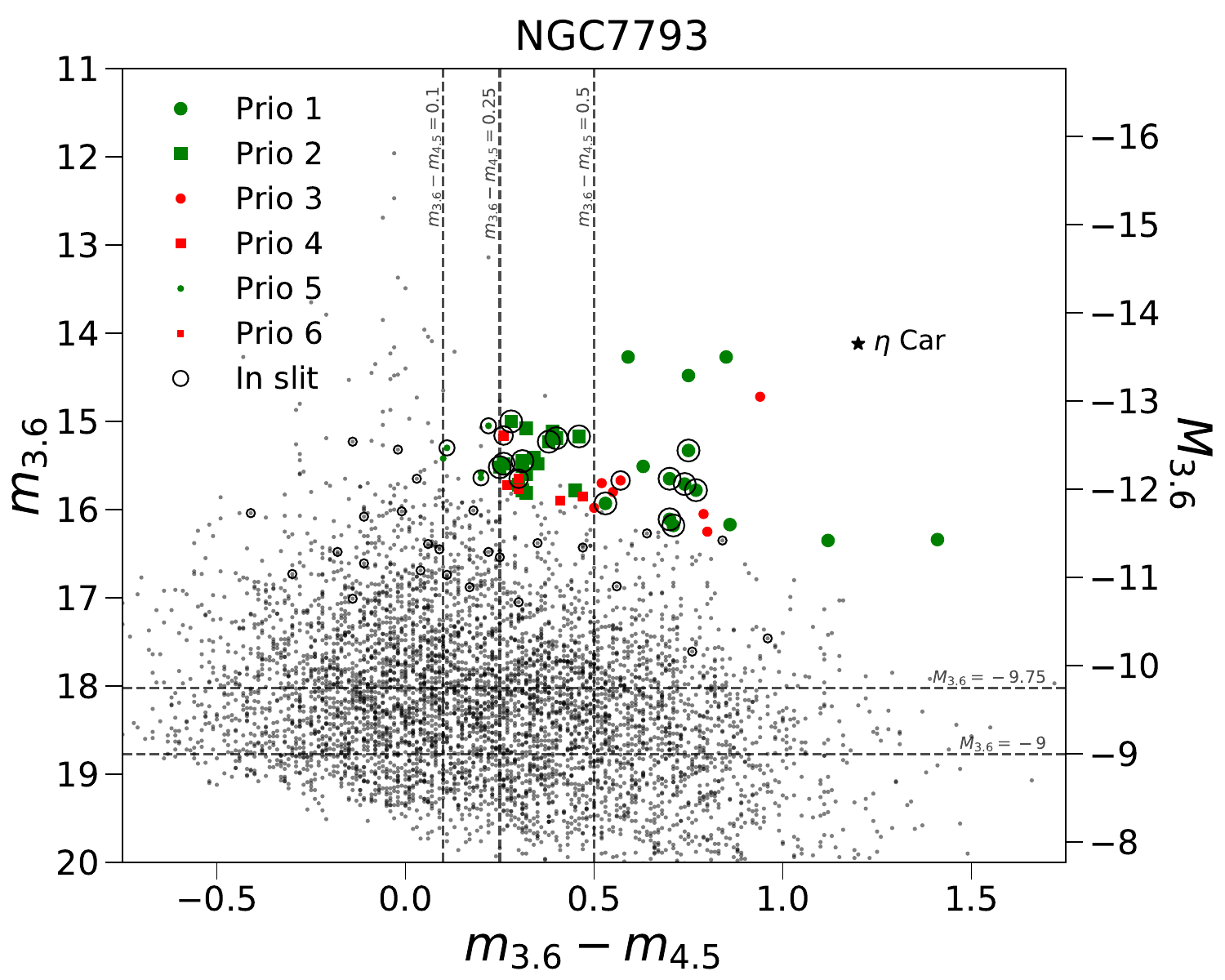}
  \captionof{figure}{{\it Spitzer} CMD of NGC 7793.}
  \label{fig:CMD_NGC7793}
\end{minipage}
\end{figure}

This appendix provides {\it Spitzer} $m_{3.6}-m_{4.5}$ vs. $m_{3.6}$ CMDs for each of the galaxies in Figures~\ref{fig:CMD_WLM} to \ref{fig:CMD_NGC7793}. All priority targets are indicated, as well as those targets that had a slit placed on them. The different magnitude cuts that were used to assign target priority are also indicated; the apparent magnitude criterion of $m_{4.5} \leq 15.5$~mag for rejecting background sources causes the diagonal cut. The location of $\eta$ Carinae is shown for reference.

\vfill

\vfill

\clearpage

\clearpage
\section{Mask design} \label{sec_ap_masks}
This appendix provides an overview of the mask layouts of each observed field. For each field an overview image is presented consisting of the pre-imaging of that field, the locations of the reference stars (double red circles), and the slit locations (black 1"$\times$6" rectangles). The target coordinates are corrected for distortions of the instrument while the pre-imaging is not, and hence the slits may appear slightly off-target in some cases. 

\subsection{WLM Field A}
\vfill
\begin{figure*}[h!]
   	\resizebox{1.0\hsize}{!}{\includegraphics{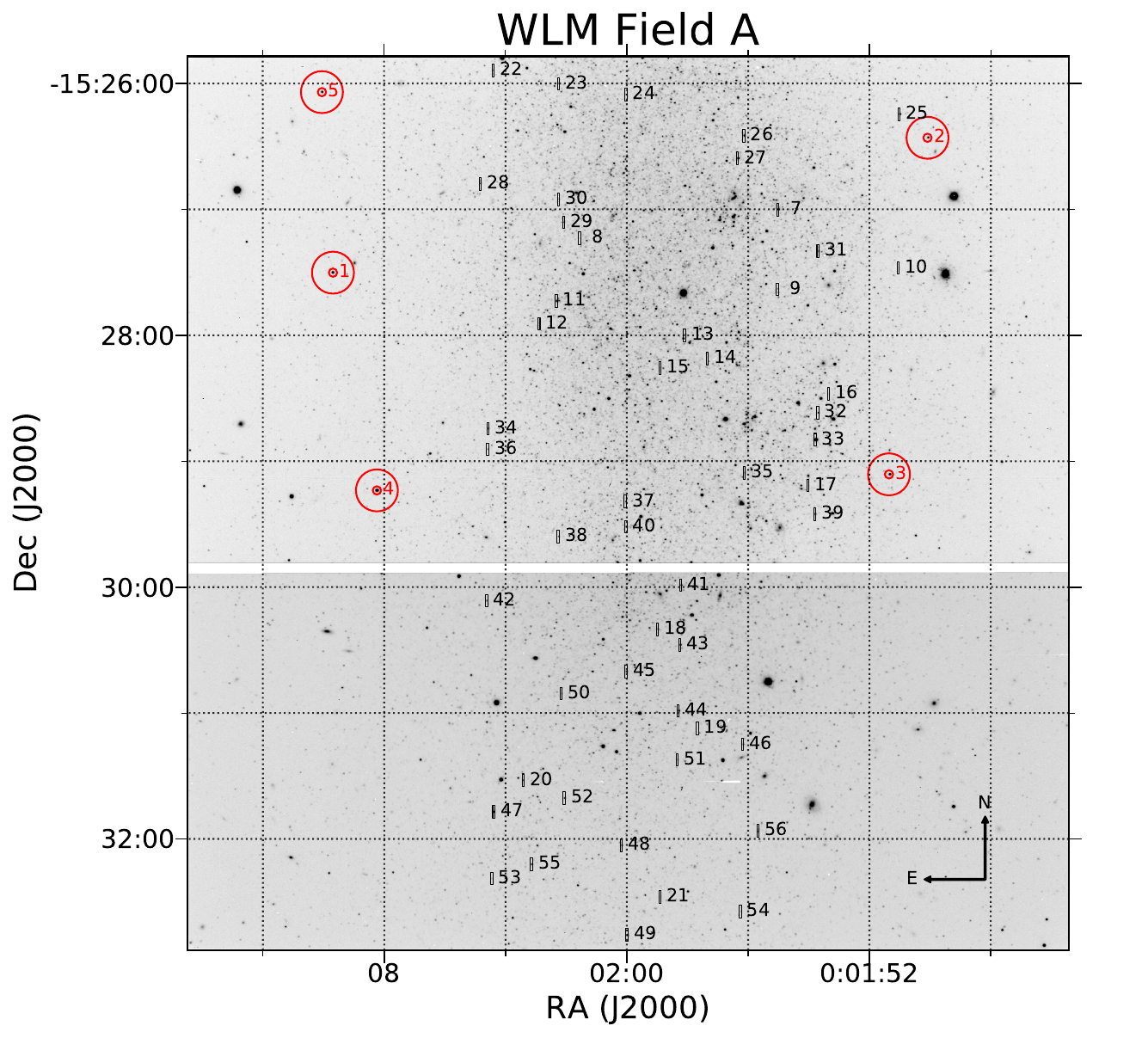}}
  	\caption{Mask layout for WLM Field A.}
\end{figure*}
\vfill

\clearpage

\section{Observation log} \label{sec:ap_obslog}
This appendix gives the observing logs of all pre-imaging (Table~\ref{tab:obslog_pre}) and MXU (Table~\ref{tab:obslog}) observations. For the pre-imaging, the mean airmass and seeing are given for the three (back-to-back) exposures, for MXU airmass and seeing are given at the start and end of the exposure. The reported seeing values come from the differential image motion monitor (DIMM) at Paranal, and refer to the seeing at 500nm at zenith. The image quality of the pre-imaging was measured by computing the median FWHM of all sources detected in the field using the Python module {\it photutils} \citep{photutils}.

\begin{table*}[ht!]
\centering
\caption{Log of pre-imaging observations.}\label{tab:obslog_pre}
\begin{tabular}{l c c c c c c c}
\hline\hline
Galaxy	& Field\tablefootmark{a}	& Date & MJD\tablefootmark{b}	& $T_{\mathrm{exp}}$	& Airmass	& Seeing  & IQ\tablefootmark{c}\\
			 & & 	& & \small{(s)}	& & \small{(\arcsec)}  & \small{(\arcsec)}\\
\hline \\[-9pt]
WLM & A & 21 Sep 2020 & 59113.2272413 &  3 $\times$ 10 & 1.030 & 0.64 & 0.45 \\
NGC 55 & A & 20 Sep 2020 & 59112.2433908 &  3 $\times$ 10 & 1.052 & 0.58 & 0.50 \\
NGC 55 & B & 20 Sep 2020 & 59112.2396810 &  3 $\times$ 10 & 1.048 & 0.57 & 0.51 \\
NGC 55\tablefootmark{d} & C & 20 Sep 2020 & 59112.2329451 &  3 $\times$ 10 & 1.042 & 0.58 & --- \\
NGC 55 & C & 20 Sep 2020 & 59112.2347649 &  3 $\times$ 10 & 1.043 & 0.56 & 0.49 \\
NGC 247 & A & 23 Sep 2020 & 59115.2255528 &  3 $\times$ 10 & 1.003 & 0.95 & 0.74 \\
NGC 247 & B & 24 Sep 2020 & 59116.3335631 &  3 $\times$ 10 & 1.273 & 0.64 & 0.58 \\
NGC 253\tablefootmark{d} & A & 22 Sep 2020 & 59114.3662291 &  3 $\times$ 10 & 1.442 & 0.73 & --- \\
NGC 253 & A & 22 Sep 2020 & 59114.3677868 &  3 $\times$ 10 & 1.454 & 0.74 & 0.71 \\
NGC 253 & B & 24 Sep 2020 & 59116.3265104 &  3 $\times$ 10 & 1.220 & 0.57 & 0.78 \\
NGC 253 & C & 24 Sep 2020 & 59116.3302976 &  3 $\times$ 10 & 1.239 & 0.58 & 0.77 \\
NGC 300\tablefootmark{d} & A & 18 Sep 2020 & 59110.3763705 &  3 $\times$ 10 & 1.376 & 0.88 & --- \\
NGC 300\tablefootmark{d} & A & 23 Sep 2020 & 59115.0247709 &  3 $\times$ 10 & 2.141 & 0.63 & ---  \\
NGC 300 & A & 23 Sep 2020 & 59115.0264411 &  3 $\times$ 10 & 2.110 & 0.60 & 0.80 \\
NGC 300 & B & 21 Sep 2020 & 59113.2227121 &  3 $\times$ 10 & 1.027 & 0.65 & 0.51 \\
NGC 300\tablefootmark{d} & C & 22 Sep 2020 & 59114.1461572 &  3 $\times$ 10 & 1.140 & 0.97 & --- \\
NGC 300 & C & 24 Sep 2020 & 59116.3366949 &  3 $\times$ 10 & 1.248 & 0.65 & 0.65 \\
NGC 300\tablefootmark{d} & D & 22 Sep 2020 & 59114.3604158 &  3 $\times$ 10 & 1.342 & 0.74 & --- \\
NGC 300 & D & 22 Sep 2020 & 59114.3619709 &  3 $\times$ 10 & 1.350 & 0.68 & 0.55 \\
NGC 1313 & A & 28 Jan 2022 & 59607.1852291 &  3 $\times$ 10 & 2.072 & 0.52 & 0.81 \\
NGC 3109 & A & 28 Jan 2022 & 59607.1905841 &  3 $\times$ 10 & 1.084 & 0.52 & 0.58 \\
Sextans A & A & 14 Nov 2020 & 59167.3357008 &  3 $\times$ 10 & 1.572 & 0.41 & 0.55 \\
M83 & A & 22 Jan 2021 & 59236.3553139 &  3 $\times$ 10 & 1.084 & 0.52 & 0.54 \\
M83 & B & 21 Jan 2021 & 59235.3380442 &  3 $\times$ 10 & 1.143 & 0.76 & 0.68  \\
M83 & C & 21 Jan 2021 & 59235.3414212 &  3 $\times$ 10 & 1.131 & 0.94 & 0.67  \\
NGC 7793 & A & 22 Sep 2020 & 59114.1420730 &  3 $\times$ 10 & 1.046 & 0.86 & 0.64 \\
\hline
\end{tabular}
\tablefoot{
\tablefoottext{a}{Field coordinates are listed in Table~\ref{tab:fields}.}
\tablefoottext{b}{At the start of the first exposure.}
\tablefoottext{c}{Measured from the stacked image.}
\tablefoottext{d}{Observations were graded C or X and repeated. However, the exposures were found to be of sufficient quality (IQ $\leq$ 0.8\arcsec) and were included in the stacked image.}
}
\end{table*}

\begin{longtable}{l c c c c c c c}
\caption{Log of MXU observations.}\label{tab:obslog} \\
\hline\hline
Galaxy/{\it Field}$^{a}$		& Date & MJD$^b$	& $T_{\mathrm{exp}}$	& \multicolumn{2}{c}{Airmass}	& \multicolumn{2}{c}{Seeing}	\\
			 & & 	& 	& Start & End & Start	& End \\
    & & & \small{(s)} & & & \small{(\arcsec)} & \small{(\arcsec)} \\
\hline \\[-9pt]
\endfirsthead
\caption{continued.}\\
\hline\hline
Galaxy/{\it Field}$^{a}$		& Date & MJD$^{b}$	& $T_{\mathrm{exp}}$	& \multicolumn{2}{c}{Airmass}	& \multicolumn{2}{c}{Seeing}	\\
			 & & 	&	& Start & End & Start	& End \\
& & & \small{(s)} & & & \small{(\arcsec)} & \small{(\arcsec)} \\
\hline \\[-9pt]
\endhead

\hline
\multicolumn{8}{l}{
\footnotesize{{\bf Notes. }$^{(a)}$ Field coordinates are listed in Table~\ref{tab:fields}. $^{(b)}$ At start of exposure.}}
\endfoot

WLM \\
\quad {\it Field A} & 7 Nov 2020 & 59160.0376645 & 900 & 1.030 & 1.020 & 0.59 & 0.75 \\
& & 59160.0485270 & 900 & 1.020 & 1.015 & 0.75 & 0.63 \\
& & 59160.0593837 & 900 & 1.014 & 1.013 & 0.63 & 0.49 \\
& & 59160.0707324 & 900 & 1.013 & 1.016 & 0.48 & 0.44 \\
& & 59160.0815957 & 900 & 1.016 & 1.024 & 0.44 & 0.40 \\
& & 59160.0924519 & 900 & 1.024 & 1.035 & 0.40 & 0.40 \\

NGC 55 \\
\quad {\it Field A} & 13 Nov 2020 & 59166.0331920 & 900 & 1.044 & 1.037 & 0.61 & 0.65 \\
& & 59166.0440546 & 900 & 1.037 & 1.033 & 0.65 & 0.78 \\
& & 59166.0549112 & 900 & 1.033 & 1.033 & 0.78 & 0.71 \\
& 16 Nov 2020 & 59169.1040013 & 900 & 1.074 & 1.093 & 0.83 & 0.77 \\
& & 59169.1148733 & 900 & 1.094 & 1.118 & 0.77 & 0.73 \\
& & 59169.1257297 & 900 & 1.119 & 1.147 & 0.73 & 0.77 \\

\quad {\it Field B} & 16 Oct 2020 & 59138.0150812 & 900 & 1.298 & 1.248 & 0.70 & 0.66 \\
& & 59138.0259345 & 900 & 1.246 & 1.203 & 0.66 & 0.57 \\
& & 59138.0368027 & 900 & 1.201 & 1.165 & 0.58 & 0.59 \\
& 14 Nov 2020 & 59167.1392733 & 900 & 1.139 & 1.171 & 0.46 & 0.64 \\
& & 59167.1501434 & 900 & 1.173 & 1.210 & 0.64 & 0.60 \\
& & 59167.1609999 & 900 & 1.212 & 1.256 & 0.54 & 0.49 \\

\quad {\it Field C} & 13 Dec 2020 & 59196.0380138 & 900 & 1.087 & 1.109 & 0.53 & 0.55 \\
& & 59196.0488758 & 900 & 1.109 & 1.136 & 0.55 & 0.48 \\
& & 59196.0597446 & 900 & 1.137 & 1.169 & 0.52 & 0.46 \\
& & 59196.0802301 & 900 & 1.204 & 1.248 & 0.53 & 0.49 \\
& & 59196.0910982 & 900 & 1.250 & 1.300 & 0.49 & 0.58 \\
& & 59196.1019662 & 900 & 1.302 & 1.362 & 0.52 & 0.51 \\

NGC 247 \\
\quad {\it Field A} & 14 Dec 2020 & 59197.0417297 & 900 & 1.036 & 1.054 & 0.65 & 0.65 \\
& & 59197.0525910 & 900 & 1.055 & 1.078 & 0.65 & 0.66 \\
& & 59197.0634478 & 900 & 1.079 & 1.108 & 0.58 & 0.62 \\
& 17 Dec 2020 & 59200.0466988 & 900 & 1.060 & 1.084 & 0.56 & 0.60 \\
& & 59200.0575653 & 900 & 1.085 & 1.115 & 0.64 & 0.50 \\
& & 59200.0684337 & 900 & 1.117 & 1.153 & 0.56 & 0.66 \\

\quad {\it Field B} & 17 Dec 2020 & 59200.0864641 & 900 & 1.183 & 1.231 & 0.57 & 0.46 \\
& & 59200.0973337 & 900 & 1.234 & 1.291 & 0.46 & 0.69 \\
& & 59200.1082019 & 900 & 1.294 & 1.363 & 0.68 & 0.53 \\
& & 59200.1264039 & 900 & 1.423 & 1.516 & 0.56 & 0.64 \\
& & 59200.1372646 & 900 & 1.520 & 1.634 & 0.73 & 0.79 \\
& & 59200.1481209 & 900 & 1.639 & 1.778 & 0.79 & 0.51 \\
f
NGC 253 \\
\quad {\it Field A} & 18 Dec 2020 & 59201.1037620 & 900 & 1.268 & 1.331 & 0.44 & 0.50 \\
& & 59201.1146188 & 900 & 1.334 & 1.409 & 0.50 & 0.36 \\
& & 59201.1254869 & 900 & 1.412 & 1.502 & 0.36 & 0.40 \\
& & 59201.1420009 & 900 & 1.562 & 1.681 & 0.41 & 0.35 \\
& & 59201.1528594 & 900 & 1.687 & 1.833 & 0.35 & 0.43 \\
& & 59201.1637276 & 900 & 1.840 & 2.022 & 0.43 & 0.52 \\

\quad {\it Field B} & 2 Sep 2021 & 59459.1591006 & 900 & 1.296 & 1.238 & 0.66 & 0.66 \\
& & 59459.1699647 & 900 & 1.236 & 1.187 & 0.66 & 0.64 \\
& & 59459.1808212 & 900 & 1.185 & 1.144 & 0.64 & 0.73 \\
& & 59459.2047246 & 900 & 1.101 & 1.073 & 0.73 & 0.62 \\
& & 59459.2155897 & 900 & 1.072 & 1.050 & 0.62 & 0.71 \\
& & 59459.2264577 & 900 & 1.049 & 1.032 & 0.71 & 0.73 \\

\pagebreak

\quad {\it Field C} & 13 Dec 2020 & 59196.1208609 & 900 & 1.285 & 1.351 & 0.49 & 0.61 \\
& & 59196.1317234 & 900 & 1.354 & 1.433 & 0.61 & 0.50 \\
& & 59196.1425914 & 900 & 1.437 & 1.532 & 0.50 & 0.59 \\
& 14 Dec 2020 & 59197.1212413 & 900 & 1.304 & 1.373 & 1.08 & 0.64 \\
& & 59197.1321047 & 900 & 1.376 & 1.460 & 0.68 & 0.73 \\
& & 59197.1429726 & 900 & 1.463 & 1.563 & 0.73 & 0.71 \\

NGC 300 \\
\quad {\it Field A} & 7 Dec 2020 & 59190.1414814 & 900 & 1.278 & 1.335 & 0.41 & 0.41 \\
& & 59190.1523466 & 900 & 1.338 & 1.404 & 0.41 & 0.52 \\
& & 59190.1632031 & 900 & 1.407 & 1.486 & 0.53 & 1.16 \\
& 12 Dec 2020 & 59195.0586379 & 900 & 1.064 & 1.083 & 0.48 & 0.56 \\
& & 59195.0695012 & 900 & 1.083 & 1.106 & 0.56 & 0.55 \\
& & 59195.0803575 & 900 & 1.107 & 1.135 & 0.55 & 0.57 \\

\quad {\it Field B} & 9 Dec 2020 & 59192.0550846 & 900 & 1.048 & 1.063 & 0.57 & 0.54 \\
& & 59192.0659500 & 900 & 1.063 & 1.082 & 0.54 & 0.56 \\
& & 59192.0768183 & 900 & 1.082 & 1.105 & 0.59 & 0.63 \\
& 12 Dec 2020 & 59195.1067032 & 900 & 1.188 & 1.230 & 0.49 & 0.42 \\
& & 59195.1175690 & 900 & 1.232 & 1.281 & 0.42 & 0.54 \\
& & 59195.1284372 & 900 & 1.283 & 1.340 & 0.56 & 0.64 \\

\quad {\it Field C} & 1 Sep 2021 & 59458.3693021 & 900 & 1.133 & 1.165 & 0.72 & 0.65 \\
& & 59458.3584365 & 900 & 1.105 & 1.132 & 0.55 & 0.72 \\
& & 59458.3801586 & 900 & 1.167 & 1.205 & 0.70 & 0.58 \\
& 2 Sep 2021 & 59459.1177149 & 900 & 1.640 & 1.539 & 0.75 & 0.72 \\
& & 59459.1285760 & 900 & 1.535 & 1.449 & 0.75 & 0.72 \\
& & 59459.1394324 & 900 & 1.446 & 1.373 & 0.67 & 0.93 \\

\quad {\it Field D} & 6 Nov 2020 & 59159.0776603 & 900 & 1.040 & 1.031 & 1.24 & 0.80 \\
& & 59159.0896766 & 900 & 1.031 & 1.027 & 0.80 & 0.78 \\
& & 59159.1016906 & 900 & 1.027 & 1.027 & 0.78 & 0.66 \\
& 12 Dec 2020 & 59195.1478909 & 900 & 1.393 & 1.469 & 0.61 & 0.56 \\
& & 59195.1587494 & 900 & 1.473 & 1.562 & 0.57 & 0.49 \\
& & 59195.1696058 & 900 & 1.566 & 1.673 & 0.49 & 0.50 \\

NGC 1313 \\
\quad {\it Field A} & 25 Oct 2022 & 59877.0444866 & 900 & 1.984 & 1.900 & 0.84 & 0.75 \\
& & 59877.0553492 & 900 & 1.897 & 1.822 & 0.75 & 0.75 \\
& & 59877.0662175 & 900 & 1.819 & 1.753 & 0.75 & 0.76 \\
& 19 nov 2022 & 59902.0943819 & 900 & 1.421 & 1.400 & 0.50 & 0.41 \\
& & 59902.1052468 & 900 & 1.399 & 1.382 & 0.41 & 0.44 \\
& & 59902.1161032 & 900 & 1.381 & 1.367 & 0.38 & 0.58 \\

NGC 3109 \\
\quad {\it Field A} & 1 May 2022 & 59700.0803524 & 900 & 1.094 & 1.125 & 0.58 & 0.56 \\
& & 59700.0912110 & 900 & 1.126 & 1.164 & 0.56 & 0.61 \\
& & 59700.1020792 & 900 & 1.165 & 1.210 & 0.61 & 0.70 \\
& 25 May 2022 & 59724.0691019 & 900 & 1.332 & 1.406 & 0.77 & 0.91 \\
& & 59724.0799643 & 900 & 1.410 & 1.498 & 0.91 & 0.66 \\
& & 59724.0908210 & 900 & 1.502 & 1.609 & 0.66 & 0.52 \\

Sextans A \\
\quad {\it Field A} & 21 Dec 2020 & 59204.2715527 & 900 & 1.292 & 1.239 & 0.81 & 0.81 \\
& & 59204.2824168 & 900 & 1.237 & 1.194 & 0.81 & 0.75 \\
& & 59204.2932732 & 900 & 1.193 & 1.157 & 0.83 & 0.55 \\
& & 59204.3139680 & 900 & 1.129 & 1.106 & 0.85 & 0.66 \\
& & 59204.3248241 & 900 & 1.105 & 1.087 & 0.66 & 0.57 \\
& & 59204.3356808 & 900 & 1.087 & 1.074 & 0.57 & 1.01 \\

\pagebreak

M83  \\
\quad {\it Field A} & 12 Jun 2021 & 59377.2122001 & 900 & 1.733 & 1.885 & 0.49 & 0.43 \\
& & 59377.2230679 & 900 & 1.892 & 2.080 & 0.42 & 0.54 \\
& & 59377.2013461 & 900 & 1.604 & 1.728 & 0.60 & 0.49 \\
& 30 Jun 2021 & 59395.0439106 & 900 & 1.050 & 1.071 & 0.54 & 0.58 \\
& & 59395.0547691 & 900 & 1.072 & 1.098 & 0.58 & 0.56 \\
& & 59395.0656374 & 900 & 1.099 & 1.130 & 0.55 & 0.51 \\

\quad {\it Field B} & 30 Jun 2021 & 59395.0850735 & 900 & 1.161 & 1.203 & 0.67 & 0.79 \\
& & 59395.0959379 & 900 & 1.205 & 1.255 & 0.83 & 0.90 \\
& & 59395.1067942 & 900 & 1.257 & 1.317 & 0.90 & 0.87 \\
& 15 Jul 2021 & 59410.9760713 & 900 & 1.018 & 1.029 & 0.67 & 0.99 \\
& & 59410.9869254 & 900 & 1.030 & 1.045 & 0.99 & 0.68 \\
& & 59410.9977822 & 900 & 1.046 & 1.066 & 0.69 & 0.62 \\

\quad {\it Field C} & 28 Jan 2022 & 59607.2902996 & 900 & 1.273 & 1.220 & 0.97 & 0.88 \\
& & 59607.3011618 & 900 & 1.218 & 1.174 & 0.88 & 0.60 \\
& & 59607.3120185 & 900 & 1.172 & 1.135 & 0.60 & 0.86 \\
& 30 Jan 2022 & 59609.2671984 & 900 & 1.383 & 1.314 & 0.80 & 0.67 \\
& & 59609.2563425 & 900 & 1.468 & 1.386 & 0.53 & 0.74 \\
& & 59609.2780666 & 900 & 1.312 & 1.253 & 0.57 & 0.50 \\

NGC 7793 \\
\quad {\it Field A} & 7 Nov 2020 & 59160.1110977 & 900 & 1.043 & 1.061 & 0.52 & 0.64 \\
& & 59160.1219530 & 900 & 1.062 & 1.084 & 0.64 & 0.59 \\
& & 59160.1328096 & 900 & 1.085 & 1.112 & 0.59 & 0.70 \\
& & 59160.1442436 & 900 & 1.114 & 1.147 & 0.50 & 0.73 \\
& & 59160.1551027 & 900 & 1.149 & 1.187 & 0.67 & 0.53 \\
& & 59160.1659705 & 900 & 1.189 & 1.235 & 0.53 & 0.49 \\

\end{longtable}

\clearpage

\section{Color-magnitude diagrams of classified targets} \label{sec:ap_cmdclass}
This appendix provides the mid-IR (M$_{3.6}$ vs. $m_{3.6}-m_{4.5}$ and $J-m_{3.6}$), near-IR (M$_{Ks}$ vs. $J-K_s$) and optical (M$_G$ vs. $G_{BP}-G_{RP}$, $g$ vs. $g-r$) CMDs for each galaxy of our sample (see Section~\ref{sec:discusscmd}), similar to the cumulative CMDs shown in Figure~\ref{fig:AbsMagCMDs}. The CMDs were constructed using data from various surveys (described in Section~\ref{sec:photometry}) when available and include all evolved massive stars in our sample, as well as the dominant sources of contamination.

\vfill
\begin{figure*}[ht!]
    \includegraphics[width=\columnwidth]{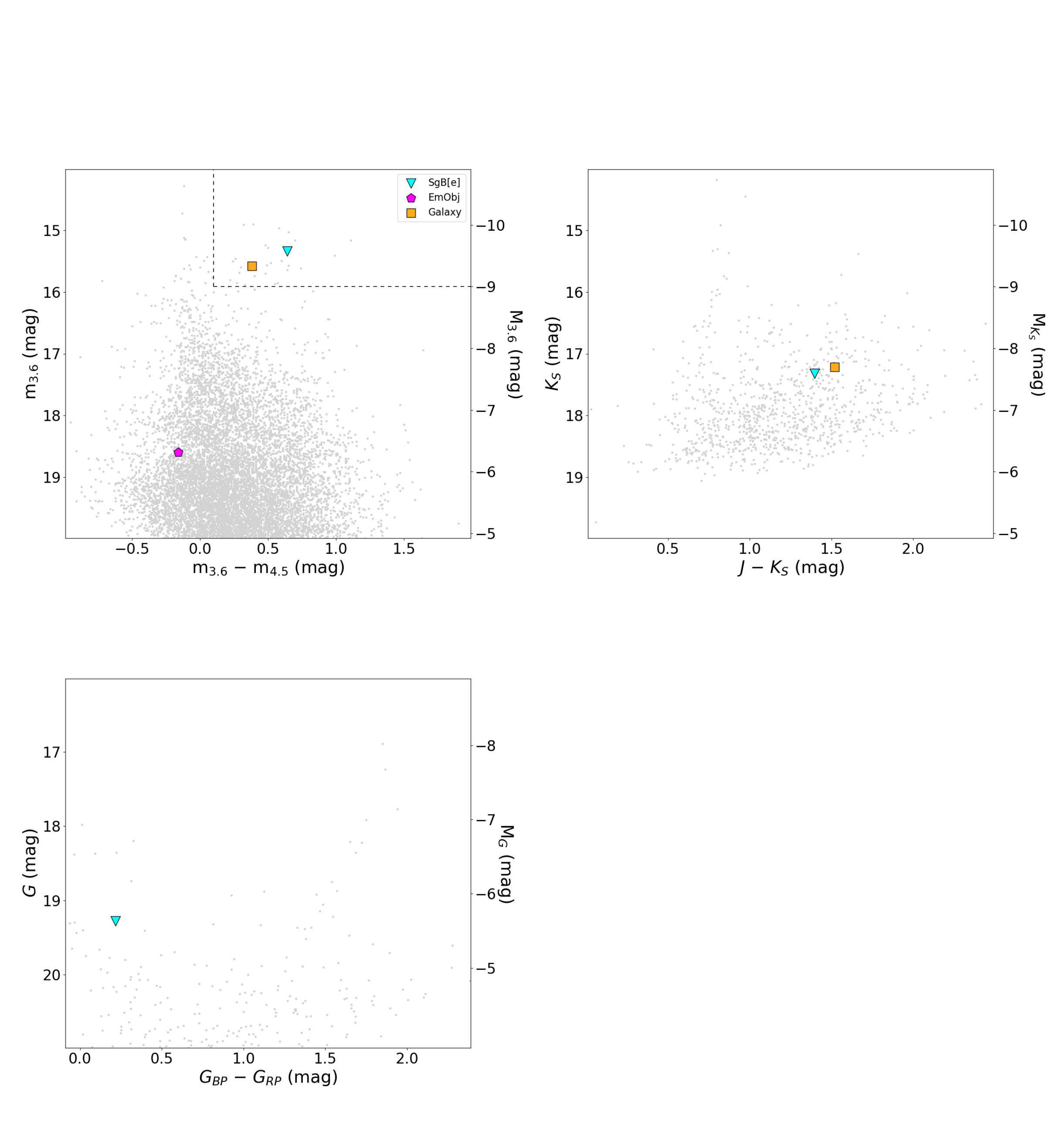}
    \caption{Infrared and optical CMDs for the classified targets of interest in WLM, i.e. all evolved massive stars, \ion{H}{ii} regions and background objects.}
    \label{fig:WLM_CMD}
\end{figure*}
\vfill
\clearpage

\vfill
\begin{figure*}[ht!]
    \includegraphics[width=\columnwidth]{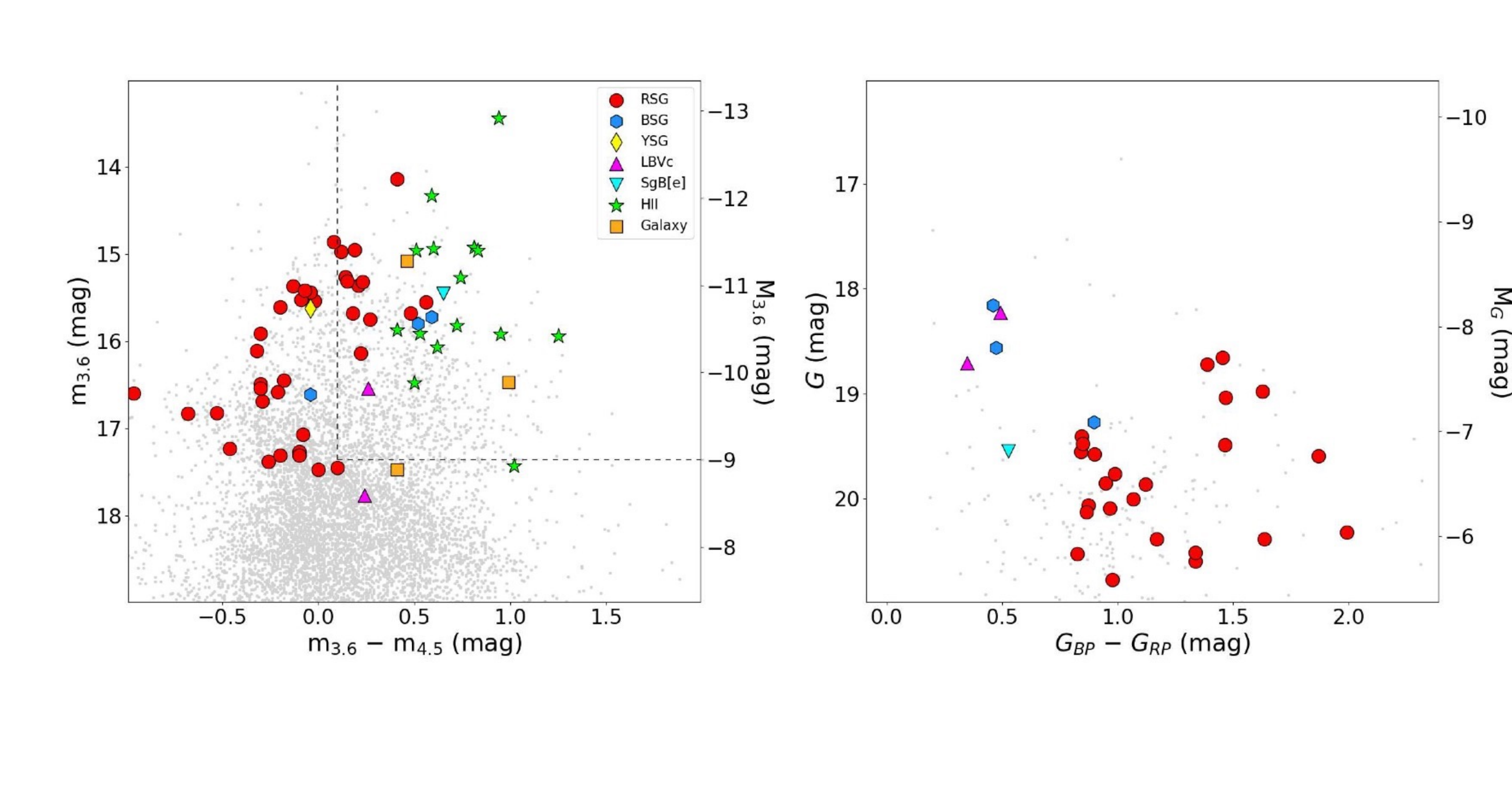}
    \caption{Same as Figure~\ref{fig:WLM_CMD}, but for NGC~55.}
    \label{fig:NGC55_CMD}
\end{figure*}
\vfill
\clearpage

\vfill
\begin{figure*}[ht!]
    \includegraphics[width=\columnwidth]{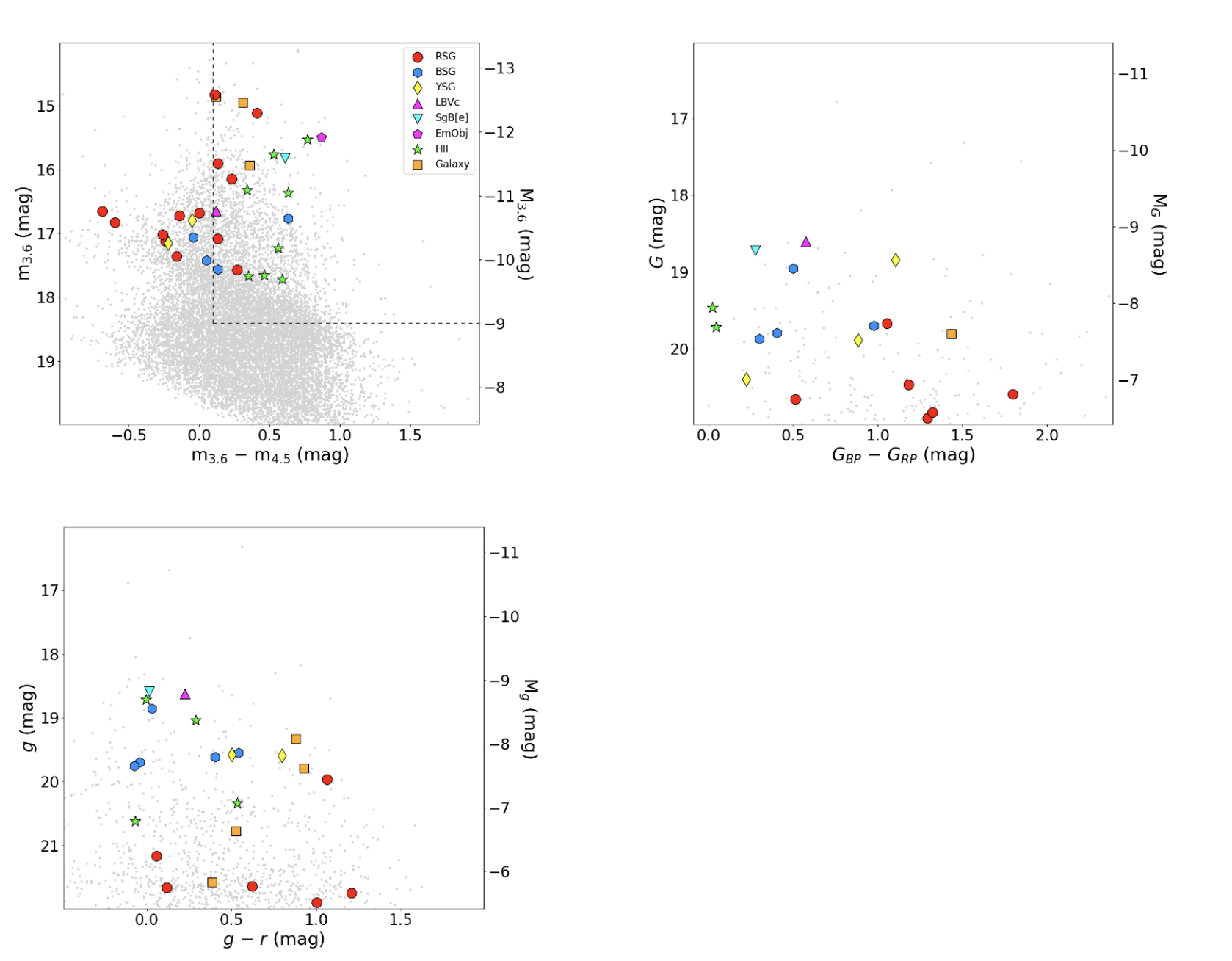}
    \caption{Same as Figure~\ref{fig:WLM_CMD}, but for NGC~247.}
    \label{fig:NGC247_CMD}
\end{figure*}
\vfill
\clearpage

\vfill
\begin{figure*}[ht!]
\includegraphics[width=\columnwidth]{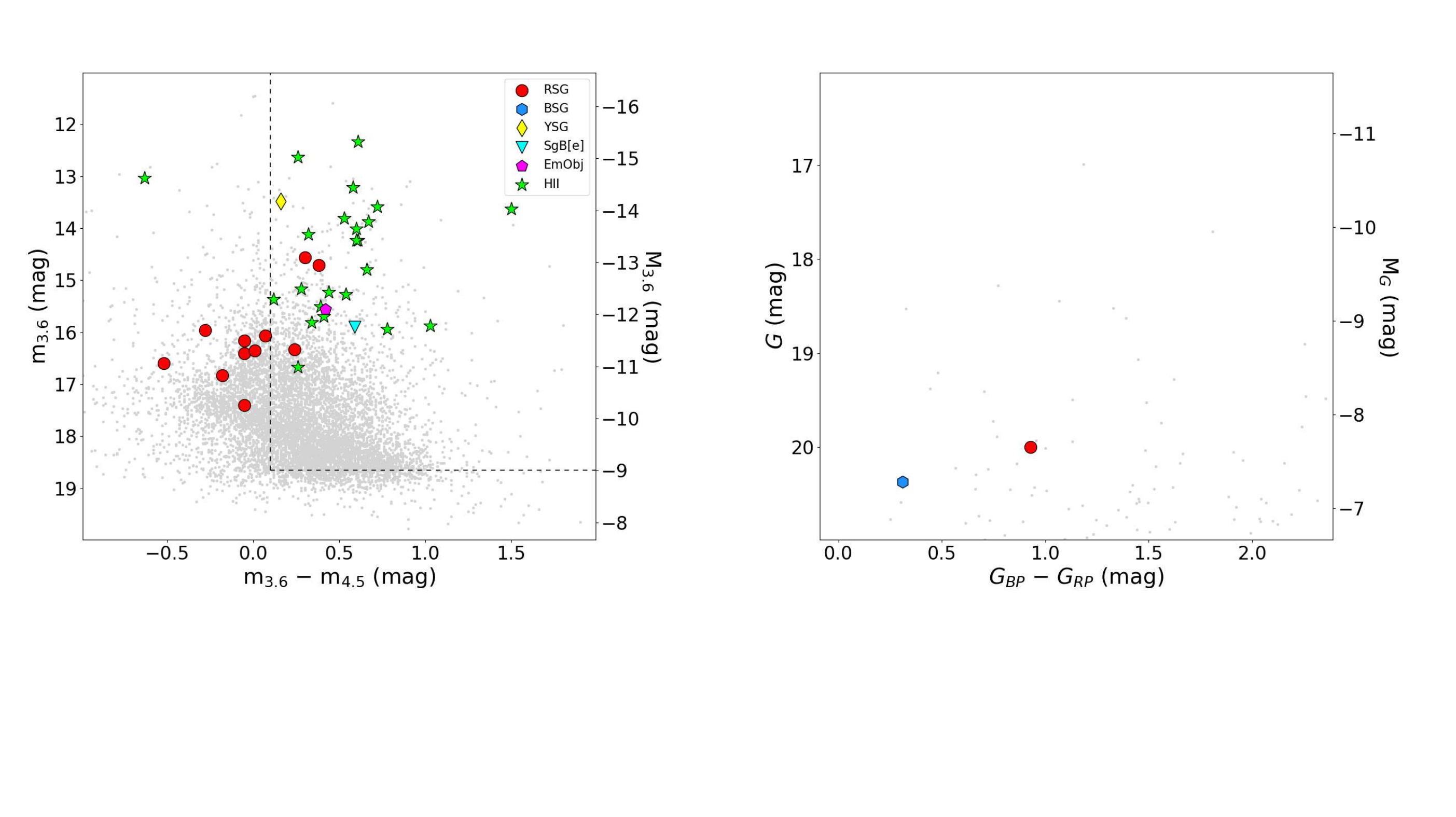}
\caption{Same as Figure~\ref{fig:WLM_CMD}, but for NGC~253.}
\label{fig:NGC253_CMD}
\end{figure*}

\begin{figure*}[ht!]
     \includegraphics[width=\columnwidth]{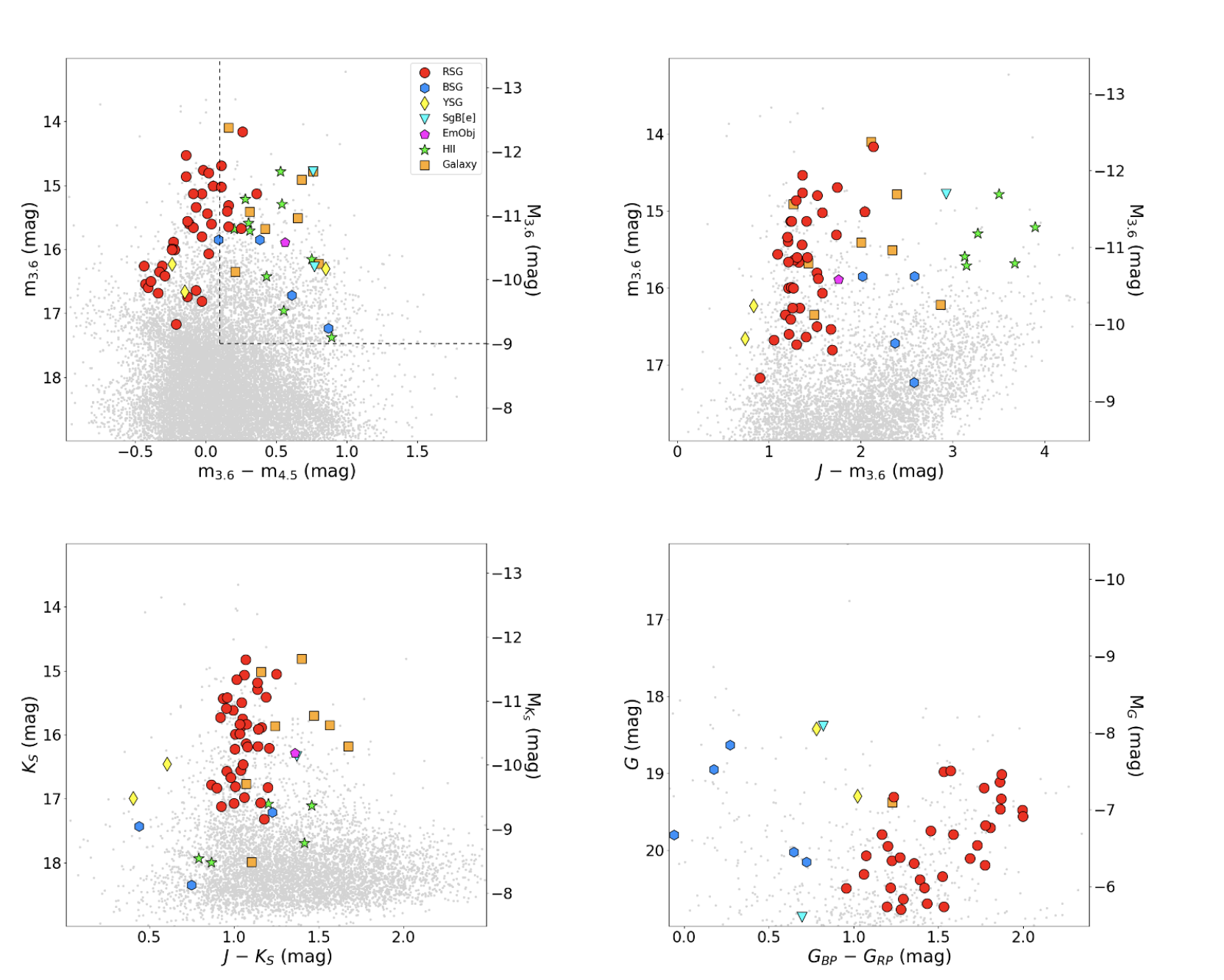}
    \caption{Same as Figure~\ref{fig:WLM_CMD}, but for NGC~ 300.}
    \label{fig:NGC300_CMD}
\end{figure*}

\begin{figure*}[ht!]
    \includegraphics[width=\columnwidth]{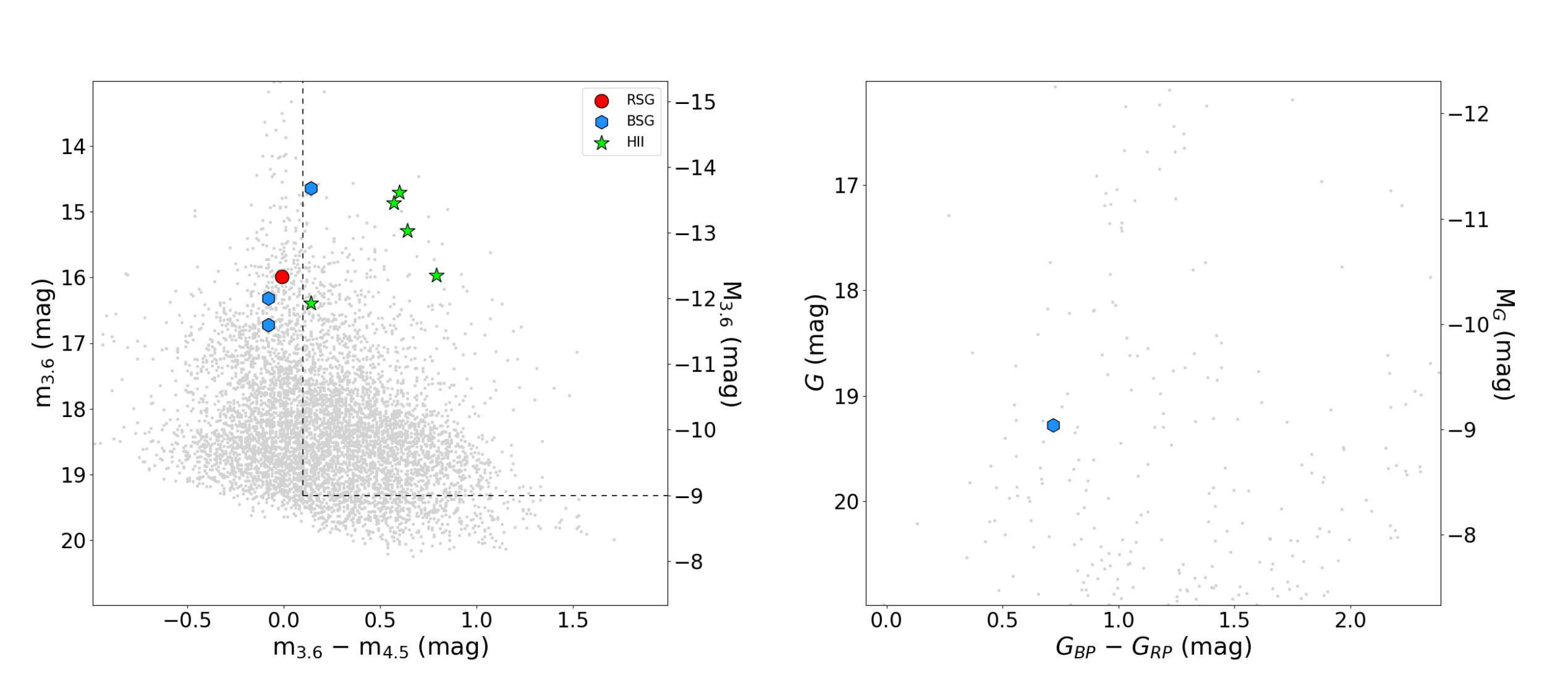}
    \caption{Same as Figure~\ref{fig:WLM_CMD}, but for NGC~1313.}
    \label{fig:NGC1313_CMD}
\end{figure*}
\vfill
\clearpage

\vfill
\begin{figure*}[ht!]
\includegraphics[width=\columnwidth]{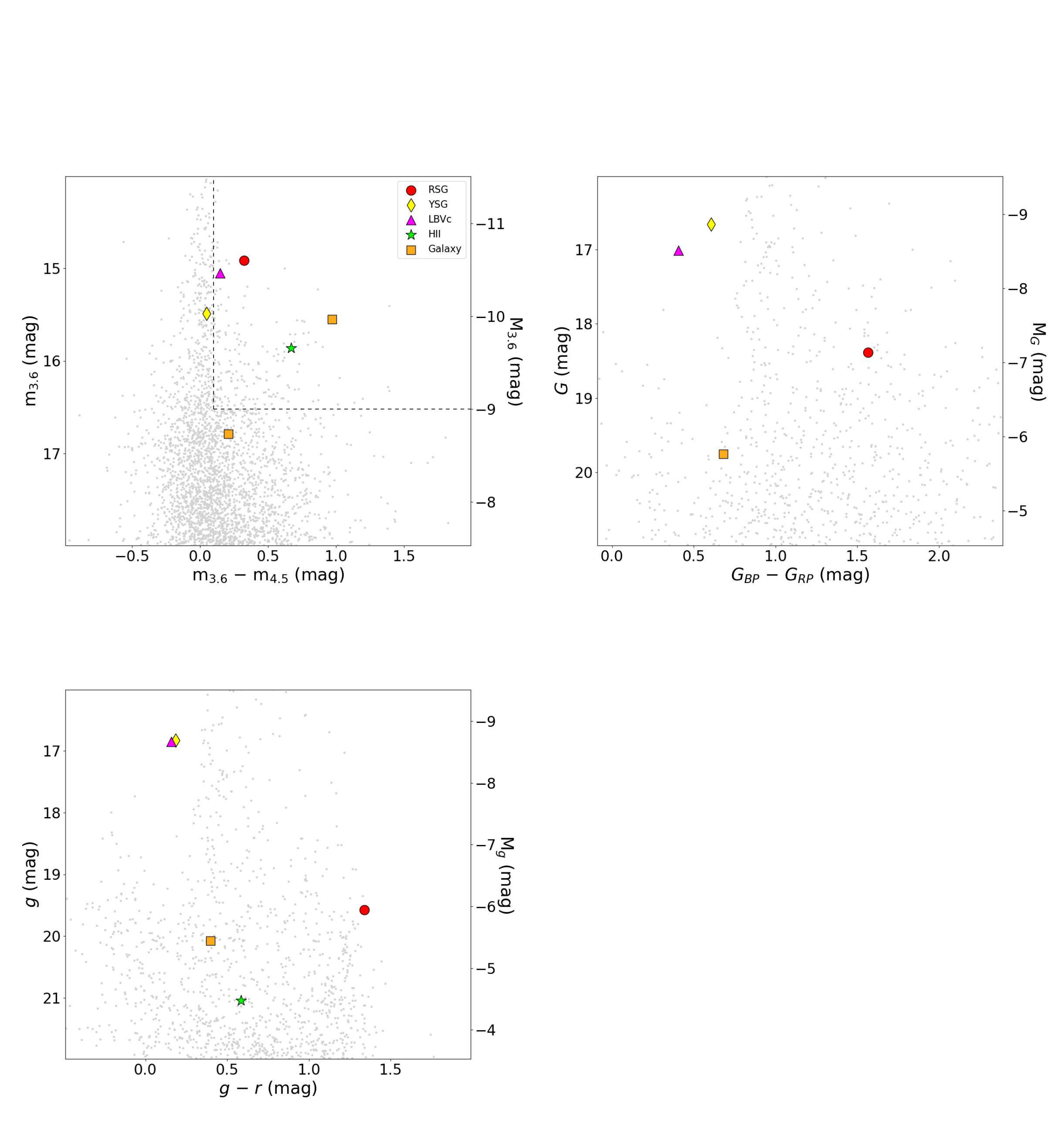}
\caption{Same as Figure~\ref{fig:WLM_CMD}, but for NGC~3109.}
\label{fig:NGC3109_CMD} 
\end{figure*}
\vfill
\clearpage

\vfill
\begin{figure*}[ht!]
\includegraphics[width=\columnwidth]{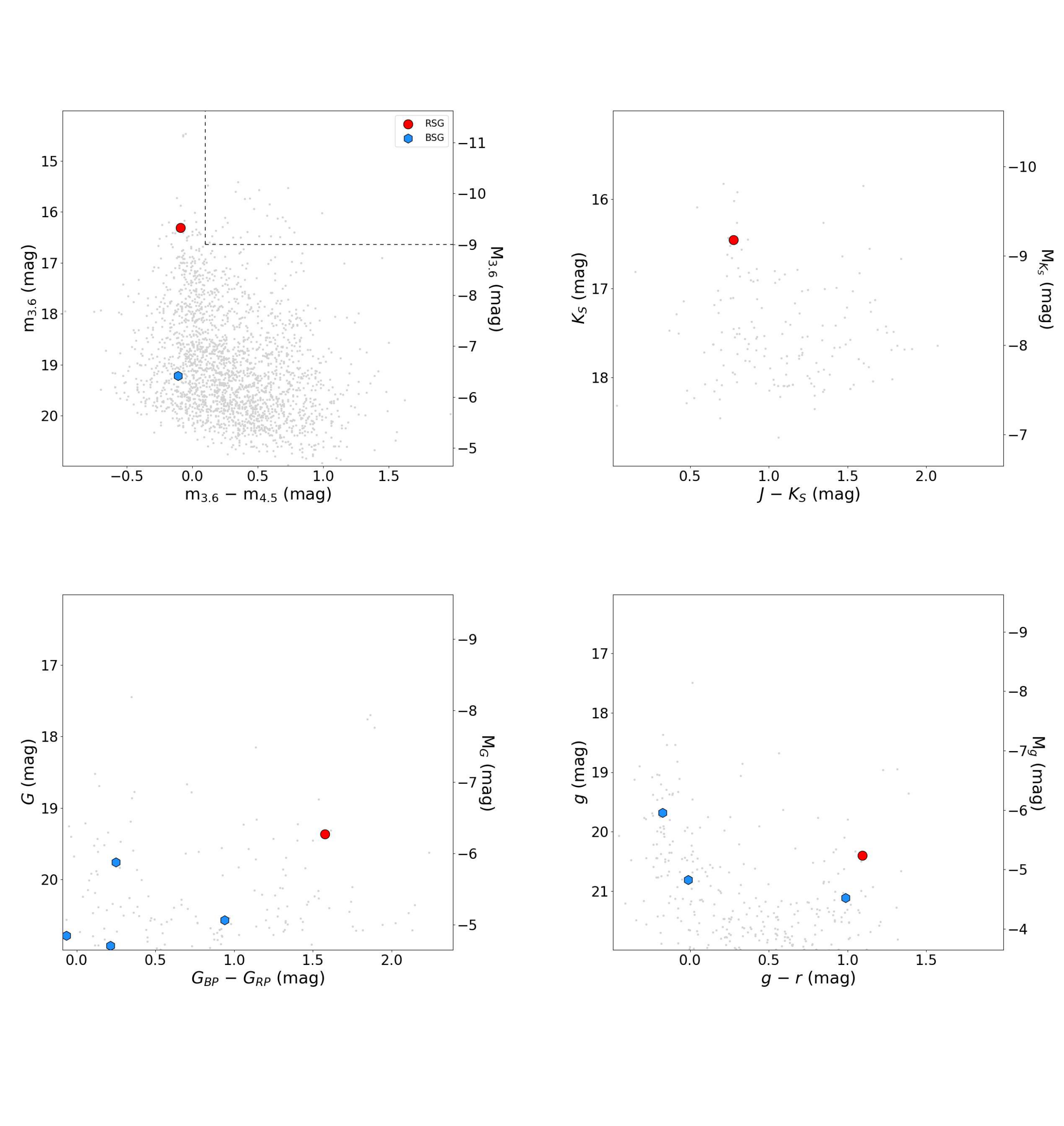}
\caption{Same as Figure~\ref{fig:WLM_CMD}, but for Sextans~A.}
\label{fig:SextansA_CMD} 
\end{figure*}
\vfill
\clearpage

\vfill
\begin{figure*}[ht!]
    \includegraphics[width=\columnwidth]{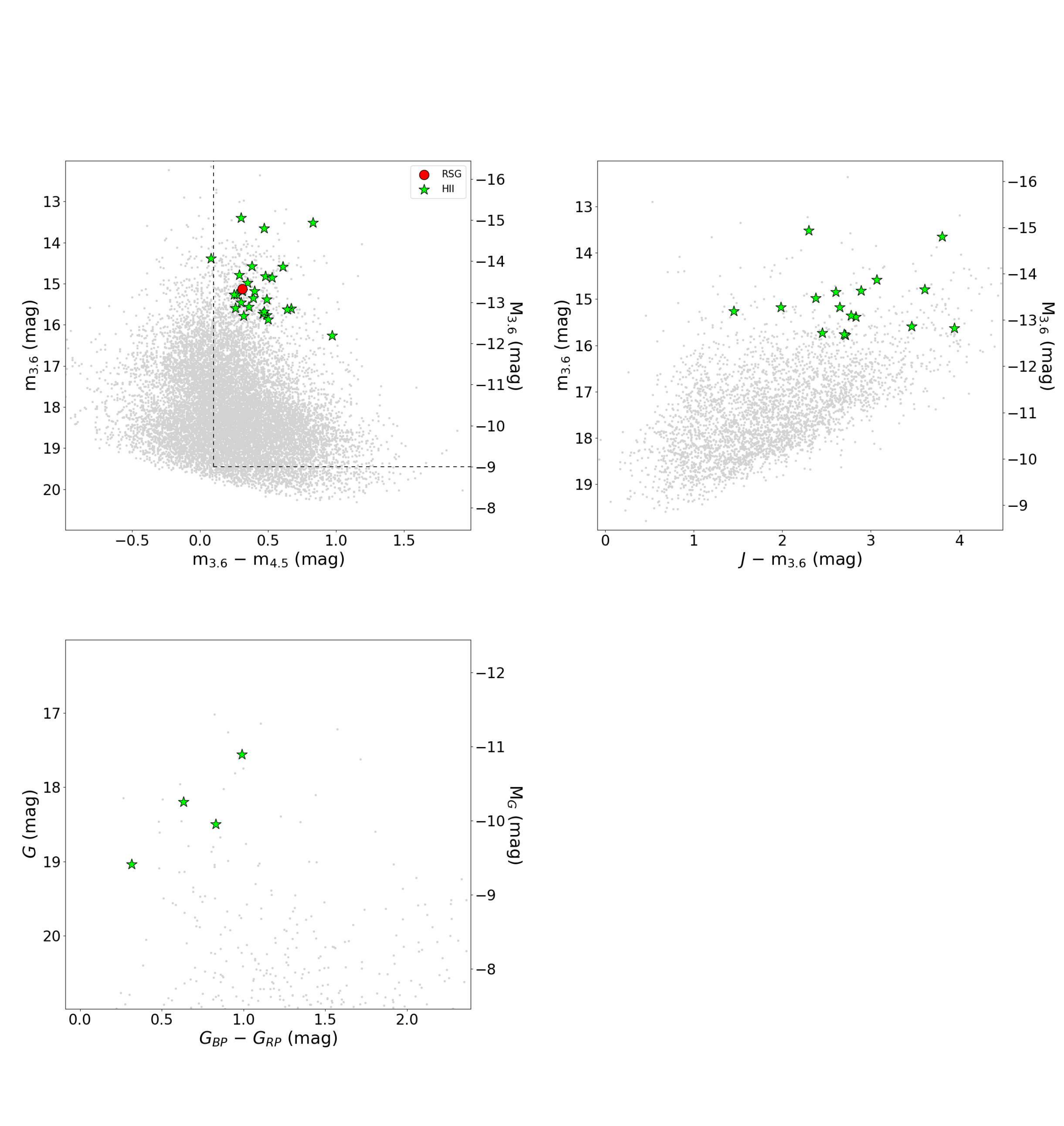}
    \caption{Same as Figure~\ref{fig:WLM_CMD}, but for M83.}
    \label{fig:M83_CMD}
\end{figure*}
\vfill
\clearpage

\vfill
\begin{figure*}[ht!]
   \includegraphics[width=\columnwidth]{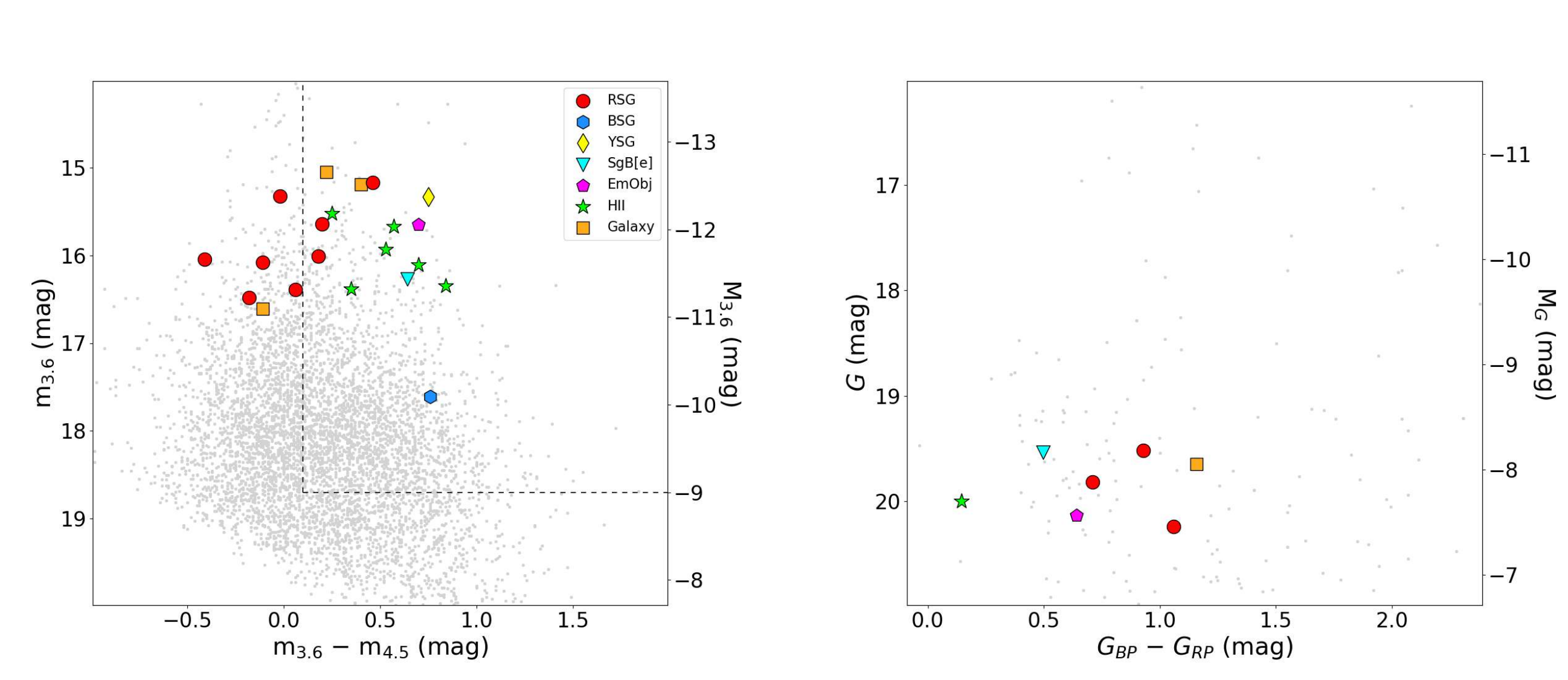}
    \caption{Same as Figure~\ref{fig:WLM_CMD}, but for NGC~7793.}
    \label{fig:NGC7793_CMD}
\end{figure*}

\clearpage

\section{Spatial distribution diagrams of classified targets}
\label{sec:ap_spclass}
This appendix provides the spatial distribution plots for all galaxies presented in this paper. These figures contain the positions of the same sources that were presented in the CMDs in Appendix~\ref{sec:ap_cmdclass}.

\begin{figure*}[ht!]
\centering    
\includegraphics[keepaspectratio,height=\linewidth, width=\linewidth]{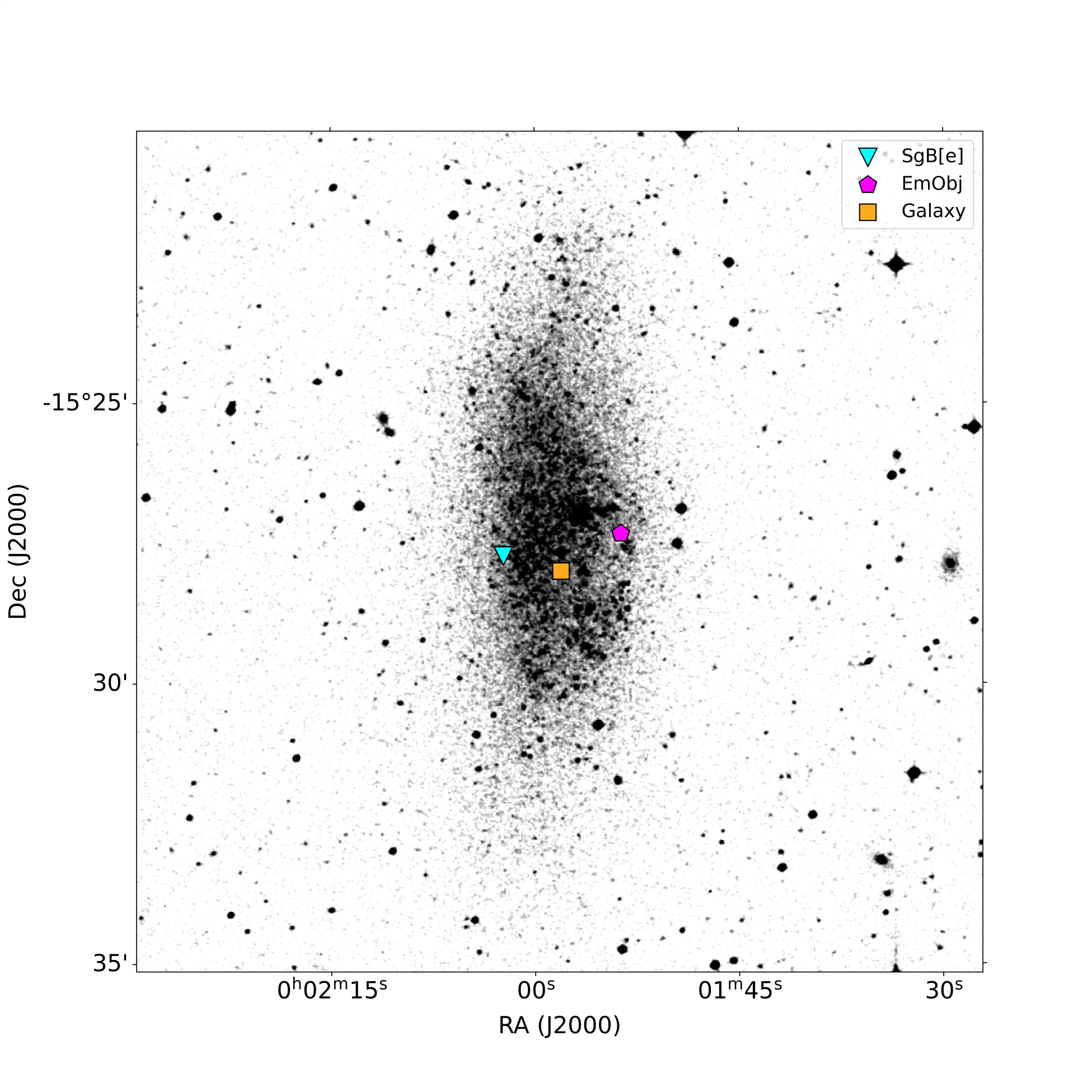}
\caption{Spatial distribution of classified targets in WLM. The background image is from DSS2.}
\label{fig:WLM_class} 
\end{figure*}

\begin{figure*}
\centering    
\includegraphics[keepaspectratio,height=\linewidth, width=\linewidth]{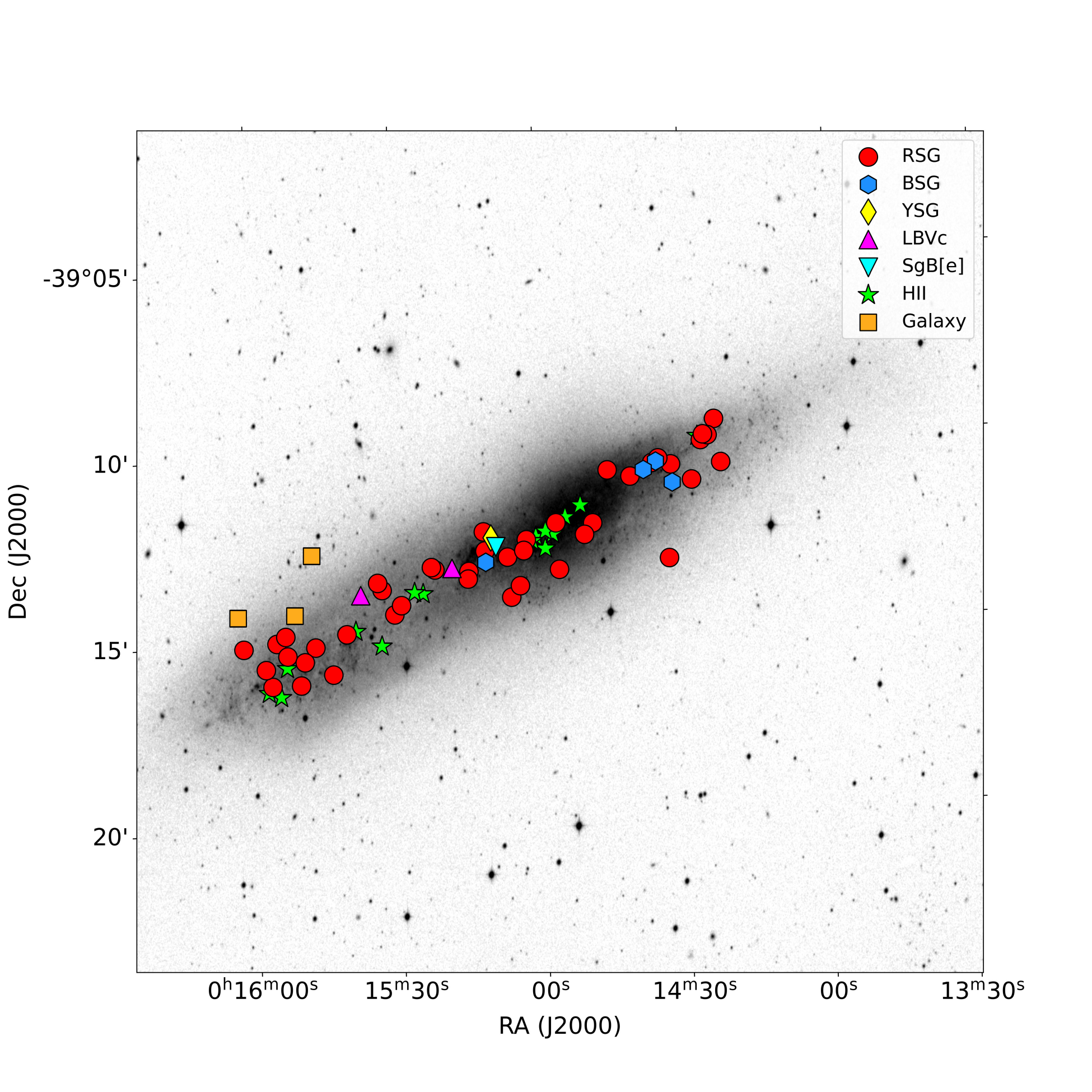}
\caption{Same as Figure~\ref {fig:WLM_class}, but for NGC~55.}
\label{fig:NGC55_class} 
\end{figure*}

\begin{figure*}
\centering    
\includegraphics[keepaspectratio,height=\linewidth, width=\linewidth]{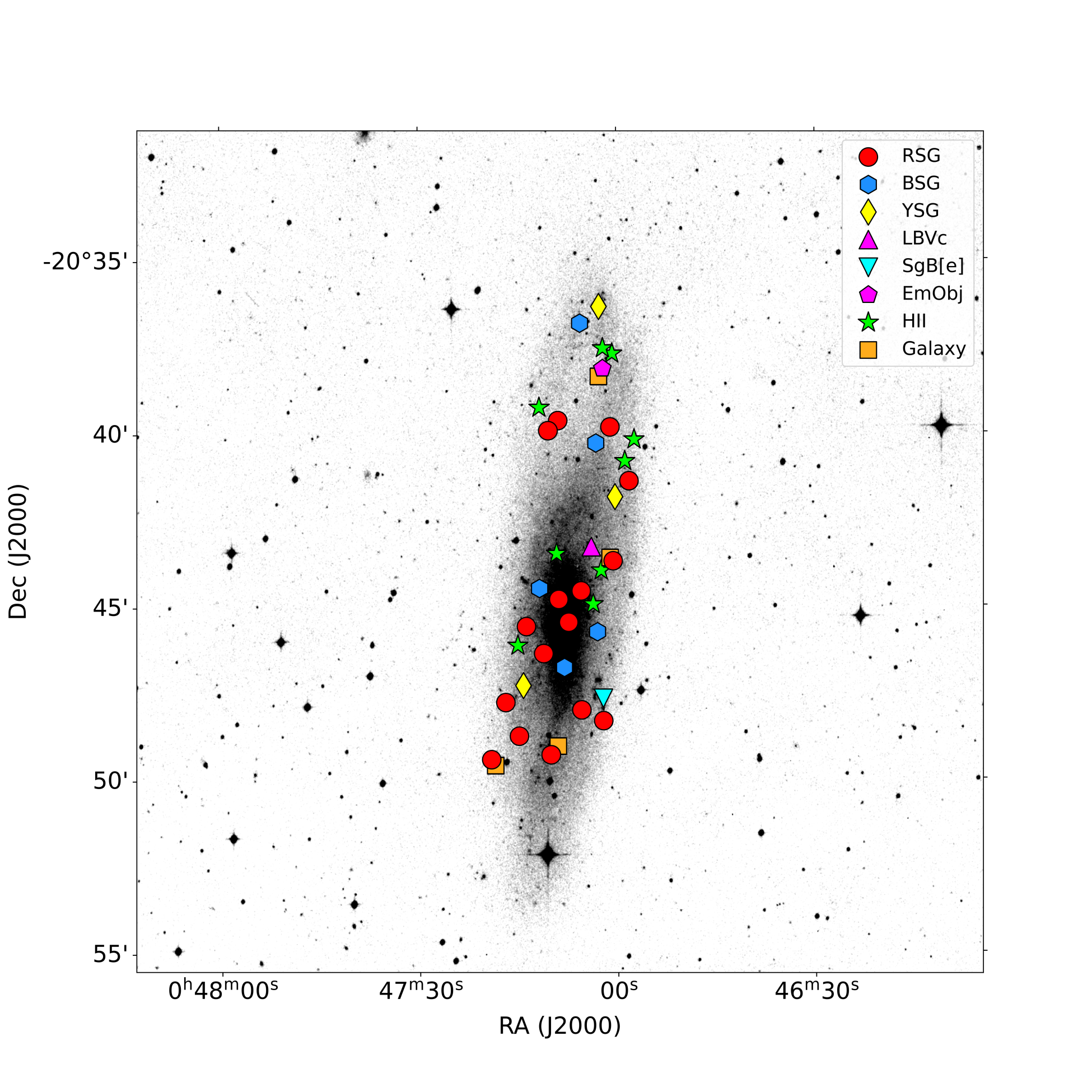}
\caption{Same as Figure~\ref {fig:WLM_class}, but for NGC~247.}
\label{fig:NGC247_class} 
\end{figure*}

\begin{figure*}
\centering    
\includegraphics[keepaspectratio,height=\linewidth, width=\linewidth]{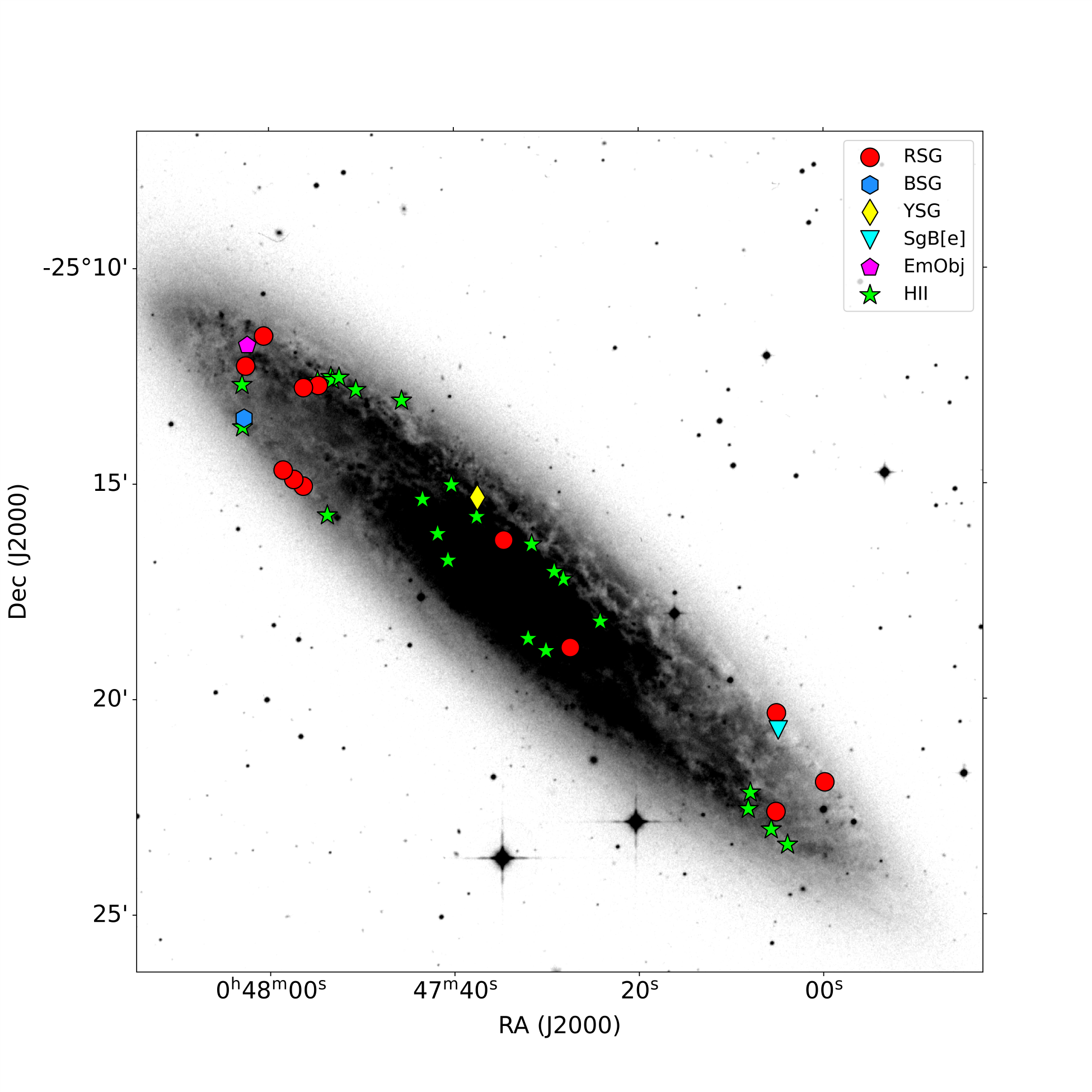}
\caption{Same as Figure~\ref {fig:WLM_class}, but for NGC~253.}
\label{fig:NGC253_class} 
\end{figure*}

\begin{figure*}
\centering    
\includegraphics[keepaspectratio,height=\linewidth, width=\linewidth]{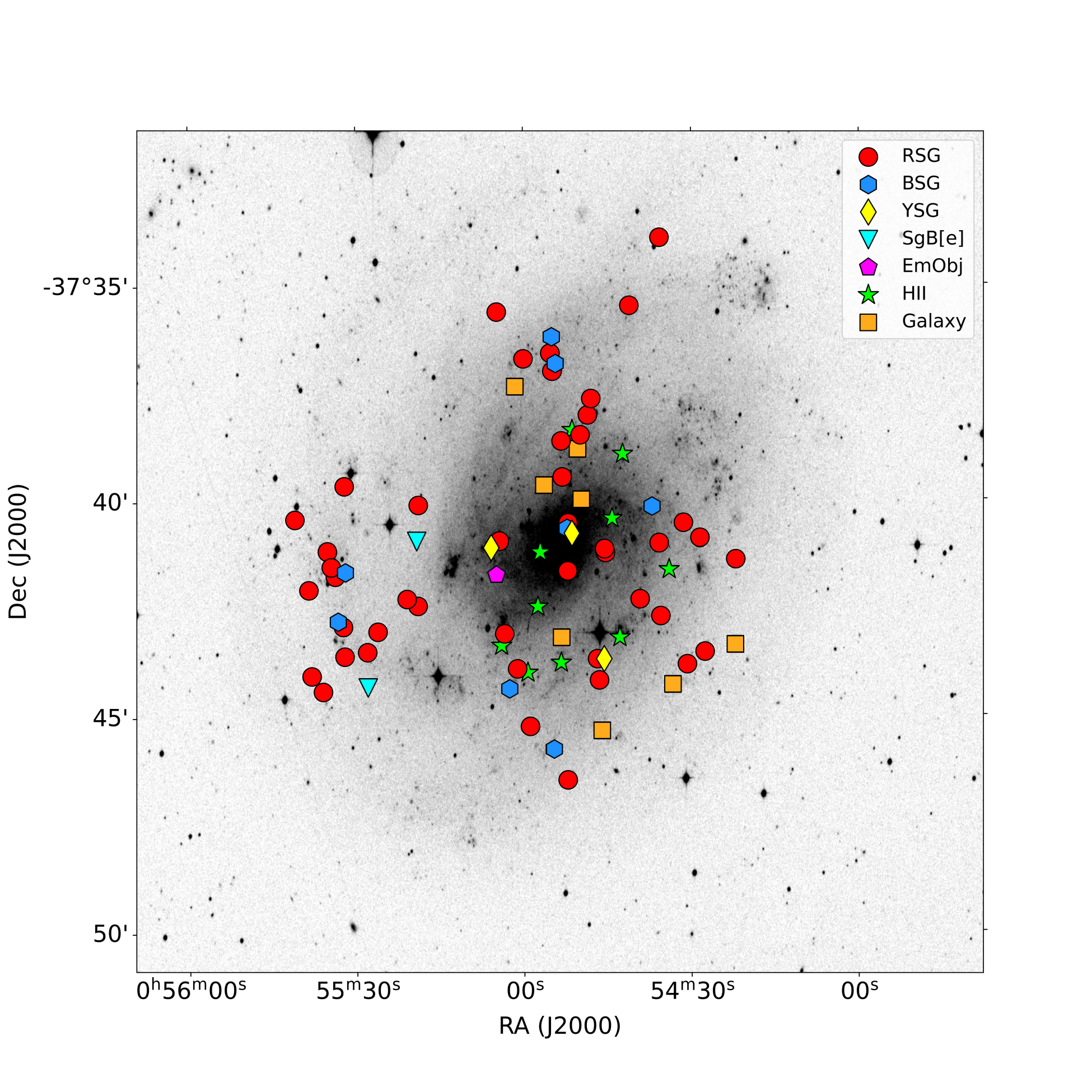}
\caption{Same as Figure~\ref {fig:WLM_class}, but for NGC~300.}
\label{fig:NGC300_class} 
\end{figure*}

\begin{figure*}
\centering    
\includegraphics[keepaspectratio,height=\linewidth, width=\linewidth]{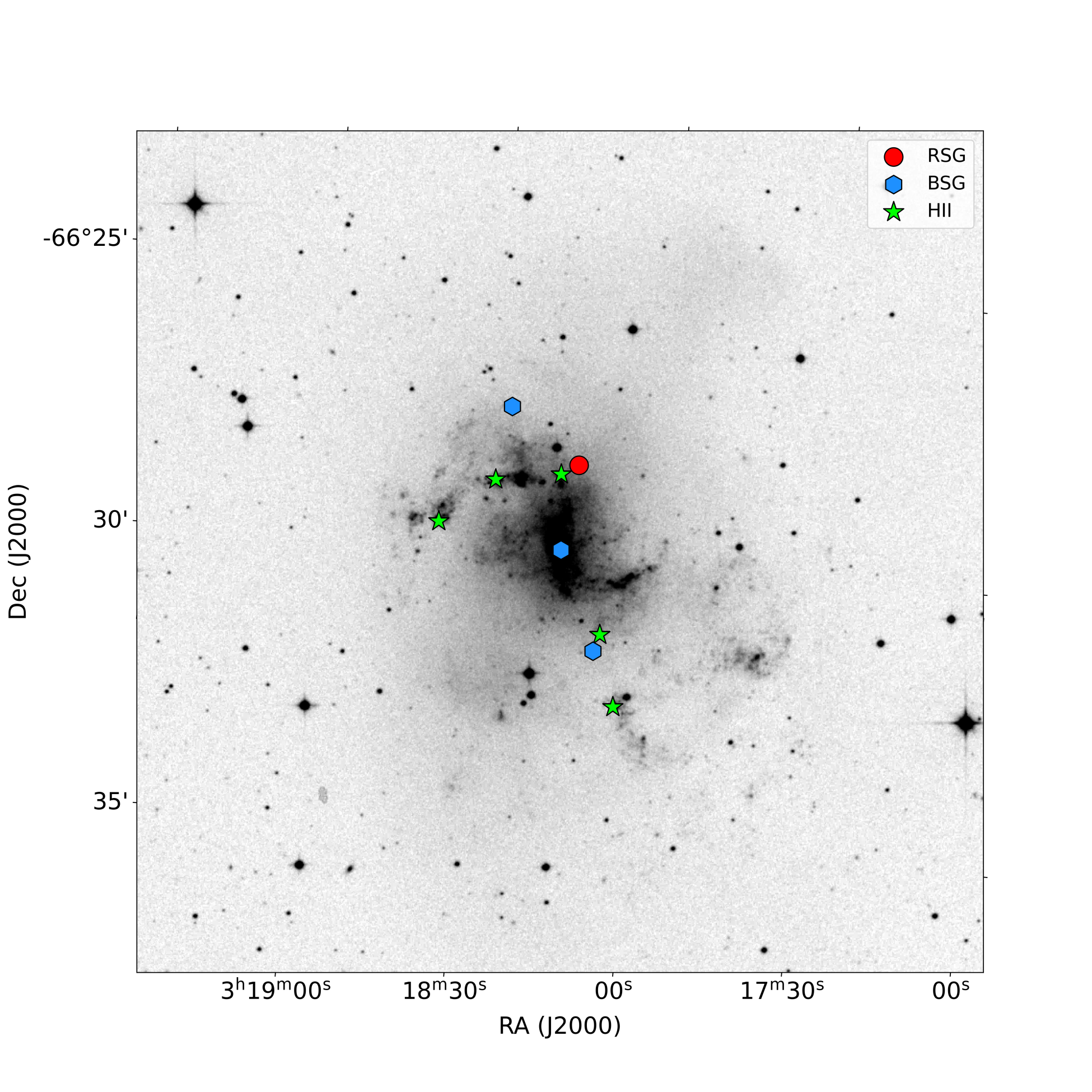}
\caption{Same as Figure~\ref {fig:WLM_class}, but for NGC~1313.}
\label{fig:NGC1313_class} 
\end{figure*}

\begin{figure*}
\centering    
\includegraphics[keepaspectratio,height=\linewidth, width=\linewidth]{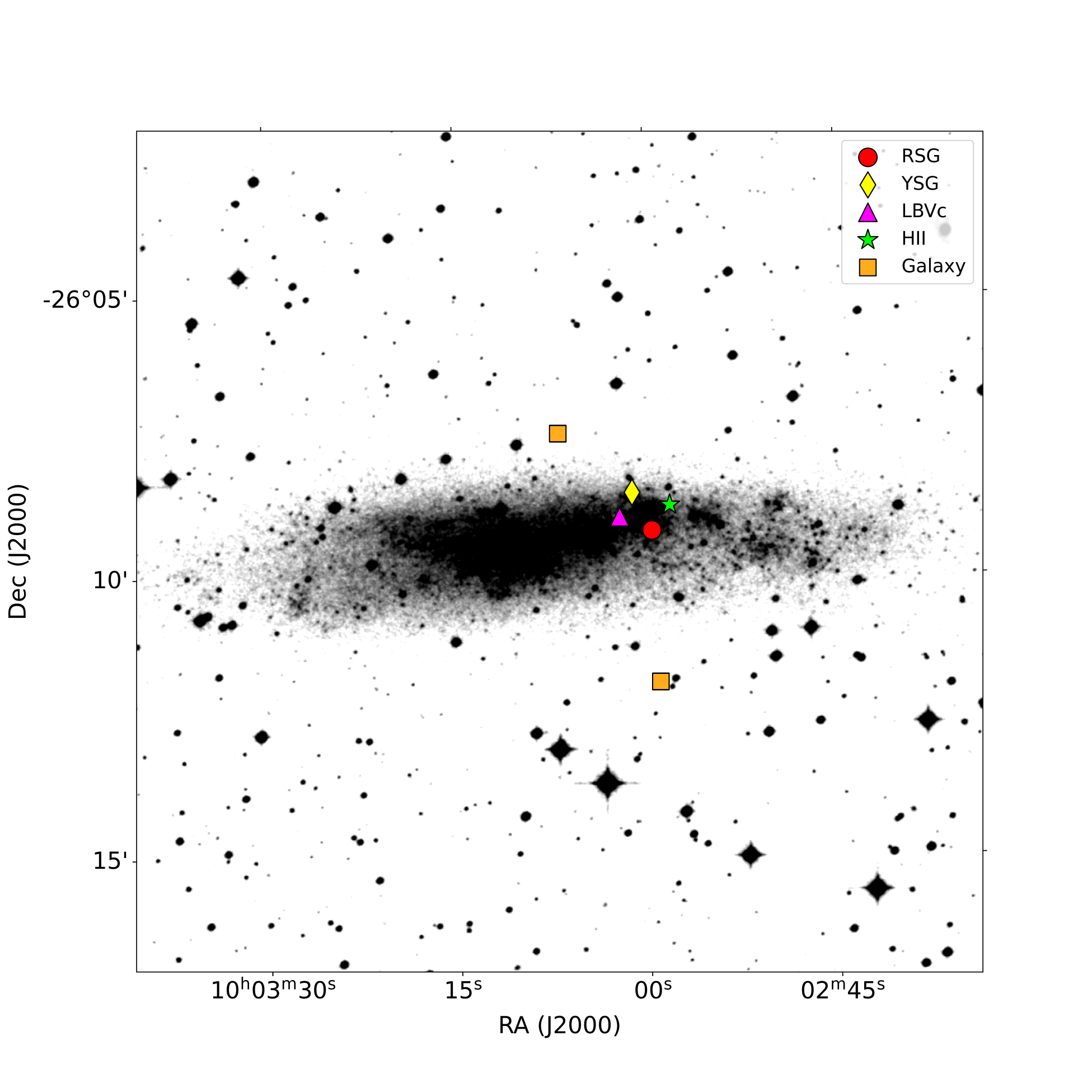}
\caption{Same as Figure~\ref {fig:WLM_class}, but for NGC~3109.}
\label{fig:NGC3109_class} 
\end{figure*}

\begin{figure*}
\centering    
\includegraphics[keepaspectratio,height=\linewidth, width=\linewidth]{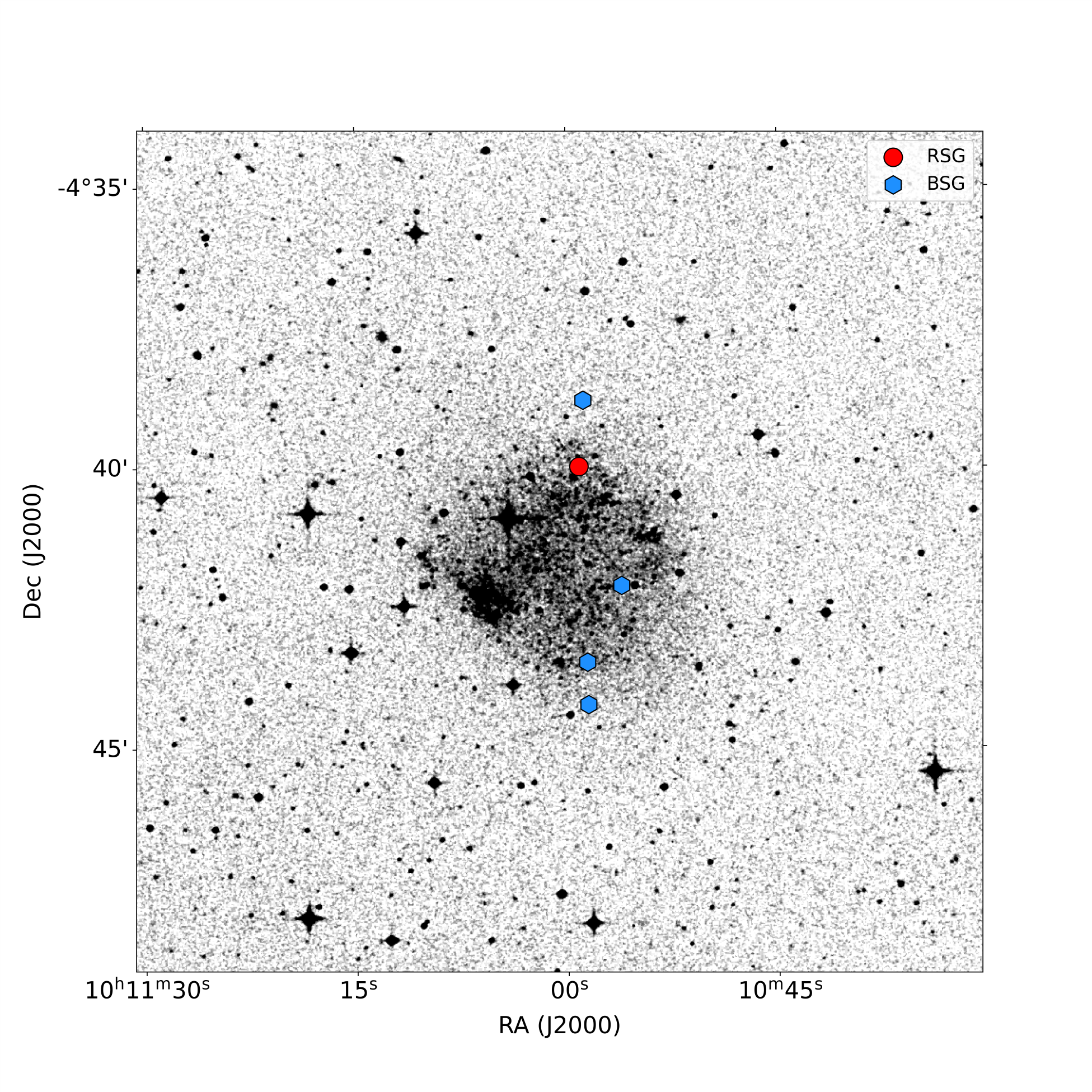}
\caption{Same as Figure~\ref {fig:WLM_class}, but for Sextans~A. }
\label{fig:SextansA_class} 
\end{figure*}

\begin{figure*}
\centering    
\includegraphics[keepaspectratio,height=\linewidth, width=\linewidth]{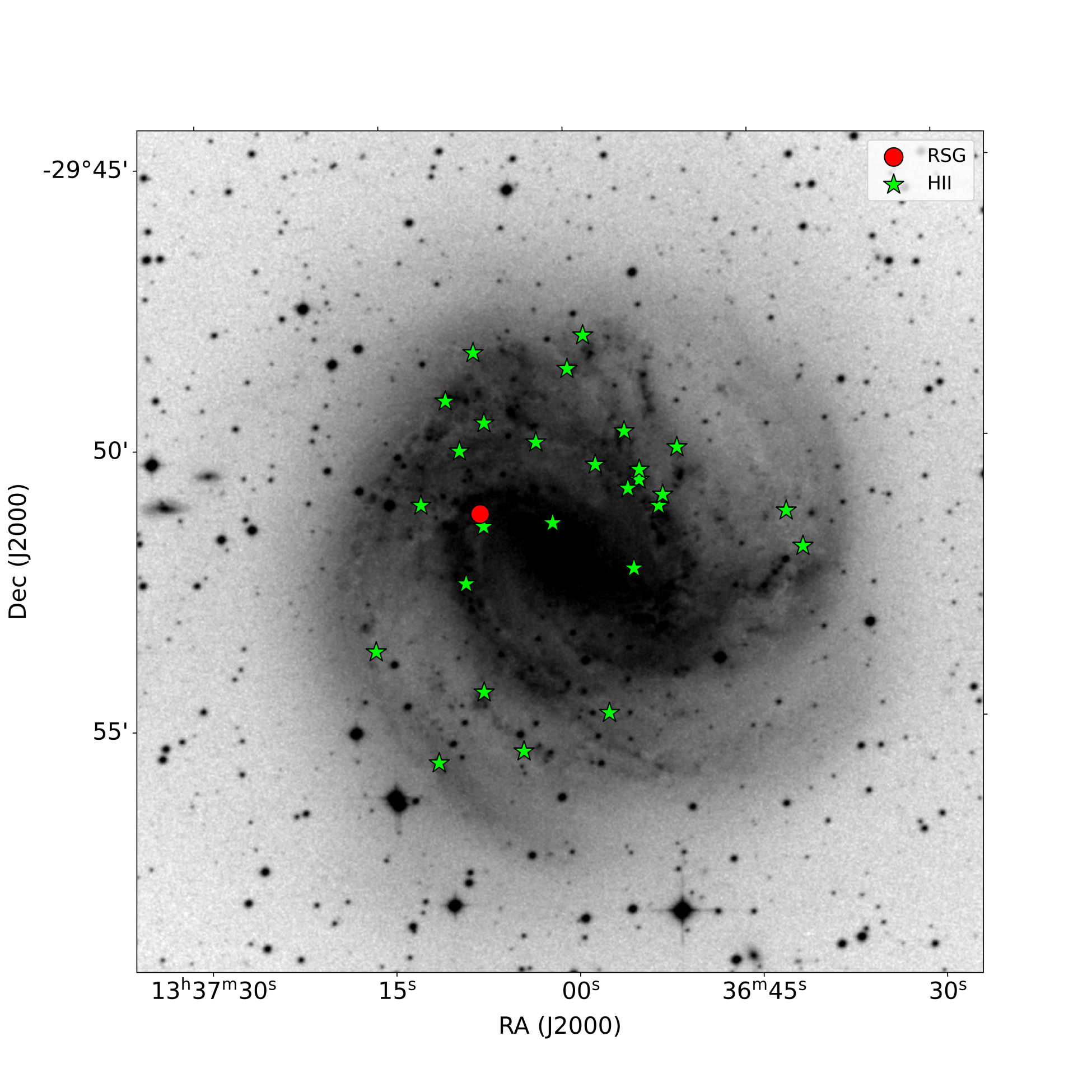}
\caption{Same as Figure~\ref{fig:WLM_class}, but for M83.}
\label{fig:M83_class} 
\end{figure*}

\begin{figure*}
\centering    
\includegraphics[keepaspectratio,height=\linewidth, width=\linewidth]{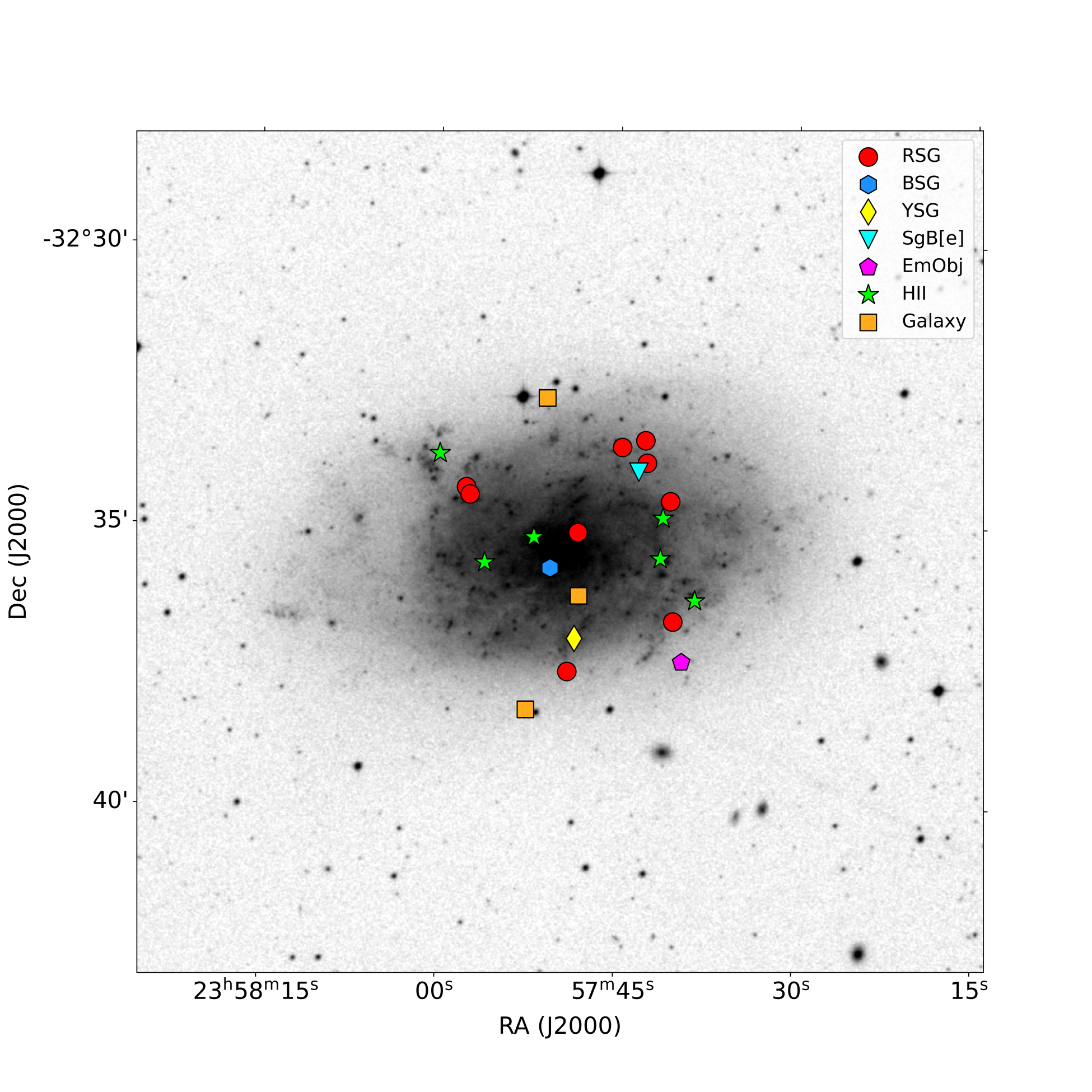}
\caption{Same as Figure~\ref {fig:M83_class}, but for NGC~7793.}
\label{fig:NGC7793_class} 
\end{figure*}

\clearpage
\section{Properties of \hii\ regions and targets with nebular emission.}
This appendix provides the flux measurements and line measurements for 76 \hii\ regions and 36 other targets with nebular emission, described in Sections~\ref{sec:class_em} and~\ref{sec:hiiregions}. We list the target ID, the spectral classification, the fluxes of the lines H$\alpha$, [\ion {N} {ii}] $\lambda$6583, [\ion {S} {ii}] $\lambda\lambda$6716 and 6731, as well as the ratios [\ion {N} {ii}]/H$\alpha$ and [\ion {S} {ii}]/H$\alpha$. 

\begin{longtable}{l l r r r r r r}
\caption{Flux measurements (10$^{-17}$ erg cm$^{-2}$ s$^{-1}$\AA$^{-1}$) and line ratios.}
\label{tab:hii_catalog}\\
\hline\hline
ID & Spectral Class. & F$_{\textsc{H}\alpha}$ & F$_{[\textsc{N~II}] 6583}$ & F$_{[\textsc{S~II}] 6716}$ & F$_{[\textsc{S~II}] 6731}$ & [$\textsc{N~II}$]/H$\alpha$ & [$\textsc{S~II}$]/H$\alpha$ \\ 
\hline 
NGC55-47 & H \textsc{ii} & 1520~$\pm$~50 & 53~$\pm$~2 & 58~$\pm$~2 & 42~$\pm$~1 & 0.035~$\pm$~0.002 & 0.066~$\pm$~0.004\\ 
NGC55-81 & H \textsc{ii}+H$\alpha$ & 2050~$\pm$~60 & 66~$\pm$~2 & 63~$\pm$~2 & 49~$\pm$~1 & 0.032~$\pm$~0.002 & 0.055~$\pm$~0.003\\ 
NGC55-84 & H \textsc{ii}+H$\alpha$ & 2920~$\pm$~90 & 130~$\pm$~4 & 119~$\pm$~4 & 87~$\pm$~3 & 0.045~$\pm$~0.003 & 0.071~$\pm$~0.004\\ 
NGC55-88 & H \textsc{ii} & 8600~$\pm$~300 & 580~$\pm$~20 & 780~$\pm$~20 & 550~$\pm$~20 & 0.067~$\pm$~0.004 & 0.154~$\pm$~0.009\\ 
NGC55-89 & H \textsc{ii} & 2640~$\pm$~80 & 97~$\pm$~3 & 86~$\pm$~9 & 66~$\pm$~7 & 0.037~$\pm$~0.002 & 0.057~$\pm$~0.007\\ 
NGC55-137 & H \textsc{ii} & 580~$\pm$~20 & 47~$\pm$~1 & 38~$\pm$~1 & 27.5~$\pm$~0.8 & 0.08~$\pm$~0.005 & 0.112~$\pm$~0.007\\ 
NGC55-149 & M I+Neb & 74~$\pm$~4 & 8.1~$\pm$~0.4 & 10.1~$\pm$~0.3 & 7.7~$\pm$~0.4 & 0.11~$\pm$~0.01 & 0.24~$\pm$~0.02\\ 
NGC55-200 & M I+Neb & 189~$\pm$~9 & 20~$\pm$~1 & 22.8~$\pm$~0.7 & 18.0~$\pm$~0.5 & 0.11~$\pm$~0.01 & 0.22~$\pm$~0.02\\ 
NGC55-295 & H \textsc{ii} & 1580~$\pm$~50 & 108~$\pm$~3 & 120~$\pm$~4 & 86~$\pm$~3 & 0.069~$\pm$~0.004 & 0.13~$\pm$~0.008\\ 
NGC55-312 & H \textsc{ii} & 340~$\pm$~20 & 13.6~$\pm$~0.4 & 17.4~$\pm$~0.5 & 12.0~$\pm$~0.4 & 0.04~$\pm$~0.003 & 0.087~$\pm$~0.007\\ 
NGC55-331 & H \textsc{ii}+H$\alpha$ & 2000~$\pm$~60 & 90~$\pm$~3 & 53~$\pm$~2 & 39~$\pm$~1 & 0.045~$\pm$~0.003 & 0.046~$\pm$~0.003\\ 
NGC55-338 & H \textsc{ii}+H$\alpha$ & 1540~$\pm$~50 & 74~$\pm$~2 & 60~$\pm$~2 & 53~$\pm$~2 & 0.048~$\pm$~0.003 & 0.073~$\pm$~0.004\\ 
NGC55-399 & H \textsc{ii} & 220~$\pm$~20 & 27.7~$\pm$~0.8 & 22.6~$\pm$~0.7 & 20.6~$\pm$~0.6 & 0.12~$\pm$~0.02 & 0.19~$\pm$~0.03\\ 
NGC55-676 & H \textsc{ii}+H$\alpha$ & 3500~$\pm$~100 & 154~$\pm$~5 & 160~$\pm$~5 & 117~$\pm$~4 & 0.044~$\pm$~0.003 & 0.079~$\pm$~0.005\\ 
NGC55-889 & M I+Neb & 23.8~$\pm$~0.7 & 6.8~$\pm$~0.3 & 9.4~$\pm$~0.5 & 7.2~$\pm$~0.4 & 0.28~$\pm$~0.02 & 0.70~$\pm$~0.06\\ 
NGC55-2117 & H \textsc{ii} & 580~$\pm$~20 & 23.2~$\pm$~0.7 & 30.0~$\pm$~0.9 & 22.0~$\pm$~0.7 & 0.04~$\pm$~0.002 & 0.089~$\pm$~0.005\\ 
NGC55-NC12 & Neb & 54~$\pm$~2 & 7.2~$\pm$~0.4 & 13.3~$\pm$~0.4 & 9.0~$\pm$~0.4 & 0.13~$\pm$~0.01 & 0.41~$\pm$~0.03\\
NGC247-162 & H \textsc{ii}+H$\alpha$ & 2690~$\pm$~80 & 133~$\pm$~4 & 96~$\pm$~3 & 75~$\pm$~2 & 0.05~$\pm$~0.003 & 0.063~$\pm$~0.004\\ 
NGC247-249 & H \textsc{ii}+H$\alpha$ & 1580~$\pm$~50 & 72~$\pm$~2 & 69~$\pm$~2 & 57~$\pm$~2 & 0.045~$\pm$~0.003 & 0.08~$\pm$~0.005\\ 
NGC247-576 & H \textsc{ii}+star & 740~$\pm$~20 & 72~$\pm$~2 & 60~$\pm$~2 & 43~$\pm$~1 & 0.097~$\pm$~0.006 & 0.14~$\pm$~0.008\\ 
NGC247-2337 & H \textsc{ii}+H$\alpha$ & 1180~$\pm$~40 & 47~$\pm$~1 & 47~$\pm$~1 & 35~$\pm$~1 & 0.040~$\pm$~0.002 & 0.07~$\pm$~0.004\\ 
NGC247-2472 & H \textsc{ii} & 350~$\pm$~10 & 32.1~$\pm$~1.0 & 26.1~$\pm$~0.8 & 18.1~$\pm$~0.5 & 0.091~$\pm$~0.005 & 0.125~$\pm$~0.008\\ 
NGC247-2832 & H \textsc{ii}+H$\alpha$ & 360~$\pm$~10 & 46~$\pm$~1 & 35~$\pm$~1 & 25.8~$\pm$~0.8 & 0.128~$\pm$~0.008 & 0.17~$\pm$~0.01\\ 
NGC247-2966 & M I+Neb & 40~$\pm$~2 & 5.8~$\pm$~0.3 & 6.3~$\pm$~0.3 & 4.8~$\pm$~0.2 & 0.15~$\pm$~0.01 & 0.28~$\pm$~0.03\\ 
NGC247-NC5 & H \textsc{ii}+H$\alpha$ & 1510~$\pm$~50 & 54~$\pm$~2 & 66~$\pm$~2 & 48~$\pm$~1 & 0.036~$\pm$~0.002 & 0.076~$\pm$~0.005\\ 
NGC253-1 & Composite & 210~$\pm$~20 & 100~$\pm$~5 & 4.5~$\pm$~0.7 & 9~$\pm$~1 & 0.49~$\pm$~0.07 & 0.07~$\pm$~0.02\\ 
NGC253-33 & H \textsc{ii} & 310~$\pm$~30 & 150~$\pm$~4 & 28.6~$\pm$~0.9 & 26.9~$\pm$~0.8 & 0.48~$\pm$~0.06 & 0.18~$\pm$~0.02\\ 
NGC253-39 & H \textsc{ii}+H$\alpha$ & 780~$\pm$~20 & 286~$\pm$~9 & 101~$\pm$~3 & 75~$\pm$~2 & 0.37~$\pm$~0.02 & 0.23~$\pm$~0.01\\ 
NGC253-47 & Cluster & 65~$\pm$~6 & 31.8~$\pm$~1.0 & 12.1~$\pm$~0.6 & 8.9~$\pm$~0.6 & 0.49~$\pm$~0.06 & 0.32~$\pm$~0.05\\ 
NGC253-75 & H \textsc{ii} & 309~$\pm$~9 & 151~$\pm$~5 & 49~$\pm$~1 & 43~$\pm$~1 & 0.49~$\pm$~0.03 & 0.30~$\pm$~0.02\\ 
NGC253-98 & H \textsc{ii} & 630~$\pm$~20 & 340~$\pm$~10 & 57~$\pm$~2 & 49~$\pm$~1 & 0.54~$\pm$~0.03 & 0.17~$\pm$~0.01\\ 
NGC253-100 & Cluster & 174~$\pm$~5 & 54~$\pm$~2 & 16.5~$\pm$~0.5 & 12.4~$\pm$~0.4 & 0.31~$\pm$~0.02 & 0.17~$\pm$~0.01\\ 
NGC253-103 & H \textsc{ii}+H$\alpha$ & 1300~$\pm$~40 & 530~$\pm$~20 & 107~$\pm$~3 & 78~$\pm$~2 & 0.40~$\pm$~0.02 & 0.142~$\pm$~0.009\\ 
NGC253-118 & H \textsc{ii}+H$\alpha$ & 1720~$\pm$~90 & 620~$\pm$~30 & 173~$\pm$~9 & 112~$\pm$~6 & 0.36~$\pm$~0.04 & 0.17~$\pm$~0.02\\ 
NGC253-146 & H \textsc{ii}+cool star & 880~$\pm$~30 & 304~$\pm$~9 & 75~$\pm$~2 & 58~$\pm$~2 & 0.35~$\pm$~0.02 & 0.151~$\pm$~0.009\\ 
NGC253-175 & H \textsc{ii} & 1010~$\pm$~50 & 350~$\pm$~20 & 91~$\pm$~3 & 71~$\pm$~4 & 0.35~$\pm$~0.03 & 0.16~$\pm$~0.01\\ 
NGC253-185 & Cluster & 1160~$\pm$~30 & 236~$\pm$~7 & 46~$\pm$~1 & 31.8~$\pm$~1.0 & 0.20~$\pm$~0.01 & 0.067~$\pm$~0.004\\ 
NGC253-214 & Cluster & 580~$\pm$~20 & 188~$\pm$~6 & 72~$\pm$~2 & 56~$\pm$~2 & 0.32~$\pm$~0.02 & 0.22~$\pm$~0.01\\ 
NGC253-227 & Neb & 330~$\pm$~10 & 89~$\pm$~3 & 23.0~$\pm$~0.7 & 16.7~$\pm$~0.5 & 0.27~$\pm$~0.02 & 0.12~$\pm$~0.007\\ 
NGC253-261 & Cluster & 201~$\pm$~6 & 55~$\pm$~3 & 21.8~$\pm$~0.7 & 15~$\pm$~1 & 0.28~$\pm$~0.02 & 0.18~$\pm$~0.02\\ 
NGC253-378 & H \textsc{ii} & 198~$\pm$~6 & 61~$\pm$~2 & 15.7~$\pm$~0.8 & 14.9~$\pm$~0.7 & 0.31~$\pm$~0.02 & 0.15~$\pm$~0.01\\ 
NGC253-402 & Neb+star & 153~$\pm$~5 & 53~$\pm$~2 & 11~$\pm$~1 & 11~$\pm$~1 & 0.34~$\pm$~0.02 & 0.14~$\pm$~0.02\\ 
NGC253-419 & H \textsc{ii} & 1900~$\pm$~100 & 280~$\pm$~10 & 93~$\pm$~5 & 67~$\pm$~3 & 0.15~$\pm$~0.01 & 0.084~$\pm$~0.008\\ 
NGC253-433 & Neb & 860~$\pm$~40 & 220~$\pm$~10 & 64~$\pm$~6 & 46~$\pm$~5 & 0.26~$\pm$~0.03 & 0.13~$\pm$~0.02\\ 
NGC253-452 & Cool star+neb & 15~$\pm$~1 & 10~$\pm$~2 & 5~$\pm$~1 & 3.3~$\pm$~0.7 & 0.7~$\pm$~0.2 & 0.6~$\pm$~0.2\\ 
NGC253-510 & H \textsc{ii} & 312~$\pm$~9 & 113~$\pm$~3 & 32.9~$\pm$~1.0 & 26.2~$\pm$~0.8 & 0.36~$\pm$~0.02 & 0.19~$\pm$~0.01\\ 
NGC253-527 & Neb & 62~$\pm$~6 & 7~$\pm$~1 & 5.3~$\pm$~0.5 & 2.4~$\pm$~0.6 & 0.11~$\pm$~0.03 & 0.13~$\pm$~0.03\\ 
NGC253-533 & Neb & 313~$\pm$~9 & 98~$\pm$~3 & 33.1~$\pm$~1.0 & 22.8~$\pm$~0.7 & 0.31~$\pm$~0.02 & 0.18~$\pm$~0.01\\ 
NGC253-608 & H \textsc{ii} & 470~$\pm$~10 & 140~$\pm$~4 & 65~$\pm$~2 & 46~$\pm$~1 & 0.3~$\pm$~0.02 & 0.24~$\pm$~0.01\\ 
NGC253-674 & H \textsc{ii}+H$\alpha$ & 2250~$\pm$~70 & 430~$\pm$~10 & 92~$\pm$~3 & 71~$\pm$~2 & 0.19~$\pm$~0.01 & 0.072~$\pm$~0.004\\ 
NGC253-733 & H \textsc{ii}+H$\alpha$ & 1760~$\pm$~50 & 232~$\pm$~7 & 78~$\pm$~2 & 56~$\pm$~2 & 0.132~$\pm$~0.008 & 0.076~$\pm$~0.005\\ 
NGC253-769 & Cluster & 4.2~$\pm$~0.8 & 1.3~$\pm$~0.4 & 2.2~$\pm$~0.4 & 1.2~$\pm$~0.4 & 0.3~$\pm$~0.2 & 0.8~$\pm$~0.4\\ 
NGC253-992 & Cluster & 105~$\pm$~5 & 16.8~$\pm$~0.8 & 6.4~$\pm$~0.3 & 5.4~$\pm$~0.2 & 0.16~$\pm$~0.02 & 0.11~$\pm$~0.01\\ 
NGC253-1162 & M I+Neb & 18~$\pm$~2 & 8.7~$\pm$~0.9 & 5.3~$\pm$~0.4 & 3.9~$\pm$~0.3 & 0.47~$\pm$~0.09 & 0.5~$\pm$~0.09\\ 
NGC253-1657 & H \textsc{ii} & 430~$\pm$~10 & 163~$\pm$~5 & 140~$\pm$~4 & 103~$\pm$~3 & 0.38~$\pm$~0.02 & 0.57~$\pm$~0.03\\ 
NGC253-1776 & Neb & 74~$\pm$~4 & 29.0~$\pm$~0.9 & 15.0~$\pm$~0.4 & 10.0~$\pm$~0.5 & 0.39~$\pm$~0.03 & 0.34~$\pm$~0.03\\ 
NGC253-NoSp2 & Blue+Neb & 15~$\pm$~3 & 5~$\pm$~1 & 6.1~$\pm$~0.6 & 3.2~$\pm$~0.6 & 0.40~$\pm$~0.2 & 0.6~$\pm$~0.2\\ 
NGC300-42 & Neb & 29~$\pm$~1 & 2.7~$\pm$~0.3 & 4.2~$\pm$~0.2 & 4.4~$\pm$~0.2 & 0.09~$\pm$~0.01 & 0.29~$\pm$~0.03\\ 
NGC300-84 & H \textsc{ii} & 3300~$\pm$~100 & 247~$\pm$~7 & 80~$\pm$~4 & 72~$\pm$~4 & 0.074~$\pm$~0.004 & 0.046~$\pm$~0.004\\ 
NGC300-145 & H \textsc{ii} & 140~$\pm$~10 & 19.6~$\pm$~1.0 & 11.8~$\pm$~0.4 & 9.2~$\pm$~0.5 & 0.14~$\pm$~0.02 & 0.15~$\pm$~0.02\\ 
NGC300-182 & H \textsc{ii} & 540~$\pm$~20 & 137~$\pm$~4 & 52~$\pm$~2 & 38~$\pm$~1 & 0.25~$\pm$~0.02 & 0.167~$\pm$~0.01\\ 
NGC300-186 & M I+H \textsc{ii} & 1450~$\pm$~40 & 340~$\pm$~10 & 167~$\pm$~5 & 122~$\pm$~4 & 0.23~$\pm$~0.01 & 0.20~$\pm$~0.01\\ 
NGC300-288 & H \textsc{ii}+H$\alpha$ & 2880~$\pm$~90 & 400~$\pm$~10 & 179~$\pm$~5 & 139~$\pm$~4 & 0.14~$\pm$~0.008 & 0.11~$\pm$~0.007\\ 
NGC300-333 & H \textsc{ii}+H$\alpha$ & 2170~$\pm$~70 & 340~$\pm$~10 & 104~$\pm$~3 & 74~$\pm$~2 & 0.156~$\pm$~0.009 & 0.082~$\pm$~0.005\\ 
NGC300-770 & H \textsc{ii} & 810~$\pm$~20 & 98~$\pm$~3 & 44~$\pm$~1 & 33~$\pm$~1 & 0.122~$\pm$~0.007 & 0.095~$\pm$~0.006\\ 
NGC300-1381 & H \textsc{ii}+H$\alpha$ & 1330~$\pm$~40 & 184~$\pm$~6 & 99~$\pm$~3 & 70~$\pm$~2 & 0.138~$\pm$~0.008 & 0.127~$\pm$~0.008\\ 
NGC300-1528 & H \textsc{ii} & 237~$\pm$~7 & 44~$\pm$~1 & 35~$\pm$~1 & 23.6~$\pm$~0.7 & 0.18~$\pm$~0.01 & 0.25~$\pm$~0.01\\ 
NGC300-NC21 & EmObj+Neb & 560~$\pm$~60 & 420~$\pm$~40 & 15.5~$\pm$~0.5 & 24.5~$\pm$~0.7 & 0.7~$\pm$~0.1 & 0.071~$\pm$~0.009\\ 
NGC300-NC4 & M I+Neb & 7.0~$\pm$~0.7 & 1.3~$\pm$~0.2 & 2.9~$\pm$~0.3 & 2.1~$\pm$~0.2 & 0.19~$\pm$~0.05 & 0.7~$\pm$~0.1\\ 
NGC1313-40 & H \textsc{ii}+H$\alpha$ & 5000~$\pm$~200 & 171~$\pm$~5 & 127~$\pm$~4 & 102~$\pm$~3 & 0.034~$\pm$~0.002 & 0.046~$\pm$~0.003\\ 
NGC1313-108 & H \textsc{ii}+H$\alpha$ & 3180~$\pm$~100 & 120~$\pm$~4 & 139~$\pm$~4 & 96~$\pm$~3 & 0.038~$\pm$~0.002 & 0.074~$\pm$~0.004\\ 
NGC1313-92 & Cluster & 1570~$\pm$~50 & 122~$\pm$~4 & 85~$\pm$~3 & 63~$\pm$~2 & 0.077~$\pm$~0.005 & 0.094~$\pm$~0.006\\ 
NGC1313-429 & H \textsc{ii} & 720~$\pm$~20 & 39~$\pm$~1 & 35~$\pm$~1 & 24.5~$\pm$~0.7 & 0.054~$\pm$~0.003 & 0.082~$\pm$~0.005\\ 
NGC1313-48 & B I & 490~$\pm$~50 & 56~$\pm$~2 & 60~$\pm$~20 & 46~$\pm$~1 & 0.11~$\pm$~0.01 & 0.22~$\pm$~0.06\\ 
NGC3109-401 & H \textsc{ii}+H$\alpha$ & 690~$\pm$~20 & 18.8~$\pm$~0.6 & 19.8~$\pm$~0.6 & 19.9~$\pm$~0.6 & 0.027~$\pm$~0.002 & 0.058~$\pm$~0.003\\ 
M83-79 & H \textsc{ii}+H$\alpha$ & 3600~$\pm$~100 & 2570~$\pm$~80 & 327~$\pm$~10 & 306~$\pm$~9 & 0.72~$\pm$~0.04 & 0.18~$\pm$~0.01\\ 
M83-127 & H \textsc{ii}+H$\alpha$ & 3300~$\pm$~100 & 1360~$\pm$~40 & 244~$\pm$~7 & 184~$\pm$~6 & 0.41~$\pm$~0.02 & 0.129~$\pm$~0.008\\ 
M83-135 & H \textsc{ii}+H$\alpha$ & 1190~$\pm$~40 & 570~$\pm$~20 & 134~$\pm$~4 & 104~$\pm$~3 & 0.48~$\pm$~0.03 & 0.2~$\pm$~0.01\\ 
M83-296 & H \textsc{ii}+H$\alpha$ & 3280~$\pm$~100 & 1410~$\pm$~40 & 283~$\pm$~8 & 217~$\pm$~7 & 0.43~$\pm$~0.03 & 0.152~$\pm$~0.009\\ 
M83-365 & H \textsc{ii}+H$\alpha$ & 2670~$\pm$~80 & 830~$\pm$~20 & 125~$\pm$~4 & 94~$\pm$~3 & 0.31~$\pm$~0.02 & 0.082~$\pm$~0.005\\ 
M83-422 & H \textsc{ii}+H$\alpha$ & 2120~$\pm$~60 & 830~$\pm$~20 & 138~$\pm$~4 & 104~$\pm$~3 & 0.39~$\pm$~0.02 & 0.114~$\pm$~0.007\\ 
M83-433 & H \textsc{ii} & 500~$\pm$~10 & 85~$\pm$~3 & 16.9~$\pm$~0.5 & 13.3~$\pm$~0.4 & 0.17~$\pm$~0.01 & 0.061~$\pm$~0.004\\ 
M83-479 & Blue+Neb & 200~$\pm$~6 & 79~$\pm$~2 & 29.2~$\pm$~0.9 & 21.8~$\pm$~0.7 & 0.39~$\pm$~0.02 & 0.26~$\pm$~0.02\\ 
M83-499 & H \textsc{ii}+H$\alpha$ & 630~$\pm$~20 & 285~$\pm$~9 & 47~$\pm$~1 & 34~$\pm$~1 & 0.45~$\pm$~0.03 & 0.127~$\pm$~0.008\\ 
M83-528 & Blue+Neb & 490~$\pm$~10 & 190~$\pm$~6 & 57~$\pm$~2 & 41~$\pm$~1 & 0.39~$\pm$~0.02 & 0.20~$\pm$~0.01\\ 
M83-562 & H \textsc{ii} & 1340~$\pm$~40 & 600~$\pm$~20 & 129~$\pm$~4 & 99~$\pm$~3 & 0.45~$\pm$~0.03 & 0.17~$\pm$~0.01\\ 
M83-601 & Blue+Neb & 490~$\pm$~10 & 132~$\pm$~4 & 59~$\pm$~2 & 55~$\pm$~2 & 0.27~$\pm$~0.02 & 0.23~$\pm$~0.01\\ 
M83-659 & H \textsc{ii} & 153~$\pm$~5 & 45~$\pm$~1 & 22.1~$\pm$~0.7 & 15.9~$\pm$~0.5 & 0.30~$\pm$~0.02 & 0.25~$\pm$~0.01\\ 
M83-704 & H \textsc{ii} & 440~$\pm$~10 & 188~$\pm$~6 & 64~$\pm$~2 & 43~$\pm$~1 & 0.43~$\pm$~0.03 & 0.25~$\pm$~0.01\\ 
M83-721 & H \textsc{ii}+H$\alpha$ & 1550~$\pm$~50 & 510~$\pm$~20 & 108~$\pm$~3 & 81~$\pm$~2 & 0.33~$\pm$~0.02 & 0.122~$\pm$~0.007\\ 
M83-761 & H \textsc{ii}+H$\alpha$ & 2600~$\pm$~80 & 1050~$\pm$~30 & 192~$\pm$~6 & 137~$\pm$~4 & 0.4~$\pm$~0.02 & 0.127~$\pm$~0.008\\ 
M83-780 & H \textsc{ii}+H$\alpha$ & 470~$\pm$~10 & 185~$\pm$~6 & 64~$\pm$~2 & 50~$\pm$~1 & 0.4~$\pm$~0.02 & 0.24~$\pm$~0.01\\ 
M83-803 & Cluster & 86~$\pm$~3 & 32.4~$\pm$~1.0 & 15.4~$\pm$~0.5 & 9.7~$\pm$~0.3 & 0.38~$\pm$~0.02 & 0.29~$\pm$~0.02\\ 
M83-807 & Neb & 86~$\pm$~3 & 40~$\pm$~1 & 16.2~$\pm$~0.5 & 12.1~$\pm$~0.4 & 0.47~$\pm$~0.03 & 0.33~$\pm$~0.02\\ 
M83-814 & H \textsc{ii} & 560~$\pm$~20 & 212~$\pm$~6 & 39~$\pm$~1 & 31.3~$\pm$~0.9 & 0.38~$\pm$~0.02 & 0.127~$\pm$~0.008\\ 
M83-830 & H \textsc{ii} & 460~$\pm$~10 & 170~$\pm$~5 & 36~$\pm$~1 & 25.8~$\pm$~0.8 & 0.37~$\pm$~0.02 & 0.136~$\pm$~0.008\\ 
M83-885 & Neb+hot star & 218~$\pm$~7 & 64~$\pm$~2 & 20.5~$\pm$~0.6 & 15.7~$\pm$~0.5 & 0.30~$\pm$~0.02 & 0.166~$\pm$~0.01\\ 
M83-993 & H \textsc{ii} & 400~$\pm$~10 & 178~$\pm$~5 & 31.4~$\pm$~0.9 & 21.2~$\pm$~0.6 & 0.45~$\pm$~0.03 & 0.132~$\pm$~0.008\\ 
M83-999 & H \textsc{ii} & 690~$\pm$~20 & 231~$\pm$~7 & 64~$\pm$~2 & 43~$\pm$~1 & 0.33~$\pm$~0.02 & 0.154~$\pm$~0.009\\ 
M83-1048 & H \textsc{ii} & 990~$\pm$~30 & 370~$\pm$~10 & 114~$\pm$~3 & 80~$\pm$~2 & 0.37~$\pm$~0.02 & 0.20~$\pm$~0.01\\ 
M83-1067 & H \textsc{ii} & 870~$\pm$~30 & 311~$\pm$~9 & 72~$\pm$~2 & 51~$\pm$~2 & 0.36~$\pm$~0.02 & 0.141~$\pm$~0.008\\ 
M83-1104 & H \textsc{ii} & 580~$\pm$~20 & 234~$\pm$~7 & 38~$\pm$~1 & 27.0~$\pm$~0.8 & 0.41~$\pm$~0.02 & 0.112~$\pm$~0.007\\ 
M83-1137 & H \textsc{ii} & 370~$\pm$~10 & 169~$\pm$~5 & 38~$\pm$~1 & 26.4~$\pm$~0.8 & 0.46~$\pm$~0.03 & 0.17~$\pm$~0.01\\ 
M83-1247 & H \textsc{ii} & 810~$\pm$~20 & 331~$\pm$~10 & 79~$\pm$~2 & 58~$\pm$~2 & 0.41~$\pm$~0.02 & 0.17~$\pm$~0.01\\ 
M83-1724 & Neb+star & 30~$\pm$~3 & 25~$\pm$~2 & -- & -- & 0.8~$\pm$~0.2 & --\\ 
NGC7793-55 & H \textsc{ii} & 1740~$\pm$~50 & 266~$\pm$~8 & 99~$\pm$~3 & 74~$\pm$~2 & 0.153~$\pm$~0.009 & 0.100~$\pm$~0.006\\ 
NGC7793-68 & H \textsc{ii} & 3700~$\pm$~200 & 420~$\pm$~20 & 163~$\pm$~8 & 120~$\pm$~6 & 0.12~$\pm$~0.01 & 0.077~$\pm$~0.008\\ 
NGC7793-76 & K I+H \textsc{ii} & 1400~$\pm$~40 & 169~$\pm$~5 & 87~$\pm$~3 & 63~$\pm$~2 & 0.121~$\pm$~0.007 & 0.108~$\pm$~0.006\\ 
NGC7793-82 & H \textsc{ii} & 1730~$\pm$~90 & 197~$\pm$~10 & 87~$\pm$~4 & 63~$\pm$~3 & 0.11~$\pm$~0.01 & 0.087~$\pm$~0.009\\ 
NGC7793-84 & H \textsc{ii}+H$\alpha$ & 2440~$\pm$~70 & 400~$\pm$~10 & 147~$\pm$~4 & 104~$\pm$~3 & 0.16~$\pm$~0.01 & 0.103~$\pm$~0.006\\ 
NGC7793-96 & H \textsc{ii} & 3300~$\pm$~200 & 230~$\pm$~10 & 143~$\pm$~7 & 110~$\pm$~6 & 0.07~$\pm$~0.007 & 0.077~$\pm$~0.008\\ 
NGC7793-221 & H \textsc{ii}+H$\alpha$ & 4200~$\pm$~100 & 190~$\pm$~6 & 139~$\pm$~4 & 101~$\pm$~3 & 0.045~$\pm$~0.003 & 0.057~$\pm$~0.003\\ 

\hline


\end{longtable}

\end{document}